\documentclass[useAMS, usenatbib, usegraphicx]{mn2e}
\usepackage{amssymb}
\usepackage[colorlinks=TRUE, dvips]{hyperref}

\def\chisq{\hbox{$\chi^2$}}
\def\chisqr{\hbox{$\chi^2_r$}}
\def\msun{\hbox{${\rm M}_{\odot}$}}
\def\rsun{\hbox{${\rm R}_{\odot}$}}
\def\mstar{\hbox{$M_{\star}$}}
\def\rstar{\hbox{$R_{\star}$}}

\def\sn{\hbox{S/N}}

\def\RV{\hbox{${\rm RV}$}}

\def\kms{\hbox{km\,s$^{-1}$}}

\def\vsini{\hbox{$v\sin i$}}
\def\rsini{\hbox{$R\sin i$}}
\def\ptt{\hbox{$10^{-4} I_{\rm c}$}}
\def\degr{\hbox{$^\circ$}}
\def\Prot{\hbox{$P_{\rm rot}$}}

\def\d{\hbox{$\rm d$}}
\def\kG{\hbox{$\rm kG$}}

\def\dOmsun{\hbox{$d\Omega_{\odot}$}}

\def\ie{i.e. }   
\def\eg{e.g., }   

\newcommand{\aap}{A\&A}

\newcommand{\aj}{AJ}
\newcommand{\apj}{ApJ}
\newcommand{\apjl}{ApJL}
\newcommand{\apjs}{ApJS}
\newcommand{\apss}{Ap\&SS}

\newcommand{\mnras}{MNRAS}
\newcommand{\pasp}{PASP}

\newcommand\avg[1]{\left\langle #1 \right\rangle}
\begin{document}
\title[Magnetic topologies of late M dwarfs] 
{Large-scale magnetic topologies of late M dwarfs\thanks{Based on
observations
obtained at the Canada-France-Hawaii  Telescope (CFHT). CFHT is operated
by the National Research Council of Canada, the Institut National des
Science de l'Univers of the Centre National de la Recherche Scientifique
of France (INSU/CNRS), and the University of Hawaii.}}
\def\newauthor{%
  \end{author@tabular}\par
  \begin{author@tabular}[t]{@{}l@{}}}
\makeatother
\author[J. Morin et al.] {\vspace{1.7mm}
J.~Morin$^{1,\star}$\thanks{E-mail: jmorin@cp.dias.ie}
J.-F.~Donati$^1$, P.~Petit$^1$, X.~Delfosse$^2$, T.~Forveille$^2$,
M.M.~Jardine$^3$ \\
$^1$ LATT, Universit\'e de Toulouse, CNRS, 14 Av.\ E.~Belin, F--31400
Toulouse, France\\
$^\star$ \textit{Now at: }Dublin Institute for Advanced Studies, School of
Cosmic Physics, 31 Fitzwilliam Place, Dublin 2, Ireland\\
$^2$ Universit\'e. J.~Fourier (Grenoble 1)/CNRS; Laboratoire d'Astrophysique de
Grenoble (LAOG, UMR 5571); F--38041 Grenoble, France\\
$^3$ School of Physics and Astronomy, Univ.\ of St~Andrews, St~Andrews,
Scotland KY16 9SS, UK \\
}
\date{\today,~ Revision:2.02}
\maketitle

\begin{abstract}
We present here the final results of the first spectropolarimetric survey of a
small sample of active M dwarfs, aimed at providing observational constraints on
dynamo action on both sides of the full-convection threshold (spectral type M4).
Our two previous studies \cite[][]{Donati08b, Morin08b} were focused on early
and mid M dwarfs. The present paper examines 11 fully convective late M
dwarfs (spectral types M5--M8). Tomographic imaging techniques were applied to
time-series of circularly polarised profiles of 6 stars, in order to infer their
large-scale magnetic topologies. For 3 other stars we could not produce such
magnetic maps, because of low variability of the Stokes~V signatures, but were
able to derive some properties of the magnetic fields.

We find 2 distinct categories of magnetic topologies: on the one hand
strong axisymmetric dipolar fields (similar to mid M dwarfs), and on the other
hand weak fields generally featuring a significant non-axisymmetric component,
and sometimes a significant toroidal one. Comparison with unsigned
magnetic fluxes demonstrates that the second category of magnetic fields shows
less organization (less energy in the large scales), similarly to partly
convective early M dwarfs. Stars in both categories have similar stellar
parameters, our data do not evidence a separation between these 2 categories in
the mass-rotation plane.

We also report marginal detection of a large-scale magnetic field on the M8 star
VB~10 featuring a significant toroidal axisymmetric component, whereas no field
is detectable on VB~8 (M7).

\end{abstract}

\section{Introduction} 
\label{sec:intro}
Magnetic field is a key parameter in theories of stellar formation and
evolution. In cool stars, it powers several activity phenomena, observed
on a wide range of wavelengths and timescales, which provide a rough proxy
of the averaged field strength. The well-established rotation-activity
relation \cite[e.g., ][]{Noyes84} supports the idea of a
dynamo-generated magnetic field in these stars.

\cite{Larmor19} first proposed that the solar magnetic field could be
induced by plasma motions, already pointing out the importance of shear
($\Omega$ effect) to generate strong toroidal fields and thus explain
Hale's polarity law of sunspots. \cite{Parker55} completed the basic
picture of the solar dynamo by introducing the $\alpha$ effect (convection
made cyclonic by the Coriolis force) to address the issues of Cowling's
anti-dynamo theorem \citep{Cowling34} and the regeneration of a poloidal
field component from a toroidal one. The $\alpha\Omega$ dynamo has
thereafter been thoroughly debated and improved \citep[e.g. ][]{Babcock61,
Leighton69}. More recently, helioseismology has been able to probe the
solar interior and revealed the existence of a thin layer of strong shear
located at the base of the convection zone: the tachocline
\cite[e.g.,][]{Spiegel92}. Although many aspects of the solar magnetism are
still not thoroughly understood, recent theoretical and numerical studies
have pointed out the crucial role of the tachocline as the place of
storage and amplification of strong toroidal fields \cite[e.g.,
][]{Ossendrijver03, Charbonneau05}.

New insight on dynamo processes can be gained from the exploration of magnetic
fields of cool stars, probing the dynamo response in very non-solar regimes of
parameters (e.g., fast rotation, deeper or shallower convection zone). M dwarfs
are particularly interesting since those below $\sim 0.35~\msun$
\cite[e.g.,][]{Chabrier97} are fully convective and therefore do not possess a
tachocline and presumably cannot host a solar-type dynamo. Yet, many M
dwarfs are known to be active, and these stars follow the usual
rotation-activity relation \cite[e.g., ][]{Delfosse98, Reiners07, Kiraga07,
West04}. However the strong correlation between X-ray and radio luminosities
established by \cite{Guedel93} for stars of spectral types ranging form F to mid
M, is no longer valid for very low mass dwarfs which exhibit very
strong radio emission whereas X-ray emissions dramatically drop
\cite[][]{Berger06}. Magnetic fields were also directly detected at photospheric
level through Zeeman effect both in unpolarised \cite[e.g., ][]{Saar85, Johns96,
Reiners06} and circularly polarised \citep{Donati06} line profiles.

Spectropolarimetry combined with tomographic imaging techniques is the optimal
technique to investigate the magnetic topologies of M dwarfs (see
Sec.~\ref{sec:model} for more details). By recovering the large-scale component
of stellar magnetic fields, we can provide dynamo theorists with observables
directly comparable with their modeling (axisymmetry, relative importance of the
poloidal and toroidal components, characteristic scales...). Previous results
have already produced strong constraints: \cite{Donati06} and \cite{Morin08a}
(hereafter M08a) demonstrated that the fully convective fast rotator V374~Peg is
able to trigger a strong large-scale axisymmetric poloidal field steady on a
timescale of 1~year; and exhibits a very low level of differential rotation
($\sim \frac{\dOmsun}{10}$).  From the analysis of a sample of early and mid M
dwarfs  \cite{Donati08b} and \cite{Morin08b} (hereafter D08 and M08b
respectively) observed a strong change near the theoretical full-convection
threshold: while partly convective stars possess a weak non-axisymmetric field
with a significant toroidal component, fully convective ones exhibit strong
poloidal axisymmetric dipole-like topologies. Differential rotation also drops
by an order of magnitude across the boundary, the observed fully convective
stars exhibit nearly solid body rotation.
D08 and M08b also
report a sharp transition in the rotation-large-scale magnetic field 
relation close to the full-convection boundary, whereas no such gap is visible
in the rotation-X-ray relation. Considering that X-ray emission is a
good proxy for the total magnetic energy, it suggests that a sharp transition in
the characteristic scales of the magnetic field occurs near the full-convection
limit. This point was further confirmed for a few stars by \cite{Reiners09b} who
report that the ratio of the magnetic fluxes measured from circularly-polarised
and unpolarised lines dramatically changes across the fully convective limit.
Several theoretical studies have addressed the challenging issue of dynamo
action in fully convective stellar interiors. \cite{Durney93} first proposed
that without a tachocline of shear, convection and turbulence should play the
main role at the expanse of differential rotation, generating a small-scale
field. \cite{Kuker99} and \cite{Chabrier06} performed mean-field modeling of
dynamo action in fully convective stars and found purely non-axisymmetric
$\alpha^2$ solutions, indicating that these objects can sustain large-scale
magnetic fields. Subsequent direct numerical simulations by \cite{Dobler06} and
\cite{Browning08} both realized large-scale dynamo action with a significant
axisymmetric component of the resulting magnetic field. The latter also achieved
magnetic energy in equipartition with kinetic energy and therefore Maxwell
stresses strong enough to quench differential rotation, resulting in nearly
solid-body rotation. Despite these recent advances, the precise causes of
differences between dynamo in fully and partly convective stars are not
completely understood, and theoretical studies need observational guidance.

In the present paper, we extend our spectropolarimetric study to 11 late M
dwarfs (spectral types ranging from M5 to M8). After a brief presentation of the
stellar sample and of spectropolarimetric observations, we describe the main
principles of the tomographic imaging process. The Zeeman-Doppler Imaging (ZDI
hereafter) analysis is then detailed for 6 stars. For 3 other late M
dwarfs, it is not possible to derive a definitive magnetic map because of the
very low level of variability in the Stokes~$V$ signatures, but we can still
infer some information about their magnetic topologies. For the very faint
stars  VB~8 and VB~10, the noise level is too high to allow definite detection
of the circularly polarised signatures in individual LSD spectra. By averaging
the spectra of each data set, we marginally detect a large-scale magnetic field
on VB~10, and show a tentative ZDI reconstruction. We finally discuss these
results and conclude on the implications of our study for the understanding of
dynamo processes in fully convective stars.

\section{Observations}
\subsection{Presentation of the sample}
\label{sec:obs-sample}
For this first spectropolarimetric survey, we selected 23 active main-sequence
M dwarfs, mostly from the rotation-activity study by \cite{Delfosse98},
covering a wide range of masses and rotation periods (although for a given mass
the extent in rotation period is rather restricted). In the present paper we
focus on the low-mass end of the sample (0.08--0.20~\msun): GJ~51, WX~UMa,
DX~Cnc, GJ~1245~B, GJ~1156 and GJ~3622 are thoroughly studied with tomographic
imaging techniques. We also present a brief study of the large-scale magnetic
topologies of GJ~1224, GJ~1154~A, CN~Leo, VB~8 and VB~10
(section~\ref{sec:other}).

All these stars are known to show signs of activity in H$\alpha$ or X-rays
\cite[e.g.][]{Gizis02, Schmitt04}, and photospheric magnetic fields have been
measured on most of them from the analysis of Zeeman broadening in molecular
bands (see below). The main properties of the sample,
inferred from this work or collected from previous ones, are shown in
Tab.~\ref{tab:sample}.

\begin{table*}
\begin{center}
\caption[]{Fundamental parameters of the stellar sample. Spectral types
 are taken from \cite{Reid95}. Formal error bars derived from our study
are mentioned between brackets for \mstar, \rsini, \rstar\ (1$\sigma$) and
\Prot\ (3$\sigma$), they apply to the last digit of the preceding number.
See Sec.~\ref{sec:obs-sample}
for more details and a discussion about uncertainties.}
\begin{tabular}{cccccccccccc}
 \hline
 Name & ST & \mstar & \vsini & $Bf$ &  $\Prot$  & $\tau_c$ & $Ro$ & \rsini &
\rstar & $i$ \\
 & & (\msun) & (\kms) & (\kG) & (d) & (d) &
 ($10^{-2}$) & (\rsun) & (\rsun) & (\degr) \\ 
 \hline
 GJ~51 & M5 & 0.21 (3) & 12 & -- & 1.02 (1) & 83 & 1.2 & 0.24 (2) & 0.22 (3) &
60 \\
GJ~1156 & M5 & 0.14 (1) & 17$^b$ & 2.1$^b$ & 0.491 (2) & 94 & 0.5 & 0.16 
($<1$)& 0.16 (1) & 60 \\
GJ~1245~B & M5.5 & 0.12 ($<1$) & 7$^a$ & 1.7$^a$ & 0.71 (1) & 97 & 0.7 & 0.10
(2) & 0.14 ($<1$) & 40 \\
WX~UMa & M6 & 0.10 ($<1$) & $5^b$ & $> 3.9^b$ & 0.78 (2) & 100 & 0.8 & $0.07$
(1)
& 0.12 ($<1$) & 40\\
DX~Cnc & M6 & 0.10 ($1$) & $13^a$  & 1.7$^a$ & 0.46 (1) & 100 & 0.5 & $0.07$ (1)
& 0.11 ($<1$) &
60\\
GJ~3622 & M6.5 & 0.09 ($<1$) & $3^c$ & -- & 1.5(2) & 101 & 1.5 & 0.09 (6) & 0.11
($<1$) & 60 \\
\hline
GJ~1154~A & M5 & 0.18 (1) & 6$^b$ & 2.1$^b$ & $\leq1.7 $&88 &$\leq1.9$
&--& 0.20 (1)  &--\\
GJ~1224 & M4.5 & 0.15 (1) & $\leq 3^a$ & 2.7$^a$ & $\leq4.3$ & 93 &$\leq4.6$
&--& 0.17 ($1$) &--\\
CN~Leo & M5.5 & 0.10 ($<1$) & 3$^a$ & 2.4$^a$ & $\leq2.0$ & 99 &$\leq2.0$
&--& 0.12 ($<1$) &--\\
VB~8 & M7 & 0.09 ($<1$) & 5$^a$ & 2.3$^a$ & $\leq1.0$ & 101 &$\leq1.0$ &--& 
0.10 ($<1$) &--\\
VB~10 & M8 & 0.08 ($<1$) & 6$^a$ & 1.3$^a$ & $\leq0.8$ & 102 &$\leq0.8$
&--& 0.09 ($<1$) &--\\
\hline
\label{tab:sample}
\end{tabular}
\end{center}
{\flushleft
$^a$ \cite{Reiners07}\\
$^b$ \cite{Reiners09}\\
$^c$ \cite{Mohanty03}\\
}
\end{table*}

Stellar masses are computed from the mass-luminosity relation derived by
\cite{Delfosse00}, based on $J$ band absolute magnitude inferred from apparent
magnitude measurements of 2MASS \citep[][]{Cutri03} and \emph{Hipparcos}
parallaxes \citep[][]{ESA97}. Formal error bars, as derived from
uncertainties on these measurements are mentioned. The
intrinsic dispersion of the relation is estimated to be lower than 10~$\%$.
Radius and bolometric luminosity suited to the stellar mass are computed from
NextGen models \citep[][]{Baraffe98}, formal error-bars on stellar mass
are propagated. The accuracy of these models for active M dwarfs is a debated
subject \cite[\eg][]{Ribas06}, but recent studies indicate that the agreement
with observations is very good for late M dwarfs \cite[][]{Demory09}.
For all stars except GJ~51, projected rotational velocities (\vsini)
are available from previous spectroscopic studies. The uncertainties on \vsini\
are typically equal to 1~\kms. For each star, we also mention the
rotation period (\Prot) derived from our analysis (see
Sec.~\ref{sec:techniques-period}), and \rsini\ is straightforwardly deduced
(with propagated error bar). An estimate of the inclination angle of the
rotation axis on the line-of-sight ($i$) is obtained by comparing \rsini\ and
the theoretical radius. With this estimate the typical error is of the order of
10\degr\ for low and moderate inclinations, and 20\degr\ for high
inclination angles, this is precise enough for the imaging process. The effect
of these uncertainties on the reconstructed magnetic maps is discussed in
Sec.~\ref{sec:techniques-uncert}.

We also mention unsigned magnetic fluxes from the literature (whenever
available in \citealt{Reiners07} and \citealt{Reiners09}) empirically derived
from unpolarised molecular (FeH) line profiles. These estimates result from the
comparison with reference spectra of an active and an inactive star (corrected
for spectral type and rotational broadening) which are used to calibrate the
relation between broadening of magnetically sensitive lines and magnetic flux
\cite[][]{Reiners06}. The authors estimate that the precision of the
method lies in the 0.5--1~\kG\ range. As this method is not sensitive to the
vector properties of the magnetic field, the $Bf$ values reflect the overall
magnetic flux on the surface of the star. Since Stokes~$V$ signatures
corresponding to neighbouring zones of fields with opposite polarities cancel
each other, our spectropolarimetric measurements are not sensitive to tangled
fields and only recover the uncancelled magnetic flux corresponding to the
large-scale component of the magnetic topology. Therefore the ratio of both
magnetic fluxes is a clear indication of the degree of organization of the
observed magnetic field.

The convective Rossby number ($Ro$), which is the ratio of the rotation
period and the convective turnover time, is believed to be the relevant
parameter to study the impact of rotation on dynamo action in cool stars
\cite[][]{Noyes84}. In Table~\ref{tab:sample}, we mention Rossby numbers based
on empirical convective turnover times derived by \cite{Kiraga07} from the
rotation--activity relation in X-rays. These turnover times are \textit{ad-hoc}
fitting parameters that reflect more the relation between activity and rotation
at a given mass than an actual turnover time at a specific depth in the
convection zone. However, the resulting Rossby numbers allow us to compare the
effect of rotation on magnetic field generation in stars having different
masses.

\subsection{Instrumental setup and data reduction}
\label{sec:obs-red}
Observations presented here were collected between June 2006
and July 2009 with the ESPaDOnS spectropolarimeter at CFHT . ESPaDOnS provides
full coverage of the optical domain (370 to 1,\,000~nm) in a single exposure, at
a resolving power of 65,\,000, with a peak efficiency of 15\% (telescope and
detector included). 

Data Reduction is carried out with \textsc{libre-esprit}, a fully-automated
dedicated pipeline provided to ESPaDOnS and NARVAL users, that performs optimal
extraction of the spectra following the procedure described in \cite{Donati97}
that is based on \cite{Horne86} and \cite{Marsh89}. Each set of 4 individual
sub-exposures taken in different polarimetric configuration are combined
together to produce Stokes~$I$ (unpolarised intensity) and $V$ (circularly
polarised) spectra, so that all spurious polarisation signatures are cancelled
to first order \citep{Semel93, Donati97}.  In addition, all spectra are
automatically corrected for spectral shifts resulting from instrumental effects
(e.g., mechanical flexures, temperature or pressure variations) using telluric
lines as a reference. Though not perfect, this procedure allows spectra to be
secured with a radial velocity (RV) internal precision of better than
$0.030~\kms$ \citep[e.g.,][]{Moutou07}.

The peak signal-to-noise ratios (\sn ) per 2.6~\kms\ velocity bin range from 51
to 245, mostly depending on the magnitude of the target and the weather
conditions. An overview of the observations is presented in Table~\ref{tab:obs},
the full journal of observations is available in the electronic
version.
Using the Least-squares deconvolution technique
\cite[LSD, ][]{Donati97}, polarimetric information is extracted from most
photospheric atomic lines and gathered into a single synthetic profile of
central wavelength $\lambda_0 = 750~{\rm nm}$ (800~nm for VB~8 and 10). The
corresponding effective Land\'e factor $g_{\rm eff}$ (computed as a weighted
average on available lines) is close to 1.2 for all the stars of our
sample. The line list for LSD was computed from an Atlas9 local thermodynamic
equilibrium model \citep{Kurucz93}
matching the properties of our whole sample, and contains about
$5,\,000$ moderate to strong atomic lines. We notice a multiplex gain of about
10 (5 for VB~8 and 10) with respect to the peak \sn\ of the
individual spectra of our sample. Although all the stars in the sample are
active, some exhibit Stokes~$V$ LSD signatures just above noise level (e.g. 
DX~Cnc), whereas on others we detect very strong signatures, with peak-to-peak
amplitudes as high as 1.8\% of the unpolarised continuum level (for WX~UMa). 
Temporal variations, due to rotational modulation, of the Zeeman signatures is
obvious for some stars, whereas it is very weak on others, depending \eg on the
inclination angle of their rotation axis with respect to the line of sight,
the complexity and the degree of axisymmetry of their magnetic topology.

For each observation we compute the corresponding longitudinal magnetic
field (i.e. the line of sight projection) from the Stokes $I$ and $V$
LSD profiles through the relation:

\begin{equation}
 B_l({\rm G}) = -2.14 \times 10^{11} \frac{\displaystyle\int v\,V(v)
\,{\rm d}v}{\lambda_0\,g_{\rm eff}\,c \displaystyle\int
\left[I_c-I(v)\right] {\rm d}v } \, ,
\label{eq:bl}
\end{equation}
\citep[][]{Rees79, Donati97, Wade00} where $v$ is the radial velocity in
the star's rest frame, $\lambda_0$, in nm, is the mean wavelength of the
LSD profile, $c$ is the velocity of light in vacuum in the same unit as
$v$, $g_{\rm eff}$ is the value of the mean Land\'e factor of the
LSD line, and $I_c$ the continuum level.

In the rest of the paper, all data are phased according to the following
ephemeris:
\begin{equation}
{\rm HJD} = {\rm HJD}_0 + \Prot E,
\label{eq:eph}
\end{equation}
where ${\rm HJD}_0=2\,453\,850$ for WX~UMa, and ${\rm HJD}_0=2\,453\,950$ for
the other stars; and $\Prot$ is the rotational period used as an input for ZDI
and given in Table~\ref{tab:sample}.

\begin{table*}
\begin{center}
\caption[]{Synthetic journal of observations. Observation year and
number of spectra collected are given in columns 2 and 3. Columns 4
and 5 respectively list the peak signal to noise ratio (per 2.6~\kms\
velocity bin) and the rms noise level (relative to the unpolarised
continuum level and per 1.8~\kms\ velocity bin) in the average circular
polarisation profile produced by Least-Squares Deconvolution (see
text) --- we precise minimum and maximum values obtained
for each observing run. The average value and standard deviation of
the longitudinal magnetic field (see Eq.~\ref{eq:bl}) and the radial velocity
measurements are given in columns 6 and 7. The rotation cycle bounds
of column 8 are computed with the rotation periods mentioned in
Table~\ref{tab:sample}. Complete observation logs are available in the
electronic version of the article.}
\begin{tabular}{cccccccc}
\hline
Name & Year & $n_{obs}$ & \sn & $\sigma_{\rm LSD}$ & $B_{\ell}$ & RV &
Cycle \\
 &  & & & (\ptt) & (G) & (\kms) & \\
\hline 
GJ~51 & 2006 & 6 & 128--165 & 7.7--10.1 & -990 (313) & -5.52 (0.20) &
5.0--9.9\\
-- & 2007 & 9 & 159--198 & 5.7--7.0 & -1657 (280) & -6.36 (0.74) &
412.7--418.7\\
-- & 2008 & 9 & 118--181 & 6.9--10.3 & -1219 (407) & -6.60 (0.57) &
788.0--794.1\\
GJ~1156 & 2007 & 6 & 120--181 & 6.8--11.0 & 82 (72) & 5.96 (0.24) &
431.9--442.0\\
-- & 2008 & 5 & 127--158 & 10.4--8.4 & -47 (166) & 5.81 (0.13) &
1091.6--1095.9\\
-- & 2009 & 9 & 183--195 & 6.5--6.9 & 24 (111) & 5.73 (0.24) &
1812.7--1817.0\\
GJ~1245~B & 2006 & 6 & 158--191 & 7.1--8.8 & -52 (163) & 5.42(0.11) &
4.2--14.0\\
-- & 2007 & 6 & 182--226 & 5.7--7.4 & -17 (128) & 5.38 (0.10) &
597.0--601.2\\
-- & 2008 & 10 & 138--194 & 7.2--10.1 & -6 (67) & 5.46 (0.09) &
1054.6--1057.8\\
WX~UMa & 2006 & 8 & 67--142 & 19.8--9.6 & -1506 (453) & 70.25 (0.53)
& 0.5--4.4\\
-- & 2007 & 6 & 115--154 & 8.4--11.8 & -1757 (405) & 69.95 (0.06) &
341.8--349.5\\
-- & 2008 & 4 & 63--129 & 10.4--21.4 & -1811 (271) & 70.15 (0.24) &
755.8--758.6\\
-- & 2009 & 11 & 113--163 & 8.0--12.5 & -1496 (271) & 69.83 (0.21) &
1214.9--1218.9\\
DX~Cnc & 2007 & 5 & 119--179 & 8.3--12.3 & 132 (76) & 10.55 (0.07) &
460.7--471.5\\
-- & 2008 & 7 & 90--161 & 9.4--16.7 & 92 (52) & 10.44 (0.56) &
1160.6--1169.5\\
-- & 2009 & 9 & 106--187 & 7.8--14.8 & 67 (44) & 10.67 (0.10) &
2012.8--2019.6\\
GJ~3622 & 2008 & 8 & 128--167 & 8.9--11.1 & -32 (29) & 2.27 (0.28) &
397.2--402.6 \\
-- & 2009 & 6 & 101--162 & 9.3--15.0 & -26 (26) & 2.37 (0.05) &
617.3--620.0\\ 
\hline
GJ~1154~A & 2007 & 6 & 118--167 & 7.0--10.5 & -714 (76) & -12.83 (0.11) & -- \\
          & 2008 & 4 & 86--154 & 8.3--15.9 & -700 (75) & -12.92 (0.18) & -- \\
GJ~1224 & 2008 & 12 & 51--185 & 5.8--19.7 & -563 (41) &  -32.68 (0.04) & -- \\
CN~Leo & 2008 & 4 & 172--245 & 4.5--6.7 & -691 (54) & 19.62 (0.05) & -- \\
VB~8 & 2009 & 9 & 83--107 & 15.7--20.0 & 29 (53) & 15.39 (0.11) & -- \\
VB~10 & 2009 & 9 & 68--80 & 22.5--25.7 & 58 (61) & 36.23 (0.14) & -- \\
\hline 
\label{tab:obs}
\end{tabular}
\end{center}
\end{table*}

\section{Data Modelling}
\label{sec:model}
Zeeman Doppler imaging \cite[][]{Donati97b} aims at assessing stellar magnetic
topologies (at photospheric level), from the analysis of time-series of high
spectral resolution spectropolarimetric observations. In this part we briefly
remind the reader with the main properties of this technique and details of our
implementation. A more complete description can be found in M08b and references
therein.

ZDI is an inverse problem, the associated direct problem consists in computing
the Stokes~$I$ and $V$ spectra for a given magnetic map.
The stellar surface being sampled on a grid of $\sim1,000$ cells, the local
Stokes $I$ and $V$ profiles are computed from a model based on
Unno-Rachkovsky's equations, for a given magnetic map. The magnetic field is
decomposed into its poloidal and toroidal components and described as a set of
spherical harmonics-like coefficients, as implemented by \cite{Donati06b}. We
introduce two filling factors that account for subgrid cancellation of
polarised signatures corresponding to fields of opposite polarities, and allows
us to accurately reproduce the LSD line profiles (M08b). The stellar
spectrum is then obtained by a disk integration taking into account Doppler
shift and limb-darkening.

\subsection{Rotational modulation of polarised lines}
In the presence of a magnetic field one can observe the Zeeman effect
on spectral lines: (i) unpolarised lines (Stokes $I$) are broadened with respect
to the null field configuration and (ii) polarised signatures (Stokes $Q$, $U$
and $V$) show up. Here we only study circular polarisation (Stokes $V$), which
is sensitive to the strength and polarity of the line-of-sight
projection of the field. Because of the combination of stellar rotation and
Doppler effect:
\begin{itemize}
  \item The contribution of a photospheric region to the stellar spectrum is
correlated with its longitude (at first order). Thus, as a magnetic region
crosses the stellar disk under the effect of rotation, the corresponding
polarised signal migrates from the blue to the red wing of the line.
  \item The amplitude of this migration depends on the spot's latitude (no
migration for a polar spot, maximum migration from -\vsini\ to +\vsini\ for an
equatorial one).
  \item The evolution of the signature during this migration (as
the angle between the field vector and the line of sight changes with rotation)
is characteristic of the field orientation (e.g. signature of constant polarity
for radial field, and polarity reversal for azimuthal field).
\end{itemize}
Therefore, from a series of polarised spectra providing even and dense sampling
of stellar rotation, it is possible to reconstruct a map for the photospheric
magnetic field.

\subsection{Magnetic field reconstruction: inverse problem}
\label{sec:techniques-inverse}
Starting from a null-field configuration, the series of spectra computed from a
test field is iteratively compared to the observed one, until a given \chisqr\
level is reached. As the problem is partly ill-posed (several magnetic
configurations can match a data set equally well), the maximum entropy solution
is selected. The spatial resolution of ZDI depends on \vsini\ as a rule of
thumb. The highest degree available in the field reconstruction is:

\begin{equation}
  \ell_{max} \simeq \max(\frac{2\pi\vsini}{W}\,;\,\ell_{\min}) %
\label{eq:lmax}
\end{equation}
where $W$ is the unpolarised local profile width ($\sim 9~\kms$ for an inactive
non-rotating M dwarf, M08a). The first term in the max function
corresponds to the limit of high \vsini, when line broadening is mostly due to
rotation and the line profile can actually be seen as one 1-D map of the
photospheric magnetic field. The second term, $\ell_{\min}$ is the minimum
resolution available in the low \vsini\ limit, when Doppler shift is small and
the information on the field topology mostly comes from the temporal evolution
of the shape and amplitude of the polarised signatures. $\ell_{\min}$ can range
from 4 to 8 mostly depending on the signal to noise ratio of the data (D08,
M08b).

\subsection{Period determination}
\label{sec:techniques-period}
Since ZDI is based on the analysis of rotational modulation, important inputs
of the code are: the projected equatorial velocity (\vsini), the inclination
angle of the rotation axis with respect to the line of sight ($i$) and the
rotation period (\Prot). As no previous definite period measurement
exists for any star of our late M subsample, we use ZDI to provide a constraint
on this parameter. For each star, given the \vsini\ and the theoretical radius
(corresponding to the stellar mass) we can derive an estimate of the maximum
value for the rotation period as : 
\begin{equation}
  P_{\max} = 50.6145 \times \frac{\rstar}{\vsini} ,
\end{equation}
where $P_{\max}$ is expressed in days, $\rstar$ in unit of \rsun\ and \vsini\
in \kms. We test period values in a reasonable range ($<1.2\times P_{\max}$),
and derive the most probable period as the one resulting in the minimum \chisqr\
at a given informational content (i.e. a given averaged magnetic flux value).
We try to resolve aliases by comparing the multiple data sets available for each
star. 
By fitting a parabola to the resulting \chisqr\ curve close to the minimum, we
derive the optimal value of \Prot\ and the associated formal error bar (see
M08b for more details). Since several data sets are available for each star, we
mention the smallest 3$\sigma$ error bar in Tab.~\ref{tab:sample}. For all the
studied stars periods inferred from different data sets are compatible with each
other within the width of the associated error bars.

\subsection{Uncertainties on the magnetic maps}
\label{sec:techniques-uncert}
Due to the use of a maximum entropy method, magnetic maps reconstructed with ZDI
are optimal in the sense that any feature present in the map is actually
required to fit the data. However this method does not allow us to derive
formal error bars on the reconstructed maps.

Numerical experiments demonstrate that ZDI provides reliable maps and
is robust with respect to reasonable uncertainties on various parameters and 
data incompleteness \cite[e.g., ][]{Donati97b}. 
In addition, our implementation based on spherical harmonics and
poloidal/toroidal decomposition successfully reconstructs global topologies such
as low-degree multipoles, as well as more complex configurations \cite[e.g.,
][]{Donati06b}.
We also find that ZDI maps are robust with respect to the selected entropy
weighting scheme, provided that phase coverage is complete enough.

We try to assess the effects of the uncertainties on the input parameters (see
Sec.~\ref{sec:obs-sample}), on the derived magnetic quantities in a similar way
as \cite{Petit08}. For each data set we perform several reconstructions with
input parameters \vsini, $i$, and \Prot\ varying over the width of the error bars
and check the resulting map and its properties, thus providing a quantitative
analysis of the robustness of our reconstructions to these uncertainties. We
therefore obtain ``variability bars'' rather than formal error bars. Given
the small uncertainties on the rotation period, varying this parameter within
the 3$\sigma$ error bar has negligible effect on the reconstructed maps. The
magnetic quantities listed in Tab.~\ref{tab:syn} to characterize the repartition
of magnetic energy into different components are affected in different
ways by variations of the input parameters. In particular, the decomposition
between poloidal and toroidal energy is very robust, the observed variation is
less than 10~\% of the reconstructed magnetic energy. The fraction of
magnetic energy in axisymmetric modes varies by up to 20~\% due to
uncertainties on input parameters. The most important effect observed is a
cross-talk between the dipole component and higher degree multipoles (in
particular the quadrupole), the variation of the fraction of energy in the
dipole is in the 10--30\% range. The variability bars on the reconstructed
magnetic flux are of the order of 30~\%.

\section{GJ~51}
\label{sec:gj51}
\begin{figure*}
\begin{center}
  \includegraphics[height=0.40\textheight]{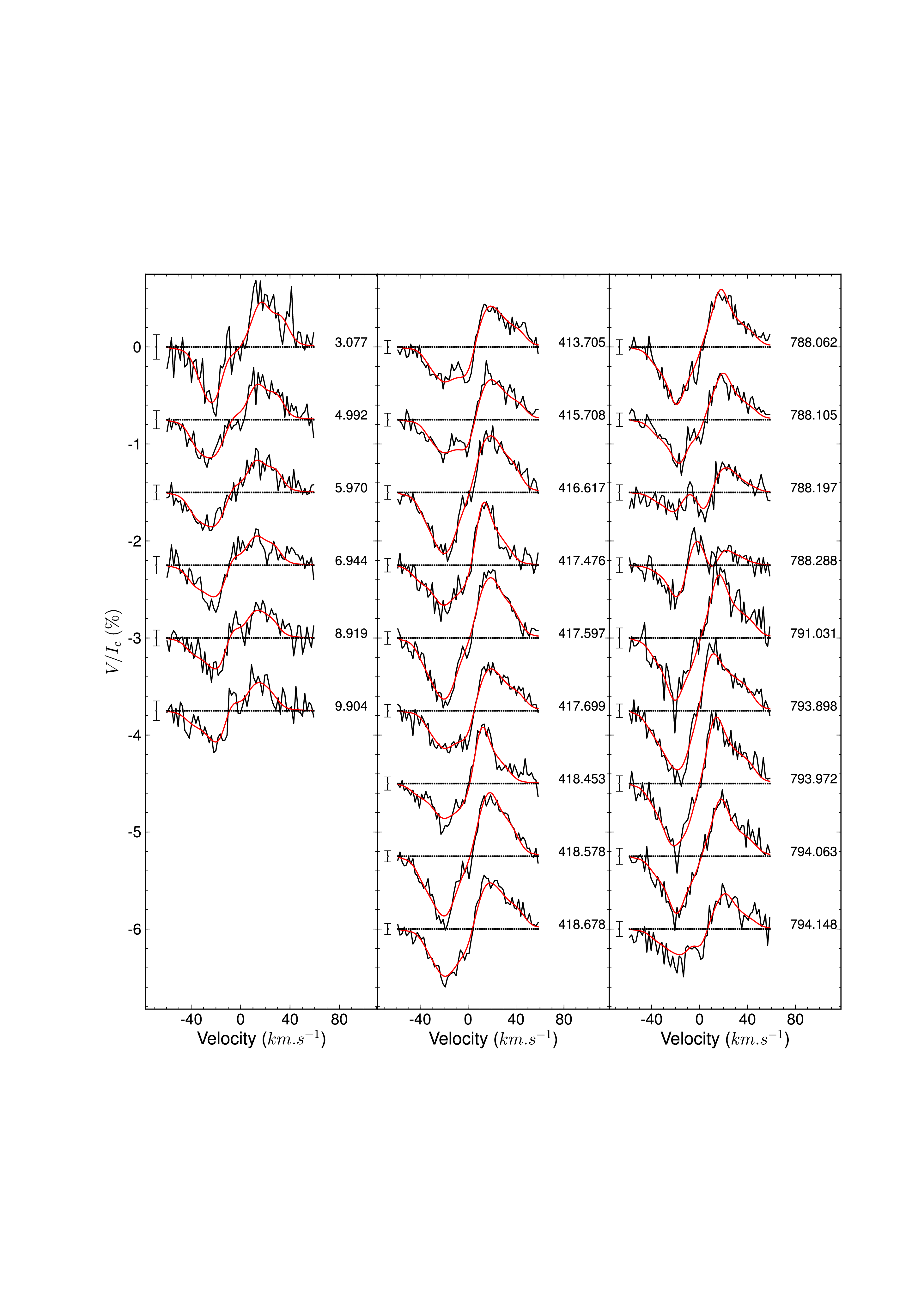}
\end{center}
 \caption[]{Time-series of Stokes $V$ profiles of GJ~51, in the
rest-frame of the star, from our 2006 (left-hand column), 2007 (middle
column) and 2008 (right-hand column) data sets. Synthetic profiles
corresponding to our magnetic models (red lines) are superimposed to the
observed LSD profiles (black lines). Left to each profile a
$\pm1\sigma$ error bar is shown. The rotational phase and cycle of each
observation is also mentioned right to each profile.
Successive profiles are shifted vertically for clarity purposes and the
associated reference levels ($V=0$) are plotted as dotted lines.}
\label{fig:gj51_spec}
\end{figure*}

\begin{figure*}
\begin{center}
\includegraphics[height=0.40\textheight]{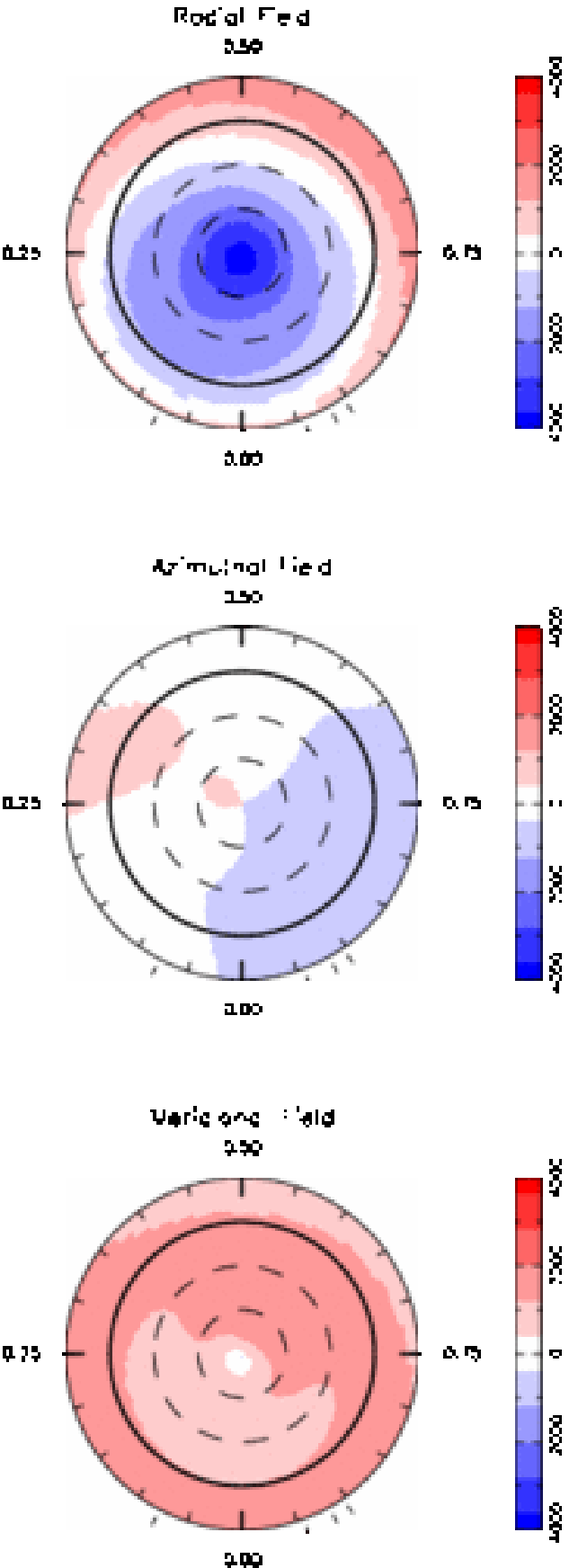}\hspace{
0.5cm }
\includegraphics[height=0.40\textheight]{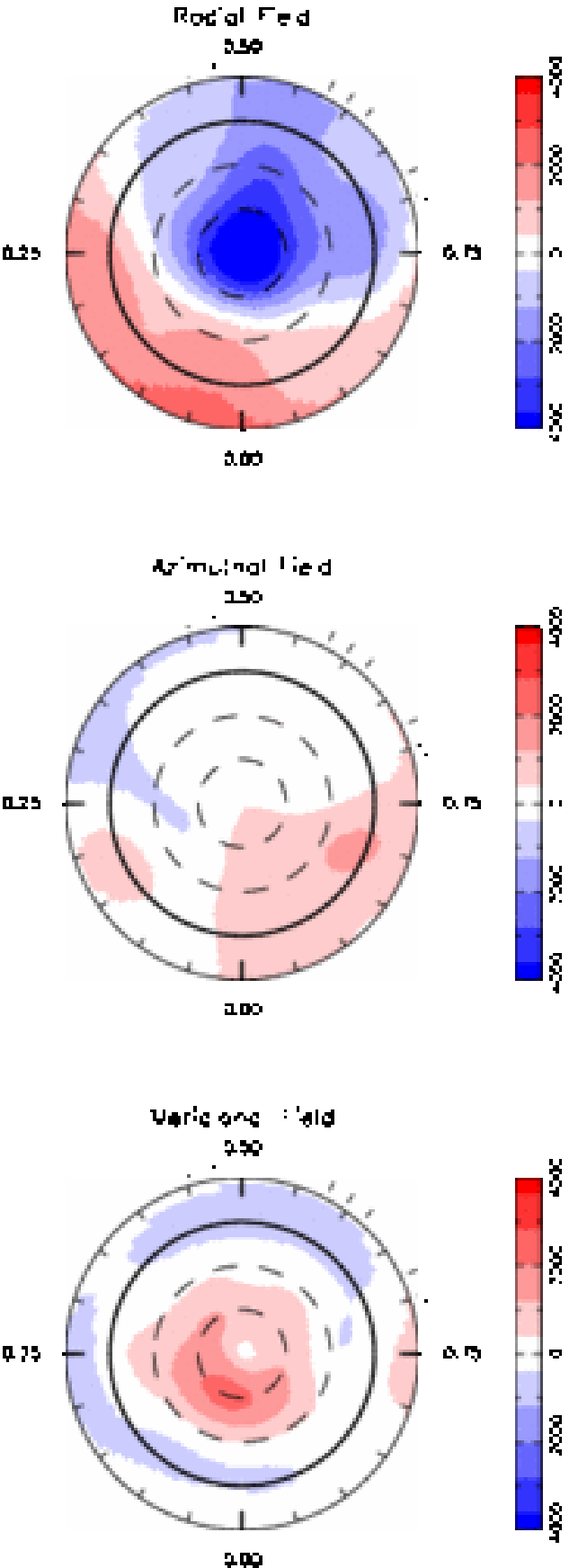}\hspace{
0.5cm }
  \includegraphics[height=0.40\textheight]{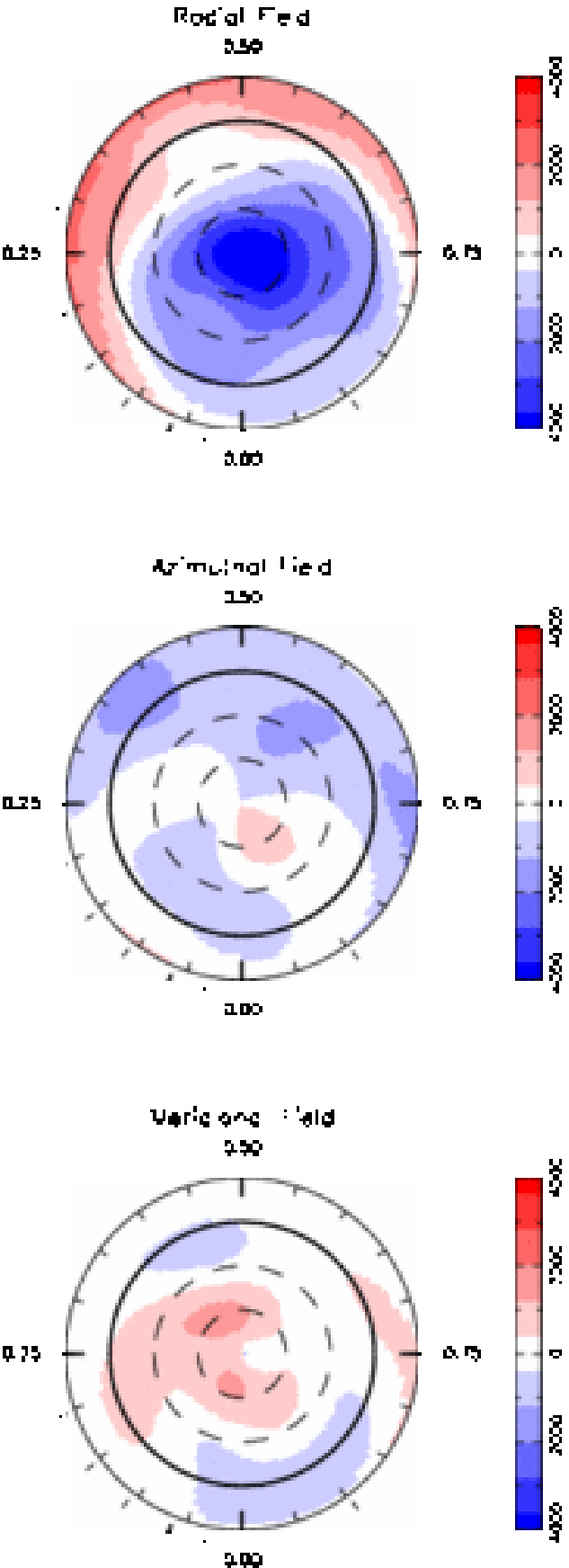}%
  \hspace{\stretch{4}}
\end{center}
\caption[]{Surface magnetic flux of GJ~51 as derived from our 2006, (left-hand
column), 2007 (middle column) and 2008 (right-hand column) data sets.
For GJ~51, the imaging process is adapted to preferentially converge
toward a mostly axisymmetric solution in order to resolve the ambiguity due to
poor phase coverage (see text). The three components of the field in spherical
coordinates are displayed from top to bottom (flux values labelled in G).  The
star is shown in flattened polar projection down to latitudes of $-30\degr$,
with the equator depicted as a bold circle and parallels as dashed circles.
Radial ticks around each plot indicate phases of observations.}
\label{fig:gj51_map}
\end{figure*}

We have acquired a total of 24 spectra on GJ~51 split in 3 series collected on 3
successive years (see Tab.~\ref{tab:obs}). All
Stokes~$V$ spectra exhibit a strong signature of constant polarity (radial field
directed toward the star). Temporal variation inside each data set is
detectable and likely due to rotational modulation. It mainly consists of an
evolution of the signature's amplitude. To our knowledge, no previously
published measurement of \Prot\ or \vsini\ exist for this star. From our LSD
Stokes $I$ profiles we measure mean RV values of $-5.52$, $-6.36$, and
$-6.60~\kms$ in
2006, 2007 and 2008, respectively. These values are in agreement with the
previously published $\RV=-7.3~\kms$ \cite[][]{Gizis02}. Our RV measurements
also reveal a drift between the 3 epochs, too large to be due to
convection and that may indicate a companion orbiting around this star. We also
observe strong RV temporal variations inside each data set presumably due to
magnetic activity (see Tab.~\ref{tab:obs}),
well above the intrinsic precision of the instrument (see
Sec.~\ref{sec:obs-red}). Our analysis of both the unpolarised and polarised
spectra leads us to $\vsini=12~\kms$ and $\Prot=1.02~\d$, this is close to the
1.06~d period derived from the MEarth\footnote{\cite{Irwin09}} photometric data
(J.~Irwin, private communication). From these values we infer
$\rsini=0.24~\rsun$, whereas for $\mstar=0.20~\msun$ evolutionary models
expect $\rstar=0.21~\rsun$. We conclude that the inclination angle of the
rotation axis with respect to the line-of-sight must be high, and set
$i=60\degr$ for ZDI. We note that for $\Prot=1.02~\d$ none of our data set
provide a complete sampling of the stellar rotation, the best one being
the 2008 data set which covers 30\% of the rotation cycle. Despite this poor
phase coverage, Stokes~$V$ profiles collected at the 3 epochs are similar,
suggesting that the magnetic field is stable and mostly axisymmetric.

\begin{table}
\begin{center}
\caption[]{Fit achieved for the 3 spectral time-series obtained on GJ~51. 
The imaging process is guided towards a mostly axisymmetric solution, see text.
In columns 2 we give the maximum degree of spherical harmonics used for ZDI
reconstruction. Columns 3--6 respectively list the initial \chisqr\ (i.e.
without magnetic field), the \chisqr\ achieved with the imaging process, and the
average and peak value of the magnetic flux on the reconstructed map.}
\begin{tabular}{cccccc}
\hline
Epoch & $\ell_{ZDI}$ & ${\chisqr}_0$ & ${\chisqr}_f$ &
$\avg{B}$ & $B_{max}$ \\
 & & & & (\kG) & (\kG) \\
\hline
2006 & 5 &  6.17 & 1.00 & 1.61 & 3.86 \\
2007 & 5 & 24.53 & 1.00 & 1.58 & 5.02 \\
2008 & 5 & 15.20 & 1.00 & 1.65 & 4.68 \\
\hline
\label{tab:gj51_fit}
\end{tabular}
\end{center}
\end{table}

Setting $\ell_{{\rm ZDI}}=5$ for the magnetic field decomposition, it is
possible to fit each of our 3 data sets with \chisqr=1. The resulting magnetic
maps feature a strong non-axisymmetric component with a dipole tilted toward the
observer, in particular those inferred from our 2006 and 2007 spectra. This is
surprising, it is indeed very unlikely that we observe three times GJ~51
at the same phase (when the magnetic pole is crossing the line of
sight).

The magnetic map reconstructed by ZDI is highly dependent on the precise
form of the entropy, whereas it is generally not the case, indicating that
this reconstruction is particularly ill-posed. We suggest this is due to
the lack of information associated with the poor phase coverage, the
reconstructed solution is strongly influenced by the maximum entropy constraint:
the resulting map is therefore mainly composed of a spot of radial field facing
the observer.

We perform another reconstruction of the magnetic field of GJ~51, with addition 
of \emph{a priori} information in the process, so that it preferentially
converges toward a mostly axisymmetric solution, as far as it allows to fit the
data at the prescribed \chisq\ level. This is done by putting a strong
entropy penalty on non-axisymmetric modes, similarly to the method used by
\cite{Donati08a} to drive the reconstructed topology towards antisymmetry with
respect to the center of the star. In these conditions, we can also fit our 3
data sets with \chisqr=1. The resulting magnetic maps for the 3 epochs are very
similar (Fig.~\ref{fig:gj51_map}). The corresponding synthetic spectra are
plotted along with the data on Fig.~\ref{fig:gj51_spec}. We find similar results
at all epochs: the magnetic topology is almost purely poloidal and axisymmetric,
it is mainly composed of a very strong dipole aligned with the rotation axis
(see tables \ref{tab:gj51_fit} and \ref{tab:syn} for more details). The
azimuthal and meridional component of the field somehow differ between the three
epochs, but we consider that this is not significant given the weak constraint
provided by our data. These maps are not the only solution permitted by our
data, but we believe that they represent the most probable one. 

Variability bars are of the same order of magnitude as for the other stars. The
almost purely poloidal nature of the magnetic field is robust to uncertainties
on the input parameters $i$ and \vsini. In addition, when varying these
parameters, the topology always features a strong purely axisymmetric component
(i.e. more than 45\% of the magnetic energy is reconstructed in $m=0$ modes) and
the main reconstructed mode is the radial component of a dipole aligned with the
rotation axis.
However, the values mentioned in Tab.~\ref{tab:syn} for GJ~51 should be
considered cautiously as our data sets provide a weak constraint and the
resulting magnetic maps are largely determined by the entropy function used.
In particular, the high values of longitudinal field measured (see
Tab.~\ref{tab:obs}) indicate that the large-scale magnetic flux is indeed higher
than those of mid-M dwarfs studied in M08b. But it is not clear whether the
large-magnetic flux of GJ~51 is actually larger than that of WX~UMa (see
Sec.~\ref{sec:wxuma}).
A definite confirmation
of the magnetic topology of GJ~51 requires multi-site observations to obtain a
complete sampling of the rotation cycle due to a period close to 1~\d. 

\section{GJ~1156}
\label{sec:gj1156}
\begin{figure*}
\begin{center}
  \includegraphics[height=0.40\textheight]{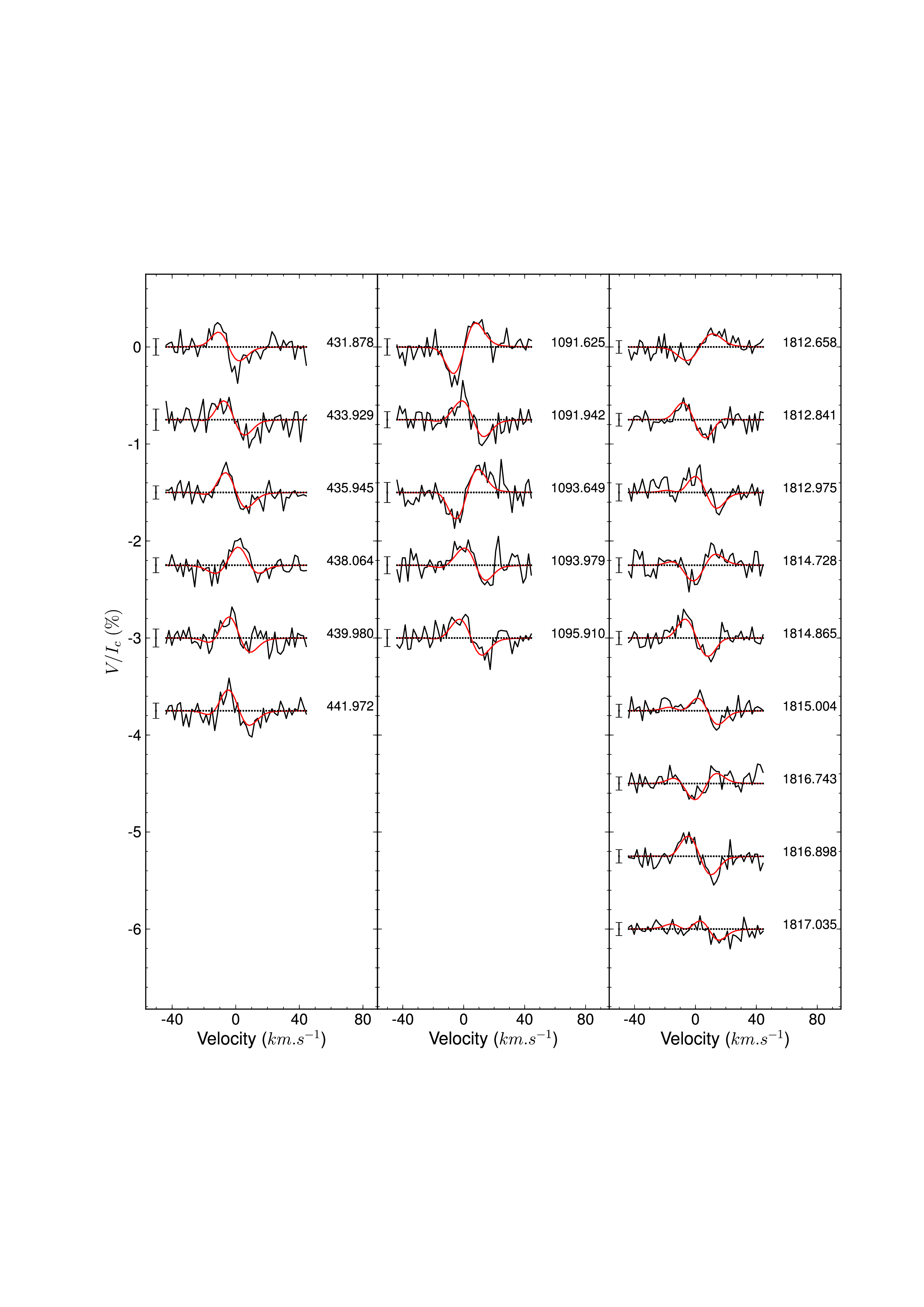}
\end{center}
\caption[]{Same as Fig.~\ref{fig:gj51_spec} for GJ~1156 2007, 2008,
and 2009 data sets (from left to right).}
\label{fig:gj1156_spec}
\end{figure*}

\begin{figure*}
\begin{center}
\includegraphics[height=0.40\textheight]{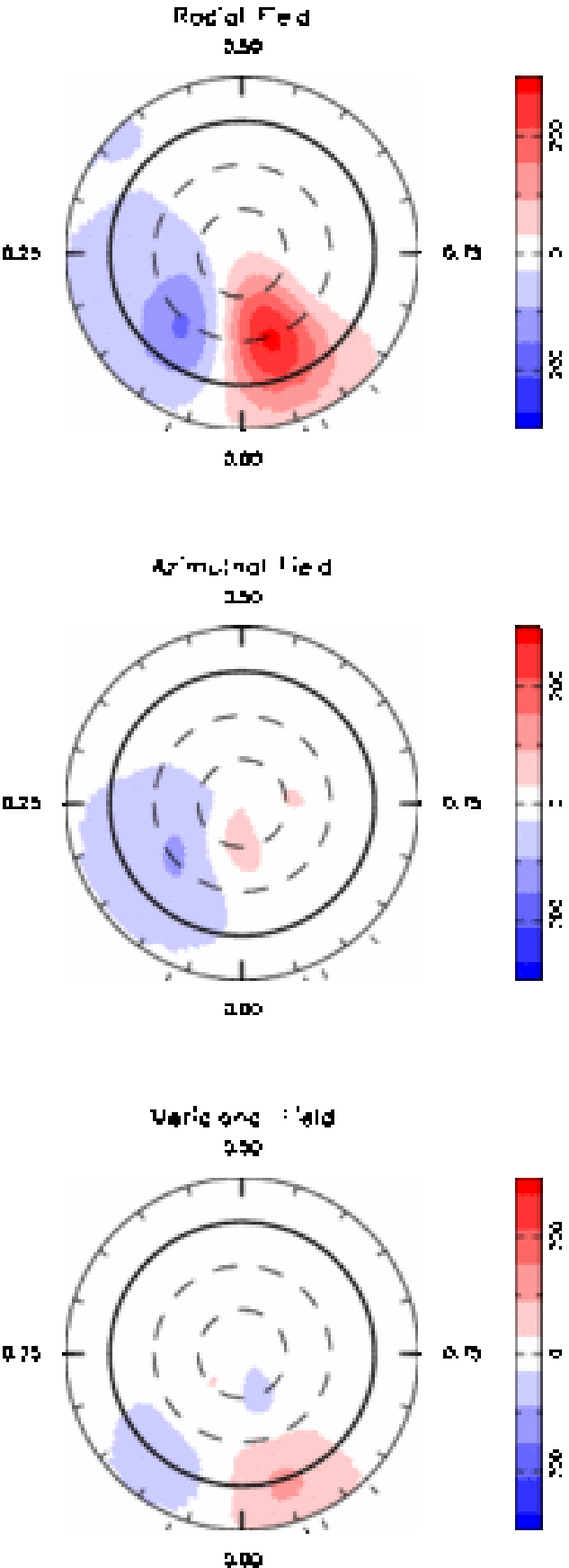}\hspace{
0.5cm}
\includegraphics[height=0.40\textheight]{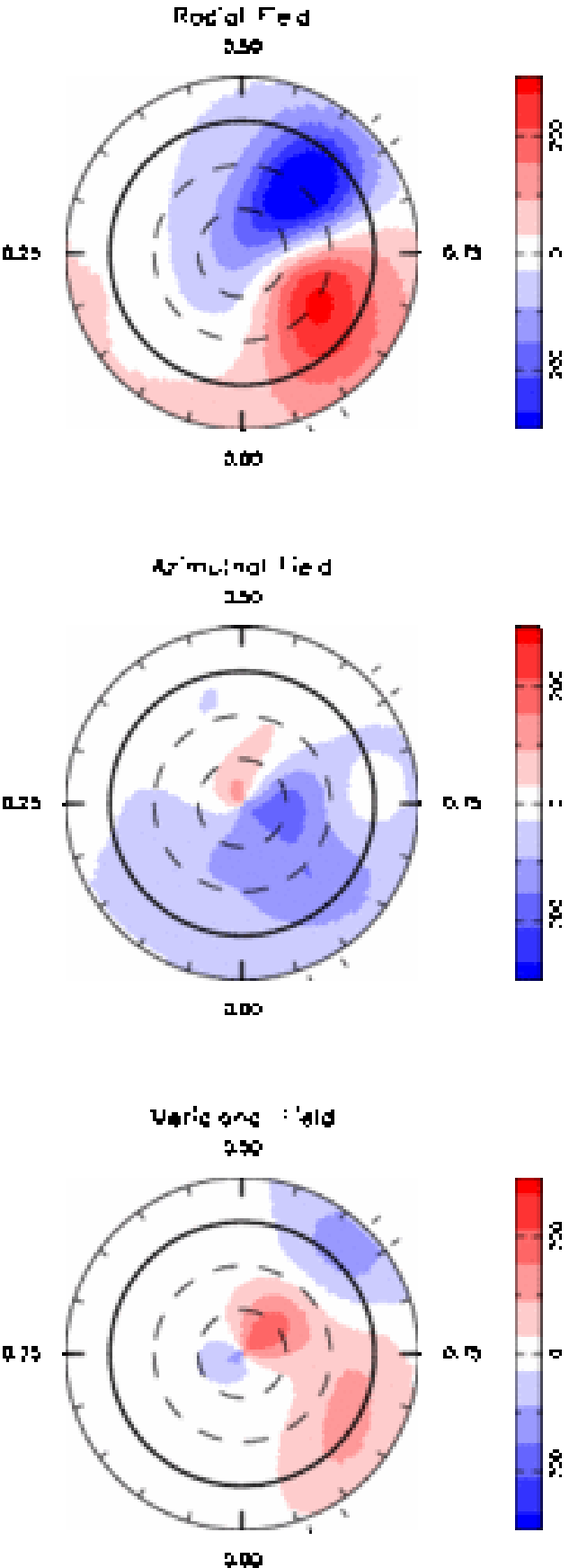}\hspace{
0.5cm
}
\includegraphics[height=0.40\textheight]{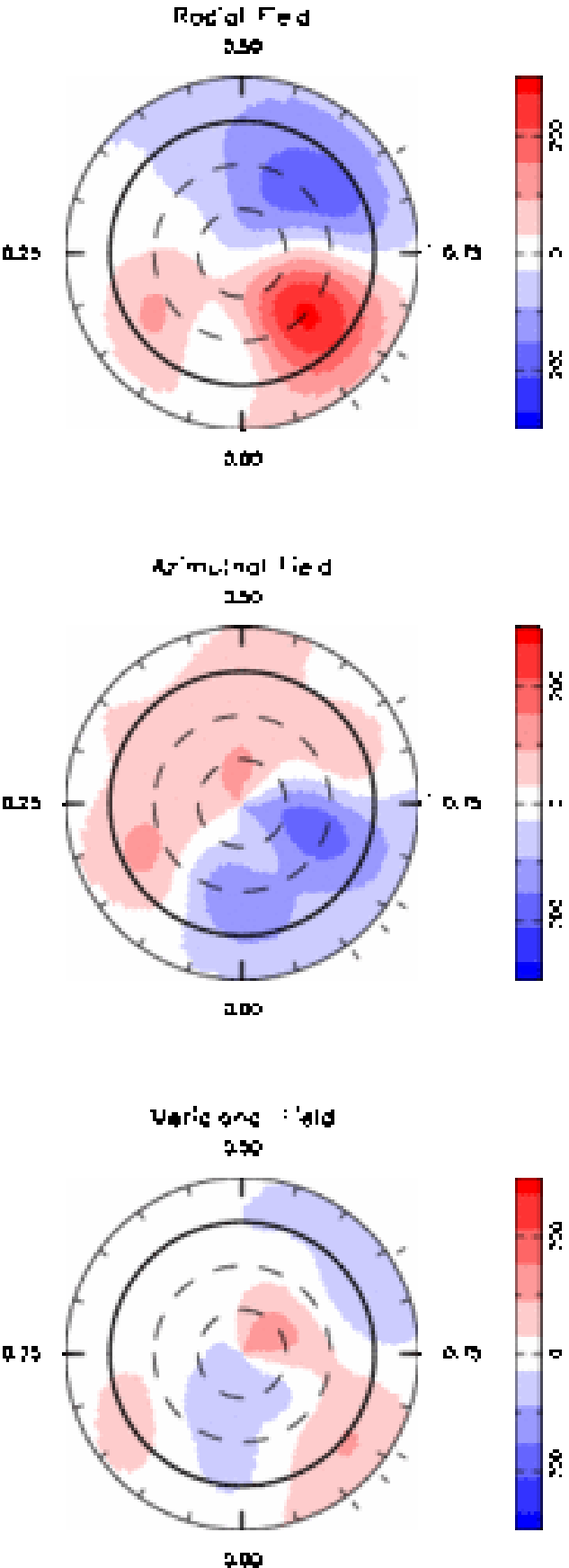}%
  \hspace{\stretch{4}}
\end{center}
\caption[]{Same as Fig.~\ref{fig:gj51_map} for GJ~1156 2007, 2008, and
2009 data sets (from left to right).}
\label{fig:gj1156_map}
\end{figure*}

We carried out 3 observing run on the flare star GJ~1156 --- in 2007, 2008 and
2009 --- and obtained 20 pairs of Stokes $I$ and $V$ spectra.  The resulting LSD
polarised signature is above noise level in nearly all observations. Temporal
variation is obvious on Figure~\ref{fig:gj1156_spec}, we observe both variations
of amplitude and polarity. We use $\vsini=17~\kms$ \cite[][]{Reiners07} which
allows us to fit the observed polarised and unpolarised profiles, as opposed to
the previously reported value \cite[$\vsini=9.2~\kms$, ][]{Delfosse98}.
We derive $\Prot=0.491~\d$, although 1/3~\d\ cannot be excluded as a
possible period. This is confirmed by MEarth photometric periodogram which also
exhibits a main peak at 0.491~\d\ and another one at 1/3~d (J.~Irwin, private
communication). The corresponding $\rsini$ is $0.16~\rsun$. As evolutionary
models predict $\rstar=0.16~\rsun$, we set $i$=60\degr\ for ZDI.
The rotation period being close to a fraction of day our observations
(especially those of 2007) do not provide an optimal sampling of rotation
phases, only our 2008 data set provides a reasonable sampling on more than half
of the rotation cycle.

\begin{table}
\begin{center}
\caption[]{Same as Tab.~\ref{tab:gj51_fit} for GJ~1156.}
\begin{tabular}{cccccc}
\hline
Epoch & $\ell_{ZDI}$ & ${\chisqr}_0$ & ${\chisqr}_f$ &
$\avg{B}$ & $B_{max}$ \\
 & & & & (\kG) & (\kG) \\
\hline
2007 & 6 & 1.95 & 0.95 & 0.06 & 0.32 \\
2008 & 6 & 2.45 & 1.00 & 0.10 & 0.36 \\
2009 & 6 & 2.07 & 1.00 & 0.09 & 0.36 \\
\hline
\label{tab:gj1156_fit}
\end{tabular}
\end{center}
\end{table}

The data sets can be fitted down to noise level with ZDI for the 3 epochs of
observation (see Tab.~\ref{tab:gj1156_fit}).  The corresponding maps of surface
magnetic flux are displayed in Figure~\ref{fig:gj1156_map}. The 3 maps exhibit
similar properties: they mainly feature two radial field spots of opposite
polarities. The magnetic topologies are thus predominantly poloidal (more than
80~\% of the overall magnetic energy in poloidal modes at all epochs) but
feature a significant toroidal component (in particular in 2007 and 2008), and
non-axisymmetric (more than 80~\% of the overall magnetic energy in modes with
azimuthal number $m > \ell/2$) as was expected from the polarity reversal of the
$V$ signature inside each data set.  The lower average magnetic flux, as well as
the weakness of the spot of negative polarity (incoming field lines, in blue),
on the 2007 map can be attributed to poor phase coverage.

The reconstructed maps are quite robust to uncertainties on the input
parameters (see Sec.~\ref{sec:techniques-uncert}). In particular, when
varying these parameters, the fraction of magnetic energy reconstructed in 
non-axisymmetric modes ($m>\ell/2$) is always higher than 70\%.
Using the alternative values \Prot=0.33~d and $i$=40\degr, the reconstructed
magnetic maps do not change significantly. All the quantities listed in
table~\ref{tab:syn} vary by less than 10~\% of the total magnetic energy. The
reconstructed magnetic flux variations range from 10 to 15~\%. Considering these
values therefore does not affect our conclusions.

\section{GJ~1245~B}
\begin{figure*}
\begin{center}
  \includegraphics[height=0.40\textheight]{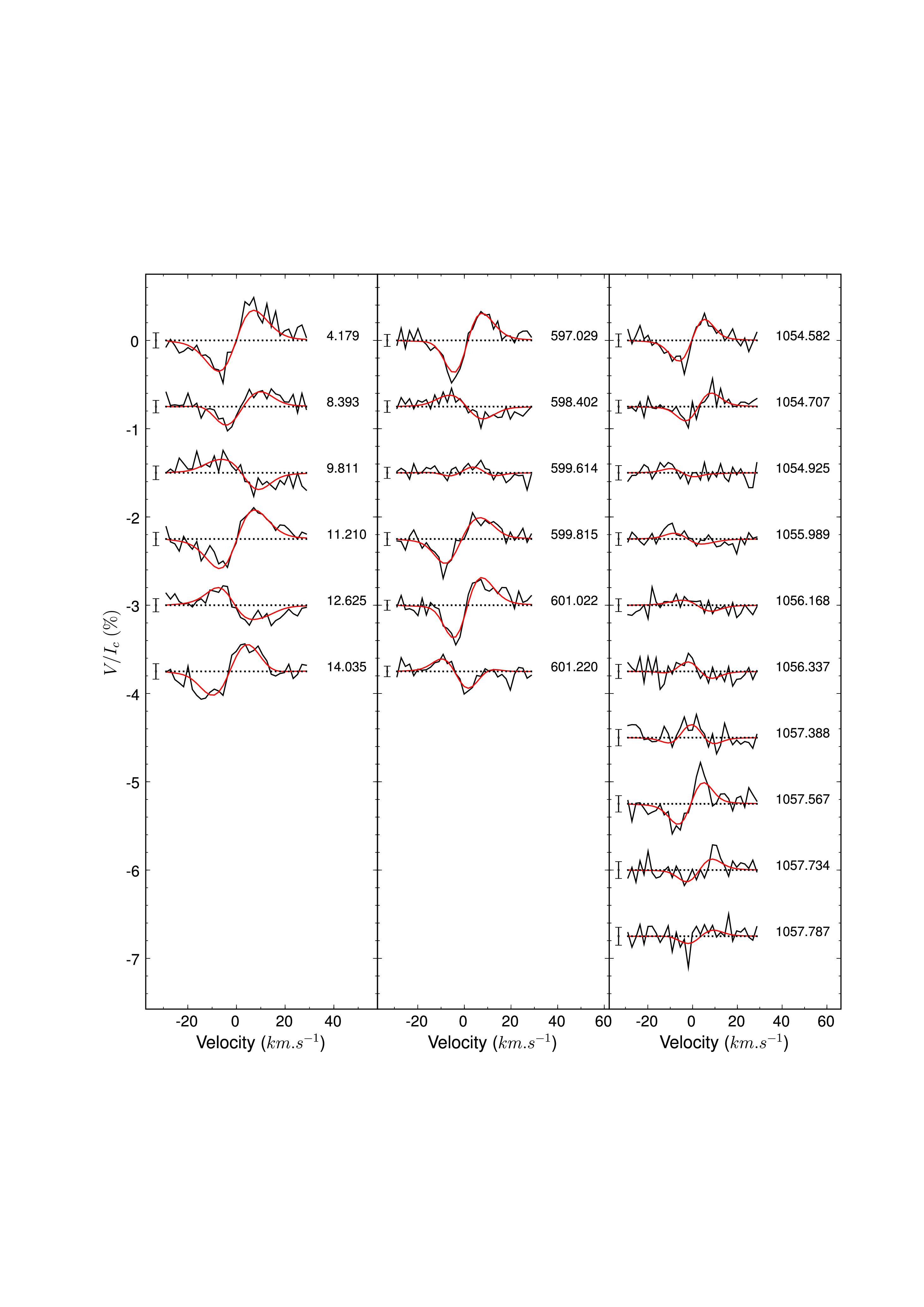}
\end{center}
\caption[]{Same as Fig.~\ref{fig:gj51_spec} for GJ~1245~B 2006, 2007, and
2008 data sets (from left to right).}
\label{fig:gj1245b_spec}
\end{figure*}

\begin{figure*}
\begin{center}
\includegraphics[height=0.40\textheight]{fig_arxiv/gj1245b_map_06c_0.06.ps}
\hspace
{ 0.5cm}
\includegraphics[height=0.40\textheight]{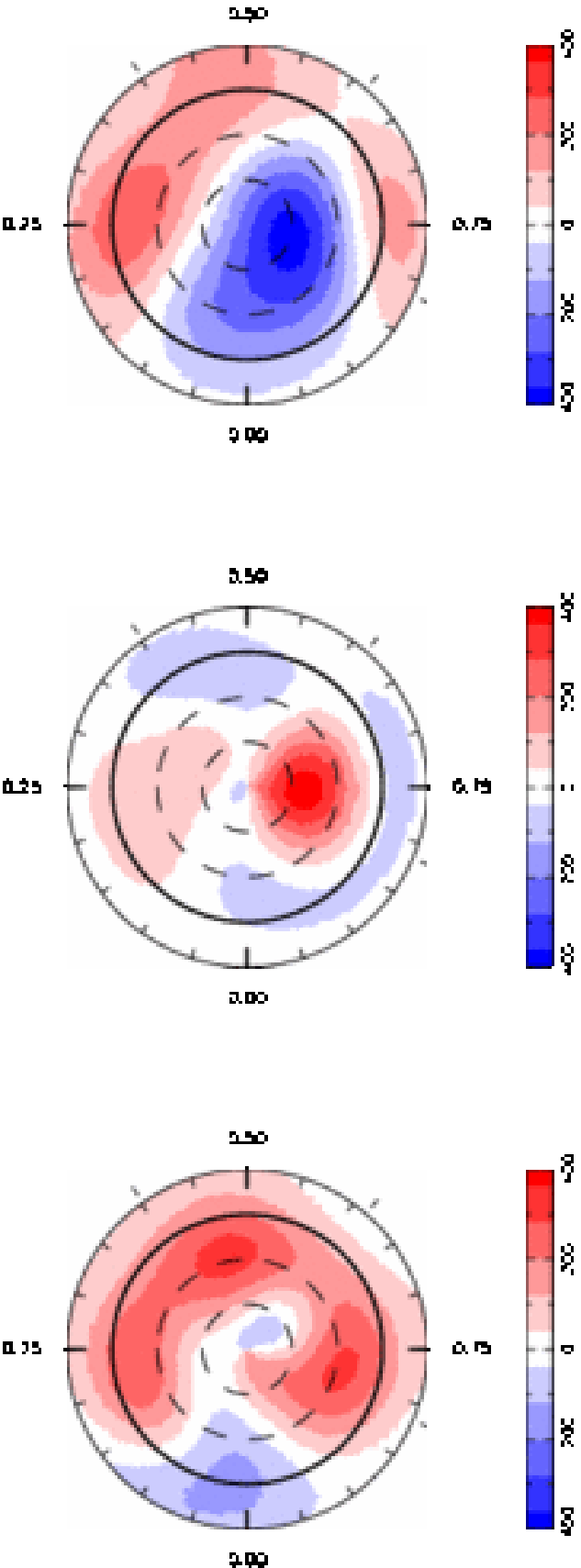}\hspace{
0.5cm }
\includegraphics[height=0.40\textheight]{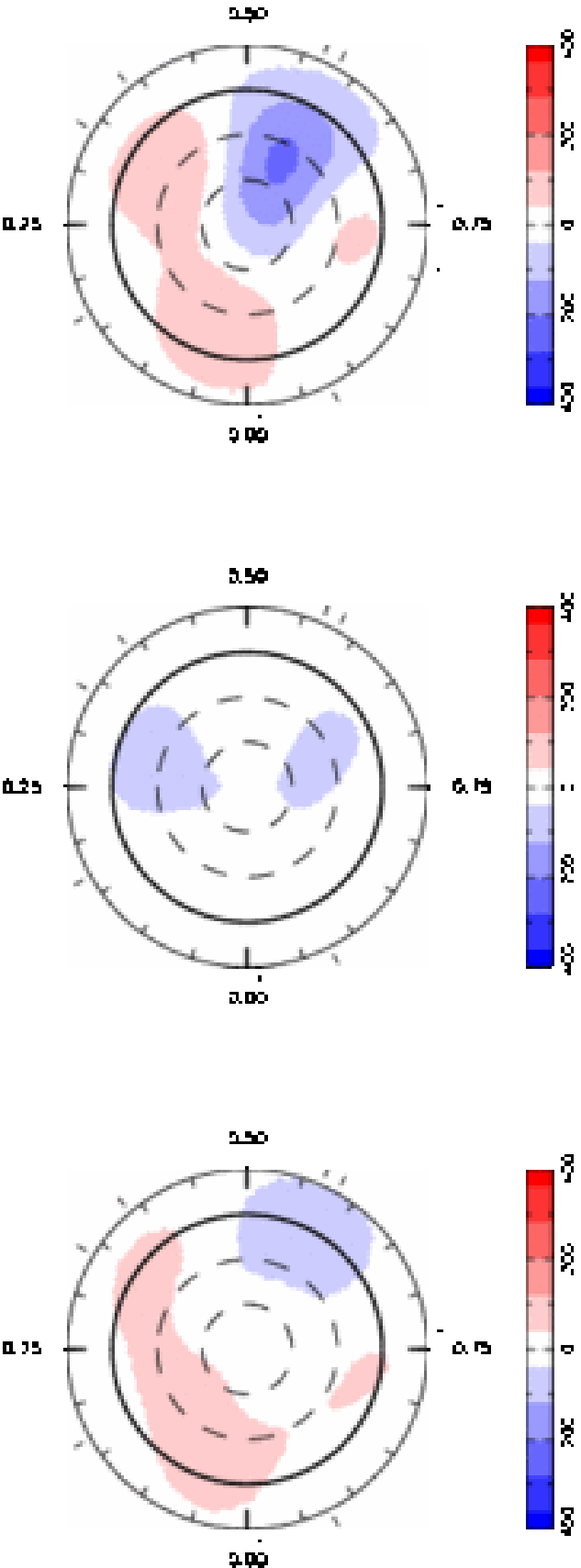}%
  \hspace{\stretch{4}}
\end{center}
\caption[]{Same as Fig.~\ref{fig:gj51_map} for GJ~1245~B 2006, 2007, and
2008 data sets (from left to right).}
\label{fig:gj1245b_map}
\end{figure*}

The M5.5 dwarf GJ~1245~B was observed during 3 successive years, for a
total of 22 pairs of Stokes $I$ and $V$ spectra. The circularly polarised LSD
signatures (see Fig.~\ref{fig:gj1245b_spec}) have a moderate amplitude and
exhibit strong variability (amplitude, shape and polarity) during an observation
run --- presumably due to rotational modulation. Variability also seems
important between the different epochs, in particular in the 2009 data set the
average amplitude of the signatures is significantly lower than at previous
epochs, this is also visible in the longitudinal field measurements (see
Tab.~\ref{tab:obs}).

We measure a mean $\RV = 5.4~\kms$ with typical dispersions of $0.1~\kms$
(see Tab.~\ref{tab:obs}). This is compatible with the mean value of $5~\kms$
reported by \cite{Delfosse98}. We use $\vsini=7~\kms$ (\citealt{Delfosse98} ;
\citealt{Reiners07}); and find a rotation period of $0.71~\d$ corresponding to a
peak in the photometric periodogram produced by the HATNet
\footnote{\cite{Bakos04}} survey (J.~Hartman, private communication). With these
values of \vsini\ and period, we find $\rsini=0.10~\rsun$, as the NextGen
evolutionary model predicts $\rstar=0.14~\rsun$ we set $i=40\degr$ for our
study.

\begin{table}
\begin{center}
\caption[]{Same as Tab.~\ref{tab:gj51_fit} for GJ~1245~B.}
\begin{tabular}{cccccc}
\hline
Epoch & $\ell_{ZDI}$ & ${\chisqr}_0$ & ${\chisqr}_f$ &
$\avg{B}$ & $B_{max}$ \\
 & & & & (\kG) & (\kG) \\
\hline
2006 & 4 & 4.41 & 1.00 & 0.17 & 0.47 \\
2007 & 4 & 4.92 & 1.10 & 0.18 & 0.58 \\
2008 & 4 & 1.81 & 1.00 & 0.06 & 0.22 \\
\hline
\label{tab:gj1245b_fit}
\end{tabular}
\end{center}
\end{table}

Running ZDI\ on the LSD time-series, with the aforementioned parameters,
we can achieve a good fit for the 3 epochs (see Fig.~\ref{fig:gj1245b_spec} and
Tab.~\ref{tab:gj1245b_fit}). The reconstructed magnetic field
significantly evolves between two successive epochs, in particular the
reconstructed magnetic flux has strongly decreased between our 2008 and 2009
observations.
However the magnetic topologies feature similar
properties: strong spots of radial field (although more than 40~\% of the
magnetic energy lies in non-radial field structures); a mostly non-axisymmetric
field (more than 50\% of the magnetic energy) and a significant toroidal
component (between 15 and 20\% of the total magnetic energy).
The mainly non-axisymmetric nature of the magnetic field in 2006 and
2008, as well as the presence of a significant toroidal component at all epochs
are robust to uncertainties on stellar parameters (see
Sec.~\ref{sec:techniques-uncert}).

\section{WX~UMA=GJ~412~B}
\label{sec:wxuma}
\begin{figure*}
\begin{center}
  \includegraphics[height=0.40\textheight]{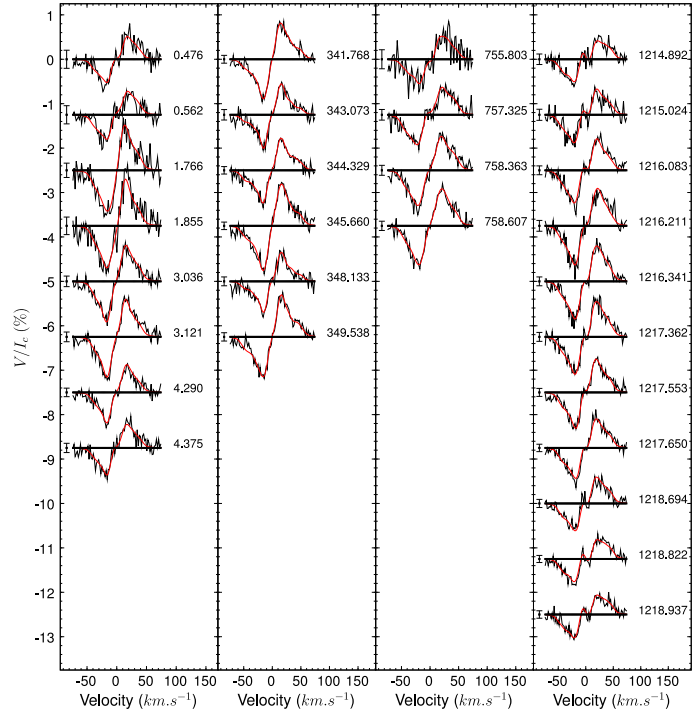}
\end{center}
\caption[]{Same as Fig.~\ref{fig:gj51_spec} for WX~UMa 2006, 2007, 2008
and
2009 data sets (from left to right).}
\label{fig:gj412b_spec}
\end{figure*}

\begin{figure*}
\begin{center}
  \includegraphics[height=0.40\textheight]{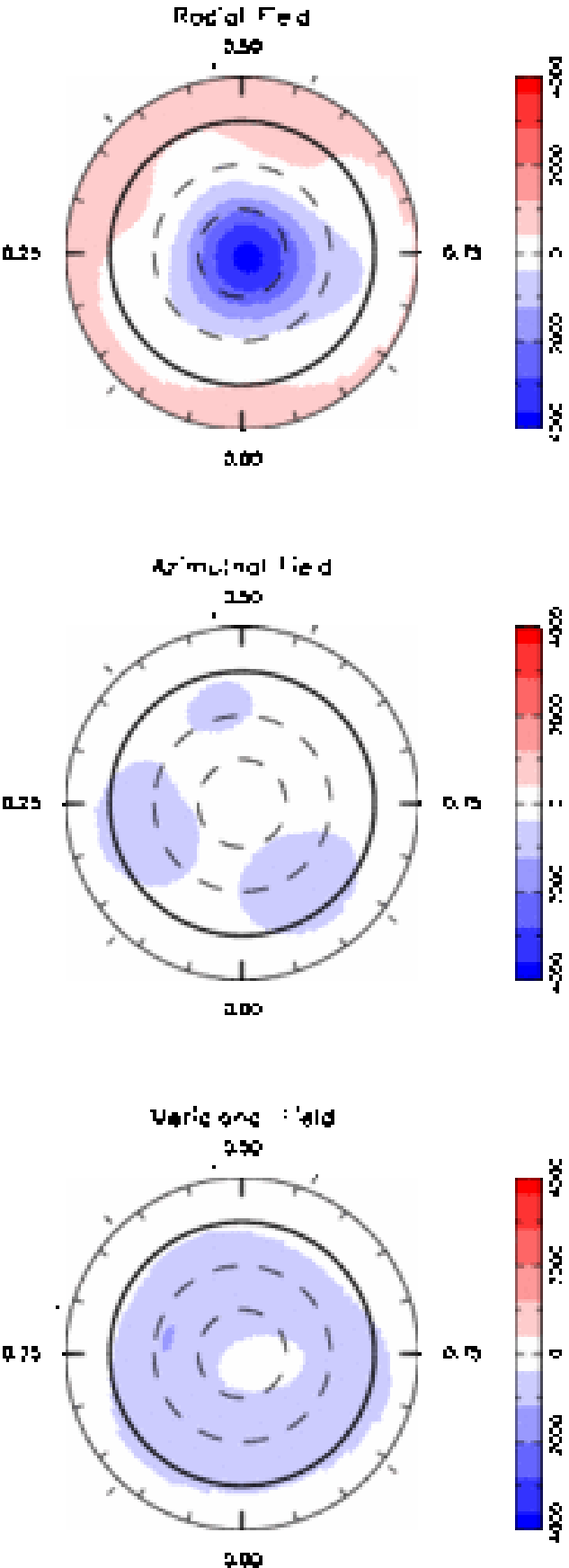}%
  \hspace{0.5cm }%
  \includegraphics[height=0.40\textheight]{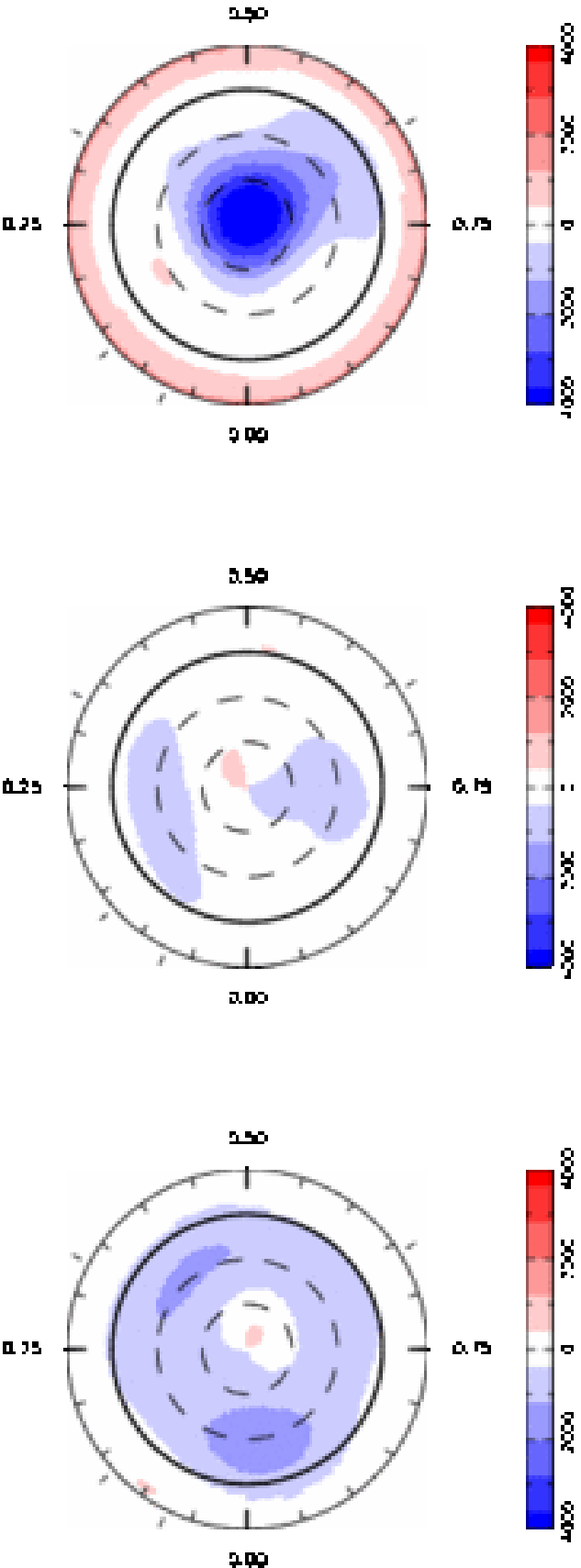}%
  \hspace{0.5cm} %
  \includegraphics[height=0.40\textheight]{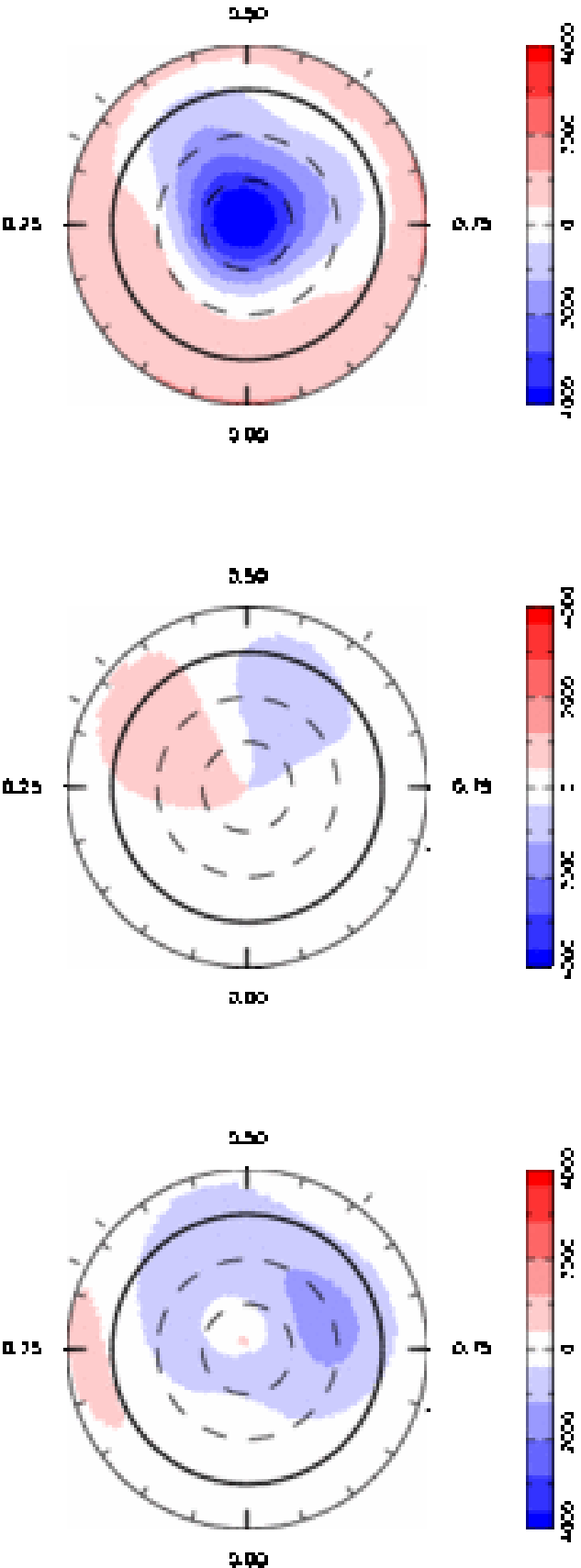}%
  \hspace{0.5cm}%
  \includegraphics[height=0.40\textheight]{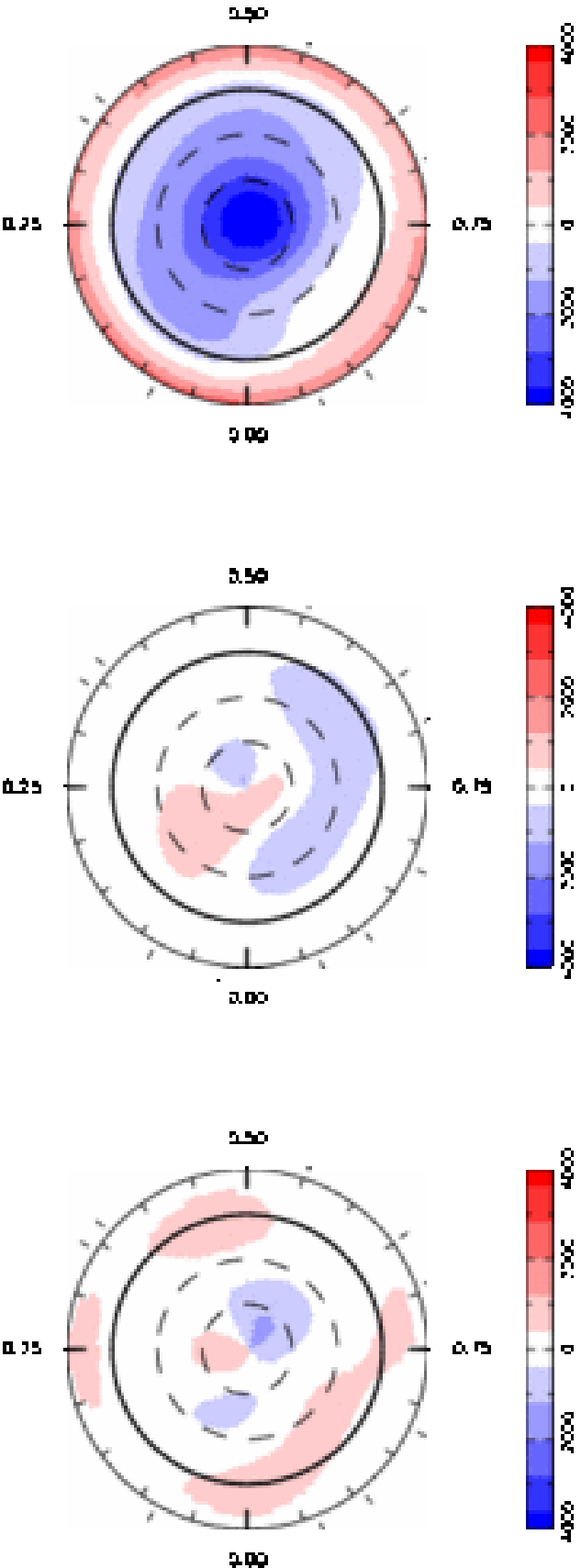}%
  \hspace{\stretch{4}}
\end{center}
\caption[]{Same as Fig.~\ref{fig:gj51_map} for WX~UMa 2006, 2007, 2008
and
2009 data sets (from left to right).}
\label{fig:gj412b_map}
\end{figure*}

Between 2006 and 2009, we observed the M6 dwarf WX~UMa during 4 runs, collecting
a total of 29 spectra. The LSD Stokes~$V$ profiles are very similar throughout
the data set: a very strong (the peak-to-peak amplitude is close to 2\% of the
unpolarised continuum level) simple two-lobbed signature of negative polarity
(i.e. corresponding to a longitudinal field directed toward the star).  Temporal
evolution inside each data set, presumably due to rotational modulation, is
noticeable though weaker than what we observe on GJ~51. We measure average
values ranging from $\RV=69.95$ to $70.25~\kms$ with a jitter ranging from 0.06
to 0.53~\kms (depending on the observation epoch, see Tab.~\ref{tab:obs}).  This
is in agreement with $\RV=68.886~\kms$ \cite[single precise measurement by ][on
GJ~412A]{Nidever02}. We use $\vsini=5.0~\kms$ \citep{Reiners09} which accounts
for Zeeman broadening and results in better agreement with our Stokes $I$ and
$V$ LSD profiles than the previous value of $\vsini=7.7~\kms$ inferred
from correlation profiles \citep{Delfosse98}. From the circularly
polarised LSD profiles we infer $\Prot=0.74~\d$. With this rotation
period, our 2007 data cover half of the rotation cycle, and the 3 other
data sets result in a good phase coverage. The comparison of the
resulting $\rsini=0.073~\rsun$ with $\rstar=0.12~\rsun$ (from
theoretical models) indicates an intermediate inclination angle, we
set $i=40\degr$.

\begin{table}
\begin{center}
\caption[]{Same as Tab.~\ref{tab:gj51_fit} for WX~UMa.} 
\begin{tabular}{cccccc}
\hline
Epoch & $\ell_{ZDI}$ & ${\chisqr}_0$ & ${\chisqr}_f$ &
$\avg{B}$ & $B_{max}$ \\
 & & & & (\kG) & (\kG) \\
\hline
2006 & 4 &  8.98 & 0.90 & 0.89 & 3.82 \\
2007 & 4 & 20.48 & 1.05 & 0.94 & 4.88 \\
2008 & 4 & 12.42 & 1.00 & 1.03 & 4.55 \\
2009 & 4 & 14.58 & 1.15 & 1.06 & 4.53 \\
\hline
\label{tab:gj412b_fit}
\end{tabular}
\end{center}
\end{table}

Although only the 2006 data set can be fitted below $\chisqr=1.0$ (see
Tab.~\ref{tab:gj412b_fit}), the ZDI synthetic Stokes~$V$ profiles 
match well the evolution of the LSD signatures for all epochs, as shown in
Fig.~\ref{fig:gj412b_spec}. The corresponding magnetic maps are presented
in Fig.~\ref{fig:gj412b_map}: they all feature a strong polar cap of radial
field (of negative polarity, i.e.  field lines directed toward the star)
reaching a maximum flux of approximately 4~\kG, whereas the magnetic flux
averaged over the visible fraction of the star is about 1~\kG. Azimuthal and
meridional field structure are much weaker.  The topology is very simple, modes
with degree $\ell > 4$ can be neglected, the dipole modes encompass more than
60\% of the reconstructed energy at all epochs.  Toroidal and non-axisymmetric
components of the field are very weak. The evolution of the magnetic field over
successive years is very weak, the maps are strikingly similar.
These conclusions are robust to uncertainties on stellar parameters (in
particular $i$ and $\vsini$). When varying these parameters over the width of
their respective error bars (see section~\ref{sec:techniques-uncert}) the
reconstructed magnetic field of WX~UMa is always almost purely poloidal, mostly
axisymmetric, and the main mode is the radial component of a dipole aligned with
the rotation axis. 

\section{DX~Cnc=GJ~1111}
\begin{figure*}
\begin{center}
  \includegraphics[height=0.40\textheight]{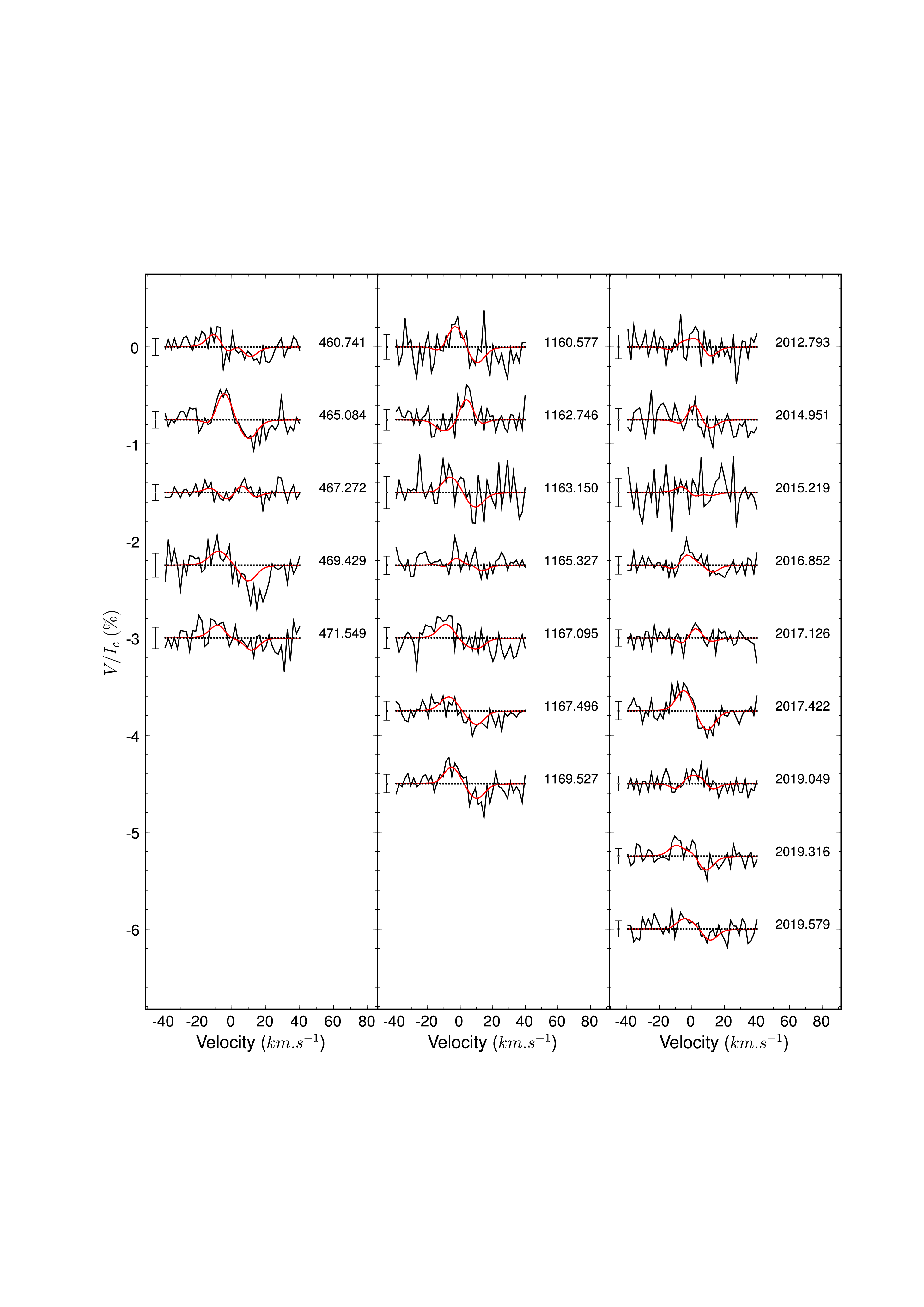}
\end{center}
\caption[]{Same as Fig.~\ref{fig:gj51_spec} for DX~Cnc 2007, 2008, and
2009 data sets (from left to right).}
\label{fig:gj1111_spec}
\end{figure*}

\begin{figure*}
\begin{center}
\includegraphics[height=0.40\textheight]{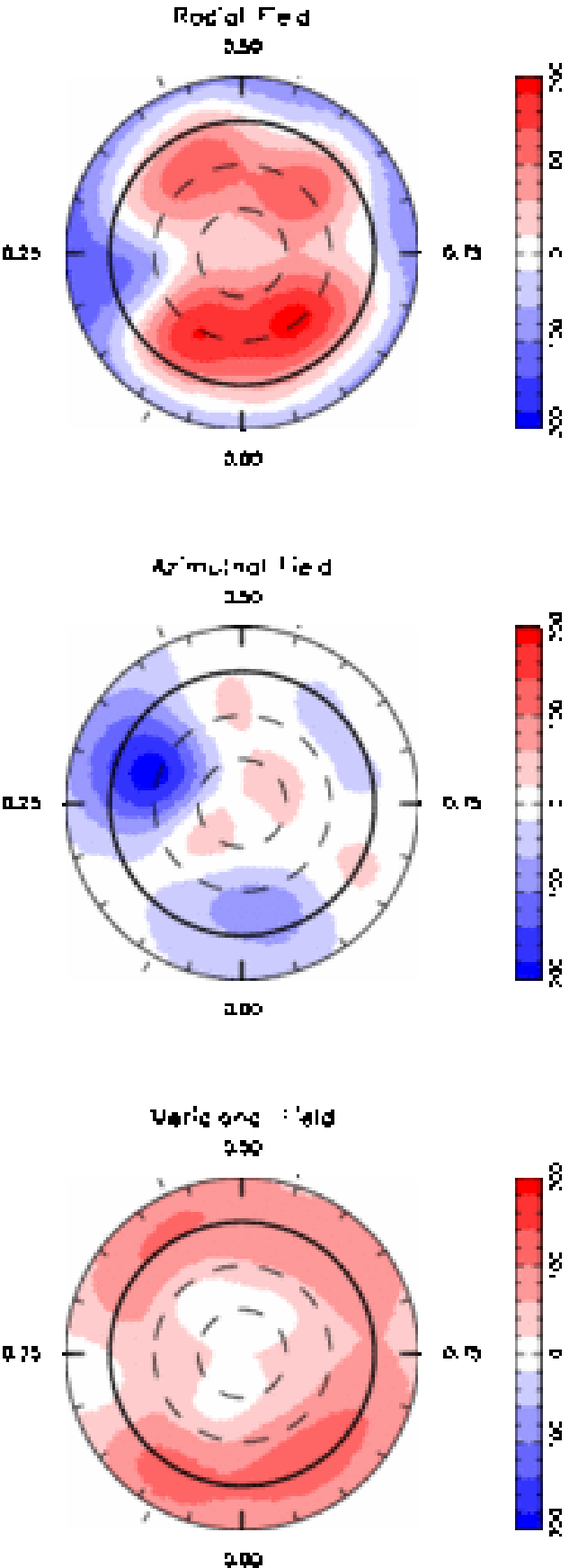}\hspace{
0.5cm
}
\includegraphics[height=0.40\textheight]{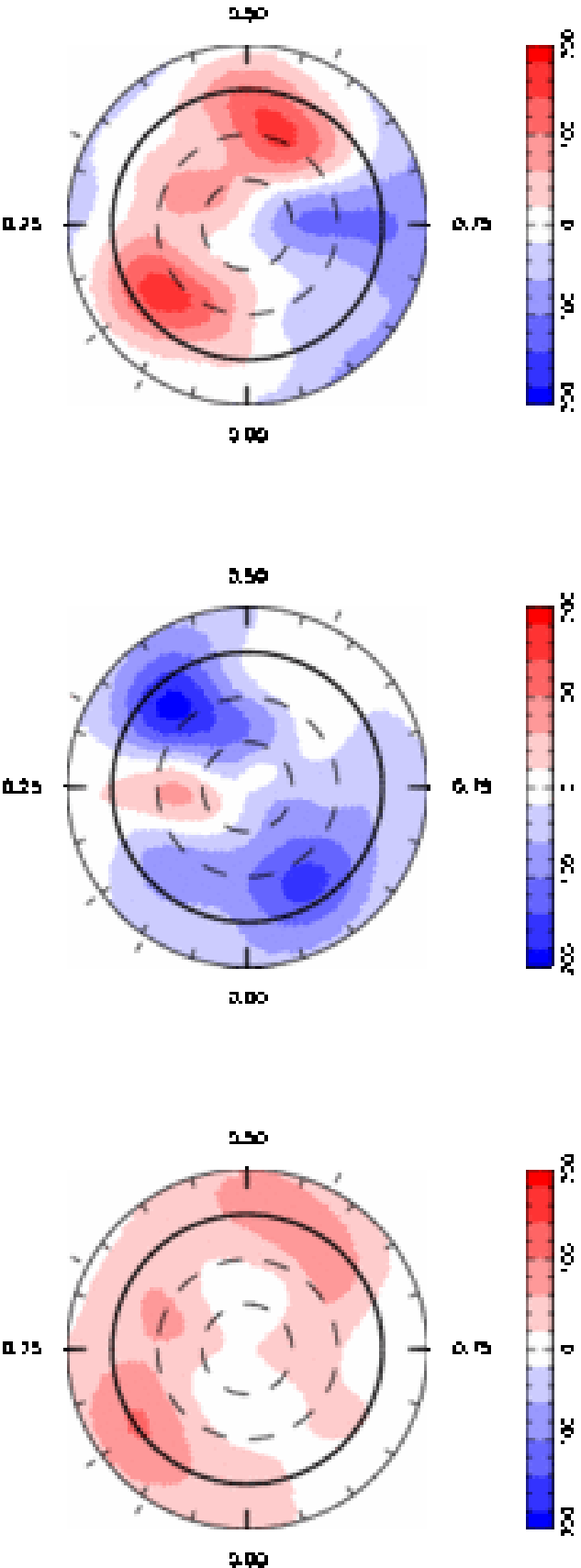}\hspace{
0.5cm
}
\includegraphics[height=0.40\textheight]{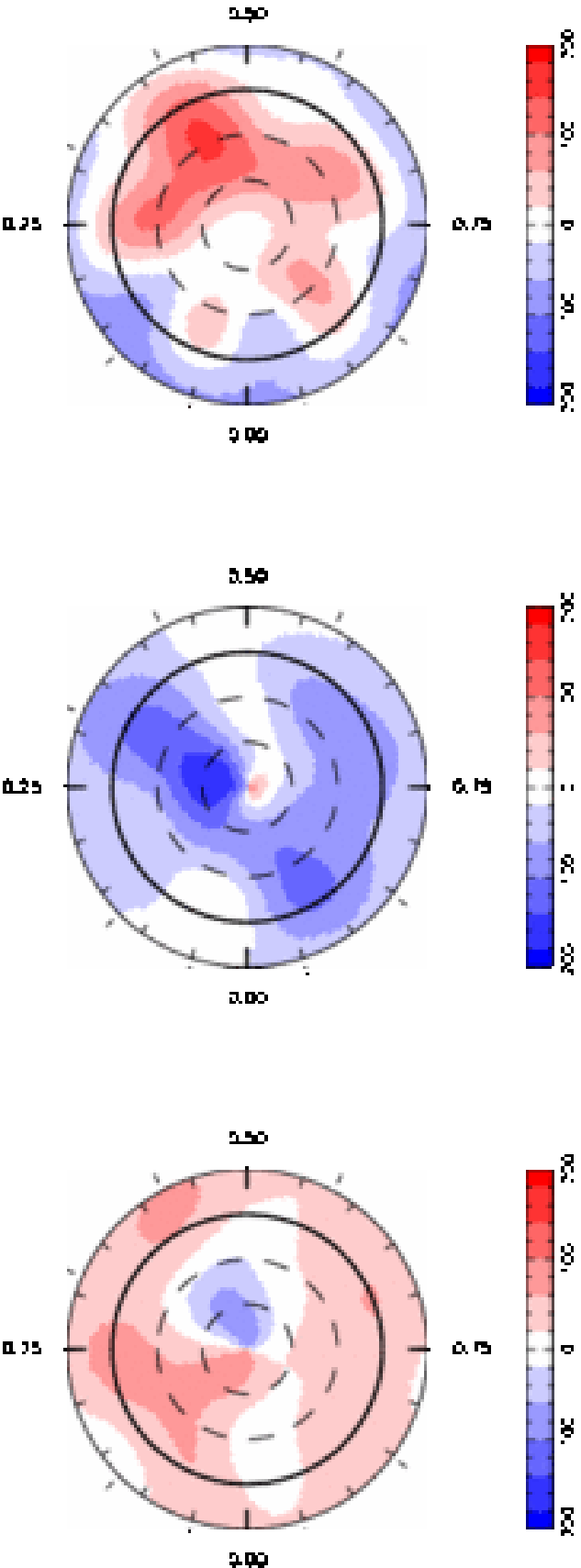}%
  \hspace{\stretch{4}}
\end{center}
\caption[]{Same as Fig.~\ref{fig:gj51_map} for DX~Cnc 2007, 2008, and
2009 data sets (from left to right).}
\label{fig:gj1111_map}
\end{figure*}

We carried out 3 observation runs on DX~Cnc between 2007 and 2009, resulting in
21 pairs of Stokes~$I$ and $V$ spectra. The LSD polarised profiles are
displayed in Fig.~\ref{fig:gj1111_spec}, the Zeeman signatures have low
amplitudes but are definitely detected in several spectra. The amplitude and
shape of the circularly polarised line dramatically evolve during each observing
run, presumably due to rotational modulation. From the LSD profiles, we measure
average RV ranging from $10.44$ to $10.67~\kms$, with a jitter strongly
depending on the observing epoch (see Tab.~\ref{tab:obs}), in agreement with the
previously published values ($\RV=9~\kms$ in \citealt{Delfosse98};
$\RV=10.1~\kms$ in \citealt{Mohanty03}). We use $\vsini=13~\kms$
\cite[][]{Reiners07}, and $\Prot=0.46~\d$ (inferred from our data). As the
resulting $\rsini=0.12~\rsun$ is already higher than $\rstar=0.11~\rsun$
predicted by theoretical models, we assume a high inclination angle of the
rotation axis and set $i=60\degr$ for the imaging process. 
Our 2007 and 2008 data sets provide a reasonable phase coverage and the 2009 one
results in a good sampling of the stellar rotation.

\begin{table}
\begin{center}
\caption[]{Same as Tab.~\ref{tab:gj51_fit} for DX~Cnc.}
\begin{tabular}{cccccc}
\hline
Epoch & $\ell_{ZDI}$ & ${\chisqr}_0$ & ${\chisqr}_f$ &
$\avg{B}$ & $B_{max}$ \\
 & & & & (\kG) & (\kG) \\
\hline
2007 & 6 & 1.70 & 1.00 & 0.11 & 0.22 \\
2008 & 6 & 1.42 & 1.00 & 0.08 & 0.20 \\
2009 & 6 & 1.37 & 1.00 & 0.08 & 0.18 \\
\hline
\label{tab:gj1111_fit}
\end{tabular}
\end{center}
\end{table}

Both the mean longitudinal field and the standard deviation from this value get
weaker from one observing run to the next one, indicating intrinsic variability
of the magnetic field.  Using the ZDI tomographic imaging code, we can fit the 3
data sets down to $\chisqr=1.0$, the resulting magnetic fields are presented in
Fig.~\ref{fig:gj1111_map}. Although for the three epochs, the reconstructed
magnetic topologies feature a significant non-axisymmetric component, they
significantly differ from each other: (i) the fraction of magnetic energy
reconstructed in the toroidal component grows from 7\% in 2007 to 38\% in 2009;
(ii) the two main spots of radial magnetic field seem to evolve between 2007 and
2008, and finally in 2009 only one region of strong radial field remains. (iii)
The averaged magnetic flux decreases from 110~G in 2007 to 80~G in 2008 and
2009.  This last point is strengthened by the fact that for a given topology,
ZDI recovers less magnetic flux for a dataset providing partial phase coverage
(due to the maximum entropy constraint), whereas here the larger flux is
inferred from the dataset providing the poorest sampling of stellar rotation.
The presence of a strong toroidal component on DX~Cnc in 2008 and 2009 is a
robust result. In particular, when varying the input parameters within their
respective error bars (see Sec.~\ref{sec:techniques-uncert}) the toroidal
component always accounts for at least 30~\% of the reconstructed magnetic
energy.

\section{GJ~3622}

\begin{figure*}
\begin{center}
  \includegraphics[height=0.40\textheight]{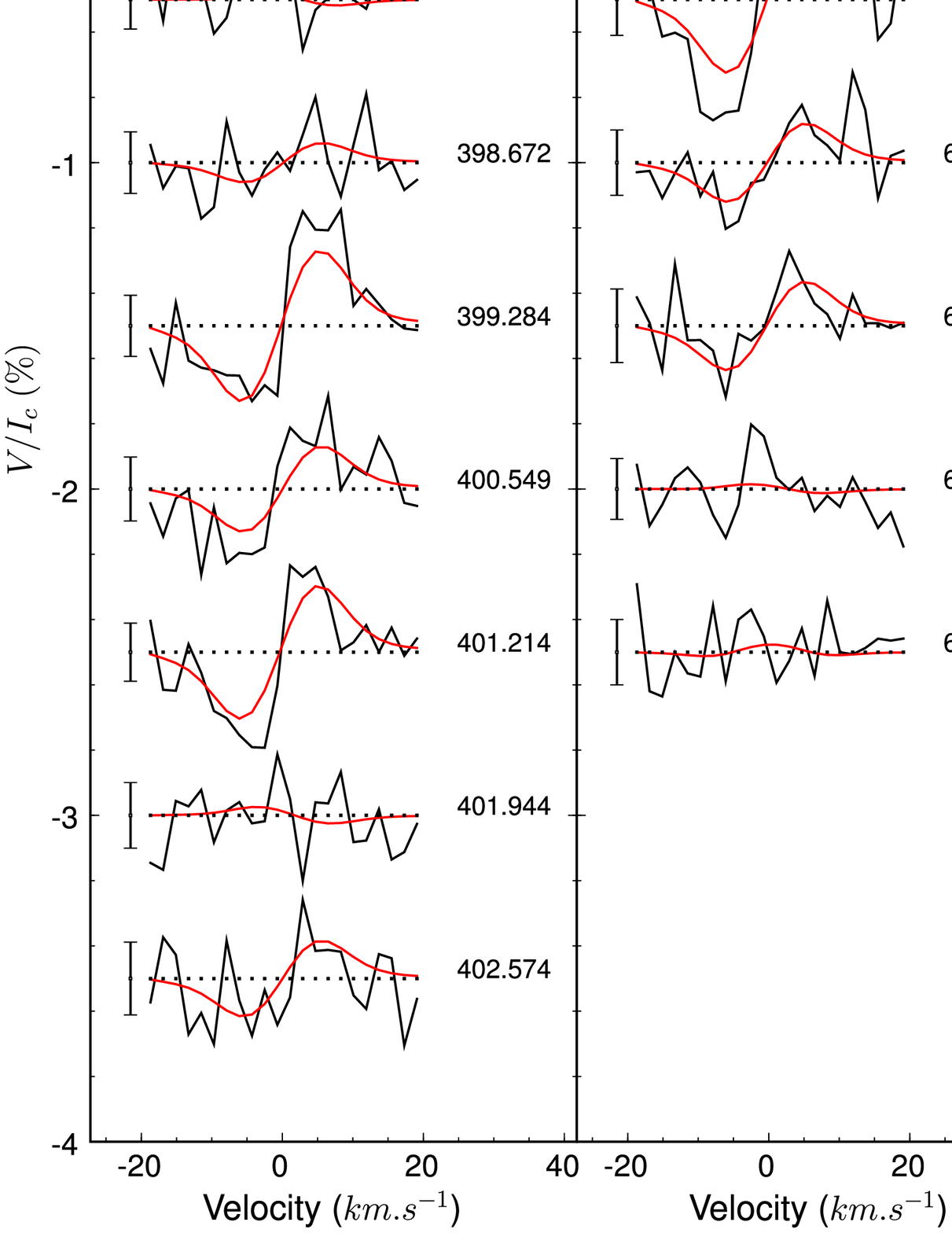}%
  \hspace{0.5cm}%
  \includegraphics[height=0.40\textheight]{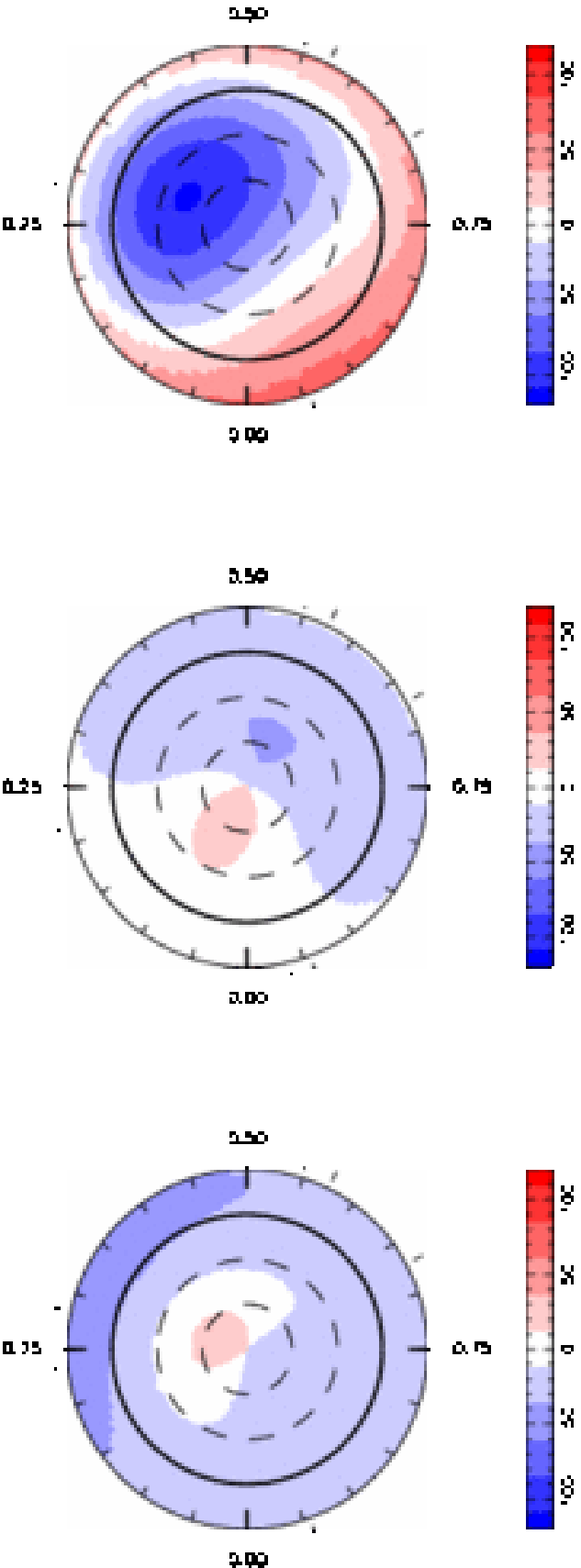}%
  \hspace{0.5cm}%
  \includegraphics[height=0.40\textheight]{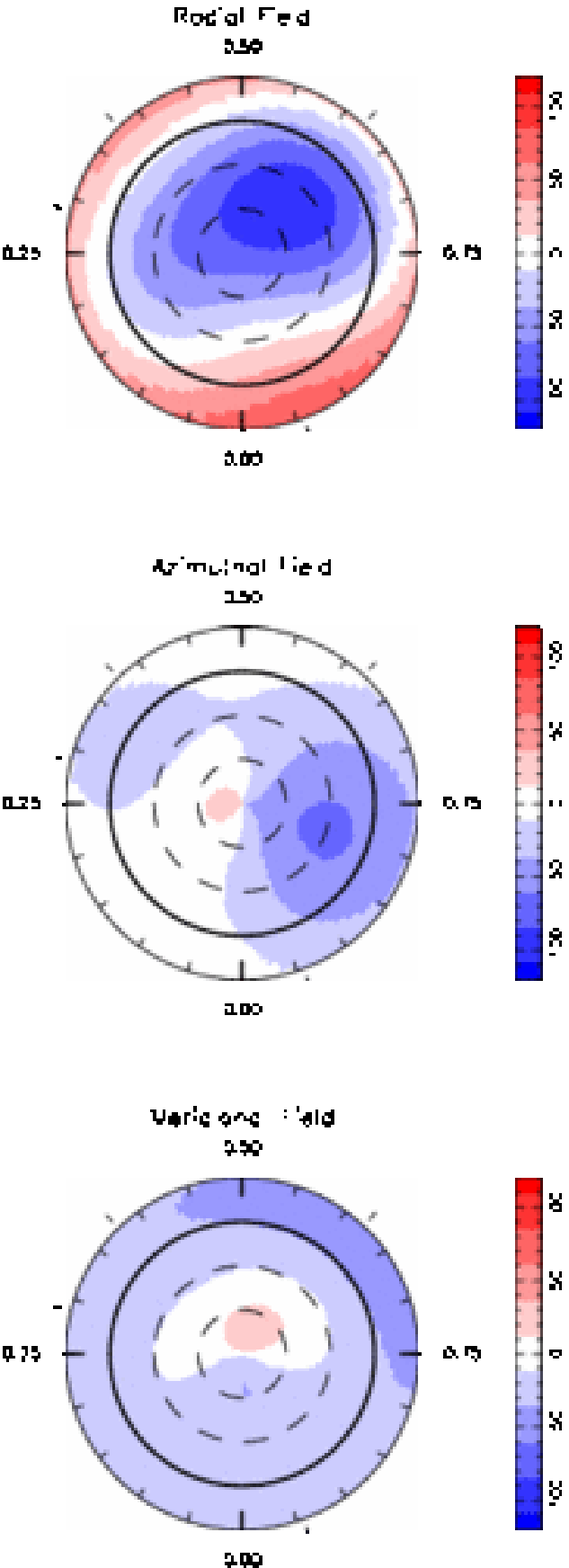}%
\end{center}
\caption[]{2008 and 2009 spectra and magnetic maps of GJ~3622. See
  Fig.~\ref{fig:gj51_spec} and \ref{fig:gj51_map} for more details.}
\label{fig:gj3622}
\end{figure*}

GJ~3622 was observed in 2008 and 2009, we collected 8 and 5
sequences, respectively. In spite of relatively low \sn\ (due to a low
intrinsic luminosity), Stokes $V$ signatures are clearly detected in
several spectra and variations are undoubtedly noticeable (see
Fig.~\ref{fig:gj3622}). From our data sets we infer $\Prot=1.5~\d$,
and use $\vsini=3~\kms$ reported by \cite{Mohanty03}, resulting in
$\rsini=0.09~\rsun$. Our data sets provide a reasonable phase coverage with
this period. As evolutionary models predict $\rstar=0.11~\rsun$
for a $0.09~\msun$ M dwarf, we set the inclination angle to 60\degr\ for
the imaging process.

Setting $\ell_{max}=2$, given the very weak signal detected, it is
possible to fit our 2 data sets down to noise level (see
Tab.~\ref{tab:gj3622_fit}). The corresponding magnetic maps (see
Fig.~\ref{fig:gj3622}) are very similar. A radial field spot of negative
polarity, i.e. field lines directed toward the star, is located at
mid-latitudes. Magnetic flux reaches up to 110~G in
this region. Azimuthal and meridional fields are much weaker, the
toroidal component represents less than 10~\% of the overall magnetic
energy at both epochs. The reconstructed topology is close to a tilted
dipole: less than 10~\% of the magnetic energy is reconstructed in
$\ell=2$ modes, and the axisymmetric component stands for more than
80~\% of the energy content. The data being very noisy the simple topology
reconstructed by ZDI likely reflects the lack of information in the
polarised spectra. 

\begin{table}
\begin{center}
\caption[]{Same as Tab.~\ref{tab:gj51_fit} for GJ~3622.}
\begin{tabular}{cccccc}
\hline
Epoch & $\ell_{ZDI}$ & ${\chisqr}_0$ & ${\chisqr}_f$ &
$\avg{B}$ & $B_{max}$ \\
 & & & & (\kG) & (\kG) \\
\hline
2008 & 2 & 2.08 & 0.95 & 0.05 & 0.11 \\
2009 & 2 & 1.68 & 0.95 & 0.06 & 0.11 \\
\hline
\label{tab:gj3622_fit}
\end{tabular}
\end{center}
\end{table}

\section{other stars}
\label{sec:other}

 \begin{figure*}
   \centering
  \includegraphics[width=0.32\textwidth]{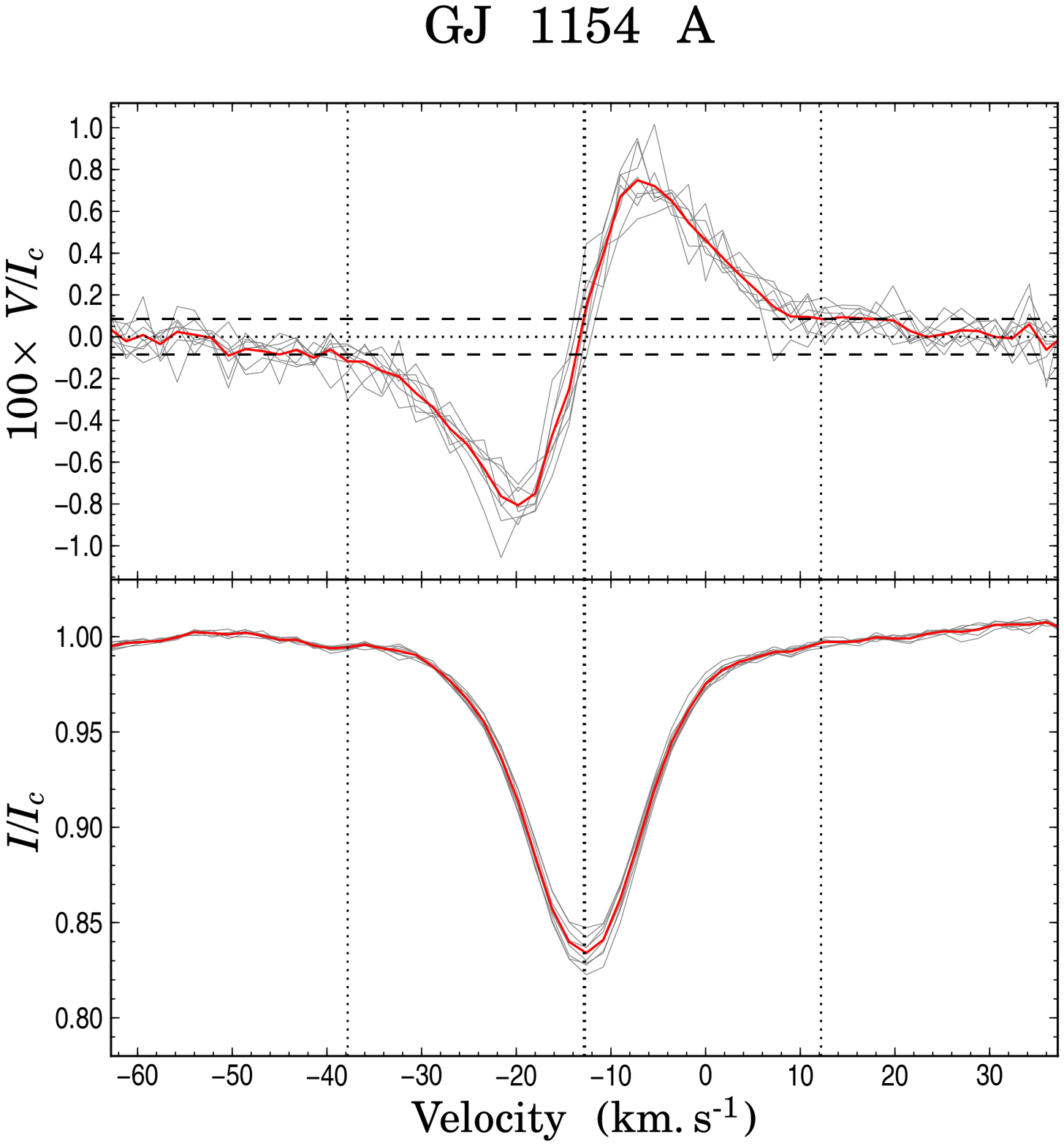} %
  \includegraphics[width=0.32\textwidth]{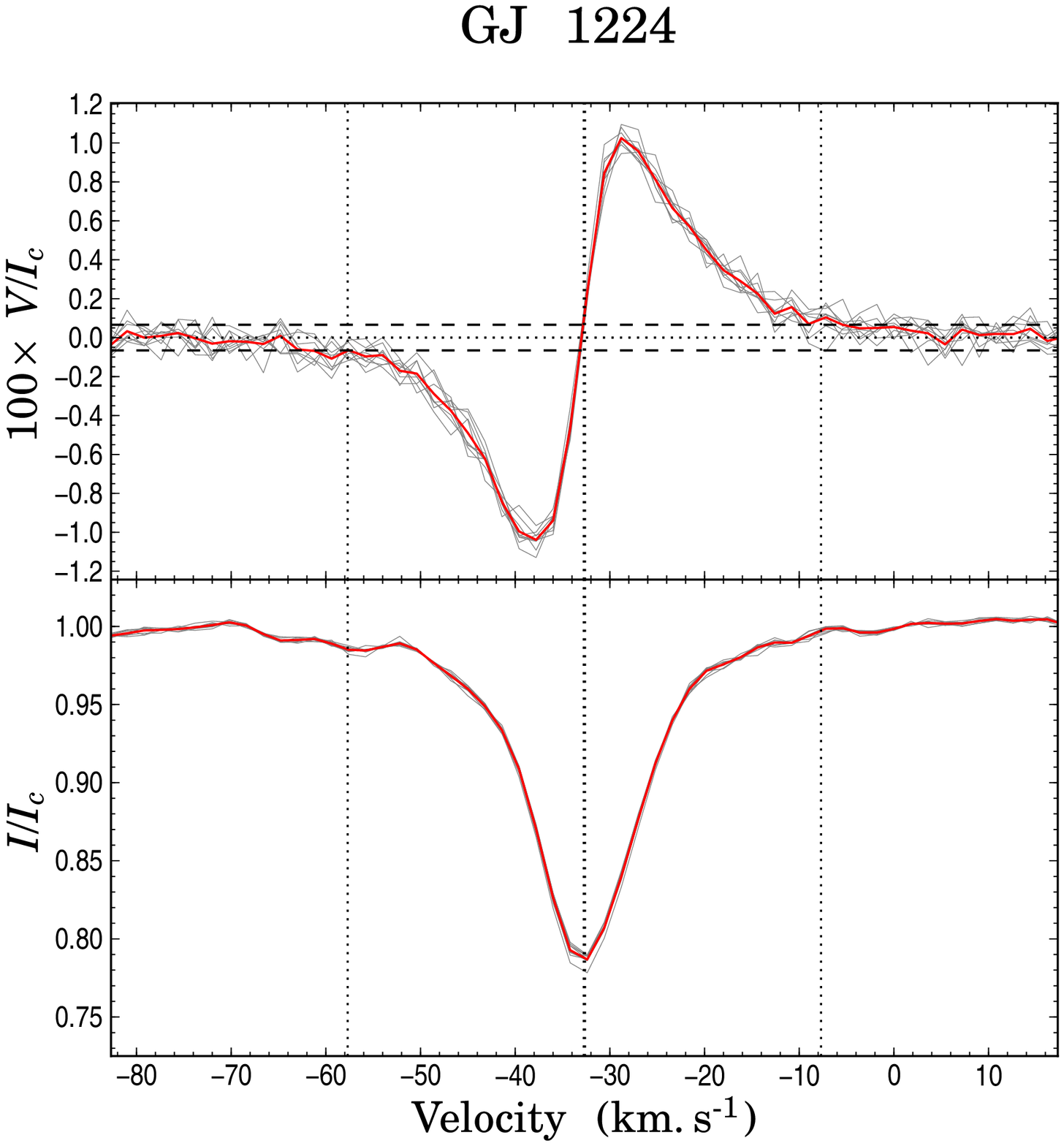} %
  \includegraphics[width=0.32\textwidth]{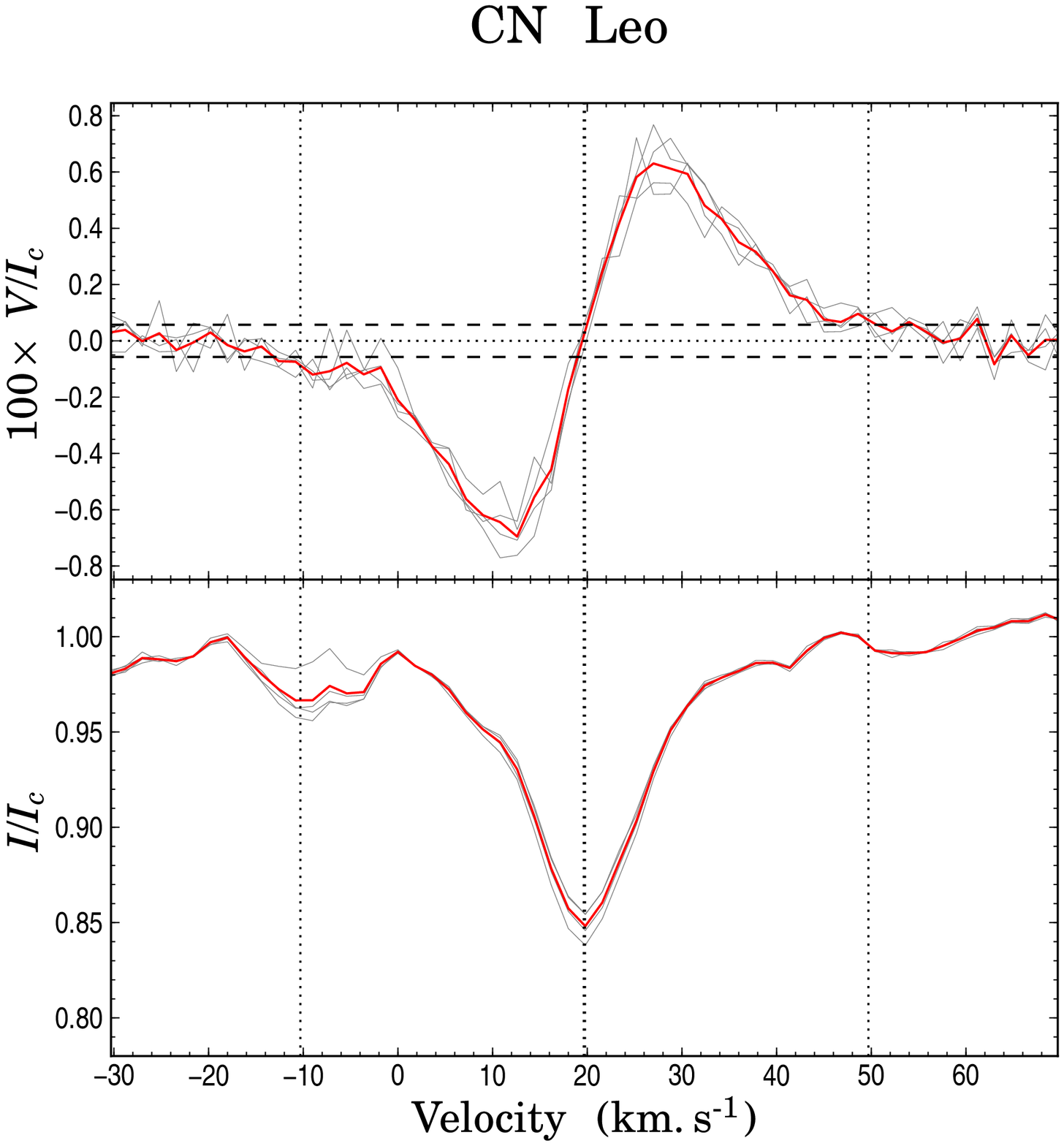} 
  \caption[]{Stokes~$I$ (lower panels) and $V$ (upper panels) LSD
signatures of
GJ~1154~A, GJ~1224 and CN~Leo, from left to right. In each panel all the
profiles of the time series are plotted as superimposed grey lines, and the
average profile is shown in red. Vertical dotted lines represent the line center
(in bold) and the approximate limits of the line. In Stokes~$V$ plots $\pm
1\sigma$ levels (corresponding the individual spectra) are shown as dashed
lines, and the reference level as a dotted line.} \label{fig:3nozdi}
\end{figure*}

For five stars of our sample we collected time series of polarised spectra but
could not produce a definitive magnetic map. For GJ~1154~A, GJ~1224 and CN~Leo
(GJ~406) we detect very strong and simple signatures (see
Fig.~\ref{fig:3nozdi}). To our knowledge, no rotation periods have been measured
for these stars and our data sets do not allow us to conclude, either because of
low intrinsic variability of the Zeeman signature or poor phase
sampling. Although we cannot compute a magnetic map for these stars, the
collected spectra unmistakably show that they host very strong
large-scale magnetic fields (longitudinal fields are about 600~G).
Strong magnetic fields
(total magnetic fluxes in the 2-3~\kG\ range) have been previously detected
on these stars by \cite{Reiners07} and \cite{Reiners09} from the analysis of
unpolarised spectra (see Tab.~\ref{tab:sample}). The simple signatures
(two-lobbed antisymmetric) featuring very low variability also clearly
suggest that these fields are mostly poloidal, strongly axisymmetric and
presumably dominated by low degree modes, similar to what we observe on
GJ~51 and WX~UMa for instance. The low dispersions of longitudinal
fields and RV values in each data set may indicate that phase sampling
is loose and thus rotation period close to a fraction of day, or/and
that these stars are observed nearly pole-on.

For the faintest stars of the sample VB~8 (GJ~644~C) and VB~10 (GJ~752~B), the
Stokes~$V$ signatures are too weak to be definitely detected in individual LSD
spectra. The initial ${\chisqr}_0$ are respectively equal to 1.089 and 1.150.
LSD signatures observed on these stars are shown on Figure~\ref{fig:vb8-10}. By
averaging all the LSD profiles of a data set, the noise level is decreased but
only features visible on all spectra --- corresponding to the axisymmetric
component of the field, if rotation sampling is even --- remain visible.  The
resulting signal shown in Fig.~\ref{fig:vb8-10} (bold red line) is processed
with a zero phase shift low-pass filter to remove the frequencies higher than
permitted by the instrumental profile (width of 4.6~\kms, or 2.5 LSD pixels).
For VB~8, the averaged LSD profile does not feature any significant signal
(\chisqr=0.99), indicating that the axisymmetric component of the magnetic field
of this star is too weak to be detected. The averaged profile of VB~10 features
a weak but distinguishable signature corresponding to $\chisqr = 1.90$.  It
suggests the presence of a large-scale magnetic field having a
significant axisymmetric component, but further observations are needed to
confirm this point. From observations in unpolarised
light, \cite{Reiners07} report total magnetic fluxes of 2.3 and 1.3~\kG\ on VB~8
and VB~10, respectively (see Tab.~\ref{tab:sample}). The very weak
Stokes~$V$ signatures observed here (with corresponding maximum longitudinal
fields of the order of 100~G) suggest that the magnetic field of these stars
is mainly structured on small spatial scales.

\begin{figure*}
  \centering
  \includegraphics[width=0.45\textwidth]{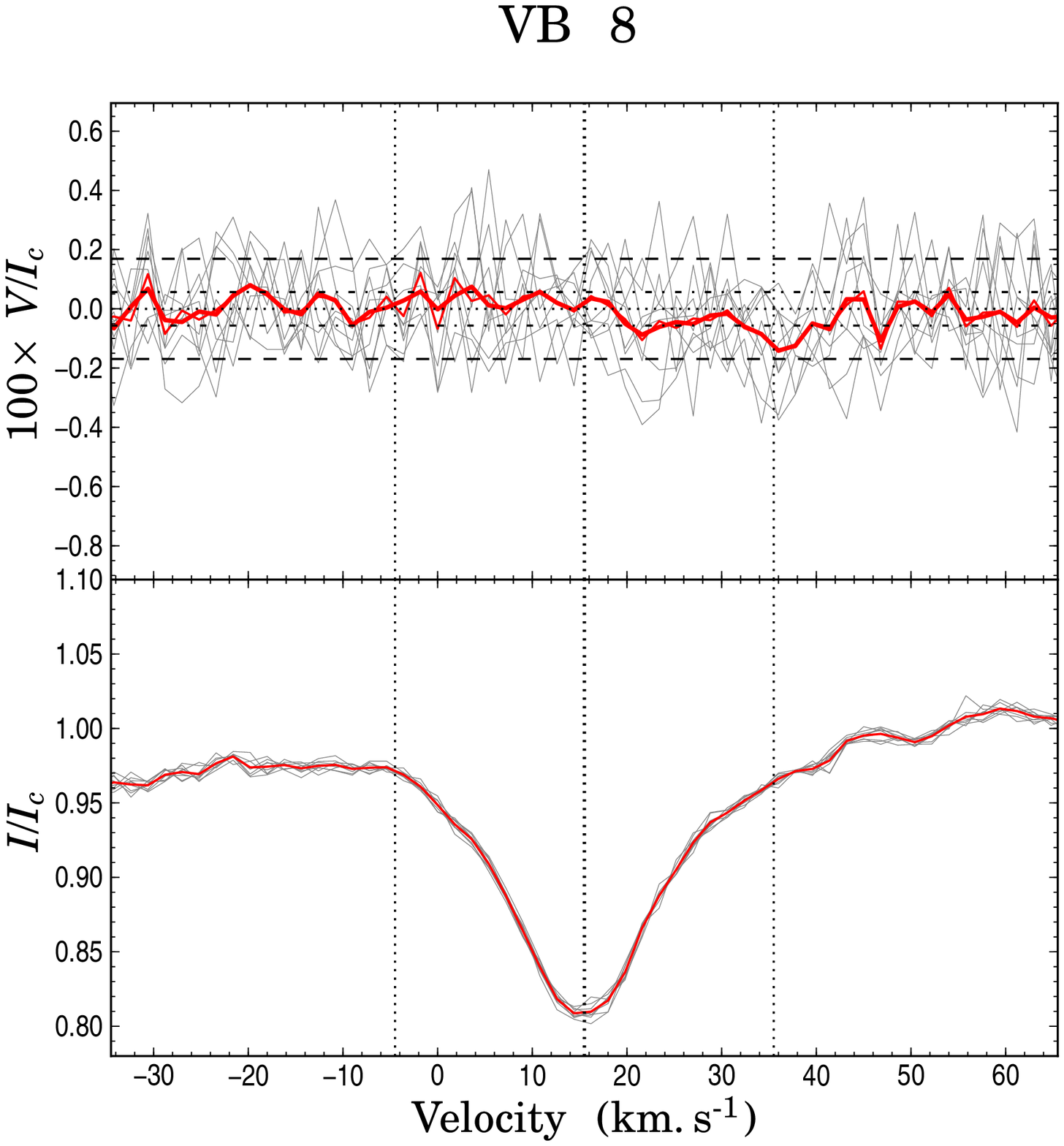} %
  \includegraphics[width=0.45\textwidth]{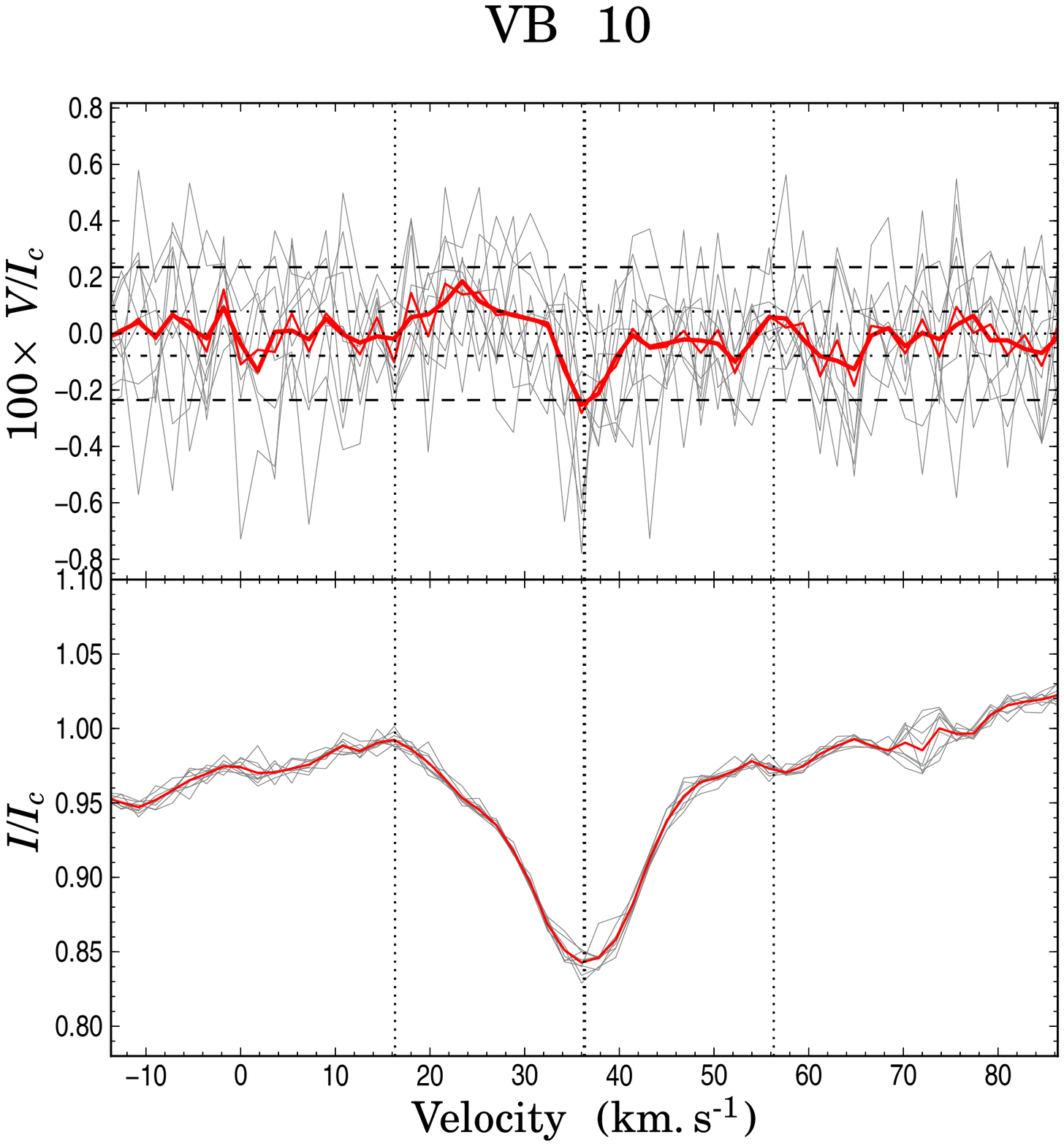} %
  \caption[]{Same as Fig.~\ref{fig:3nozdi} for VB~8 and VB~10. The bold
red line results from low-pass filtering of the Stokes~$V$ signatures and
dash-dotted lines show the $\pm 1\sigma$ levels corresponding the averaged
spectra.}
  \label{fig:vb8-10}
\end{figure*}

Performing a ZDI analysis on the VB~10 time series, we find 2 possible rotation
periods: 0.52, and  0.69~\d, the second being favored by photometric
measurements (MEarth project, J.~Irwin, private communication).
Figure~\ref{fig:vb10_spec_maps} shows the fit achieved for $\Prot=0.69~\d$,
$i=60~\degr$ and the corresponding magnetic maps. As expected from the signature
shape (non antisymmetric with respect to the line centre), the reconstructed
magnetic field exhibits a significant axisymmetric toroidal component. A
non-axisymmetric poloidal field (tilted quadrupole, mode $\alpha_{22}$) is also
reconstructed to fit the Stokes~$V$ component that varies along rotation.

\begin{figure*}
  \centering
  \includegraphics[height=0.4\textheight]{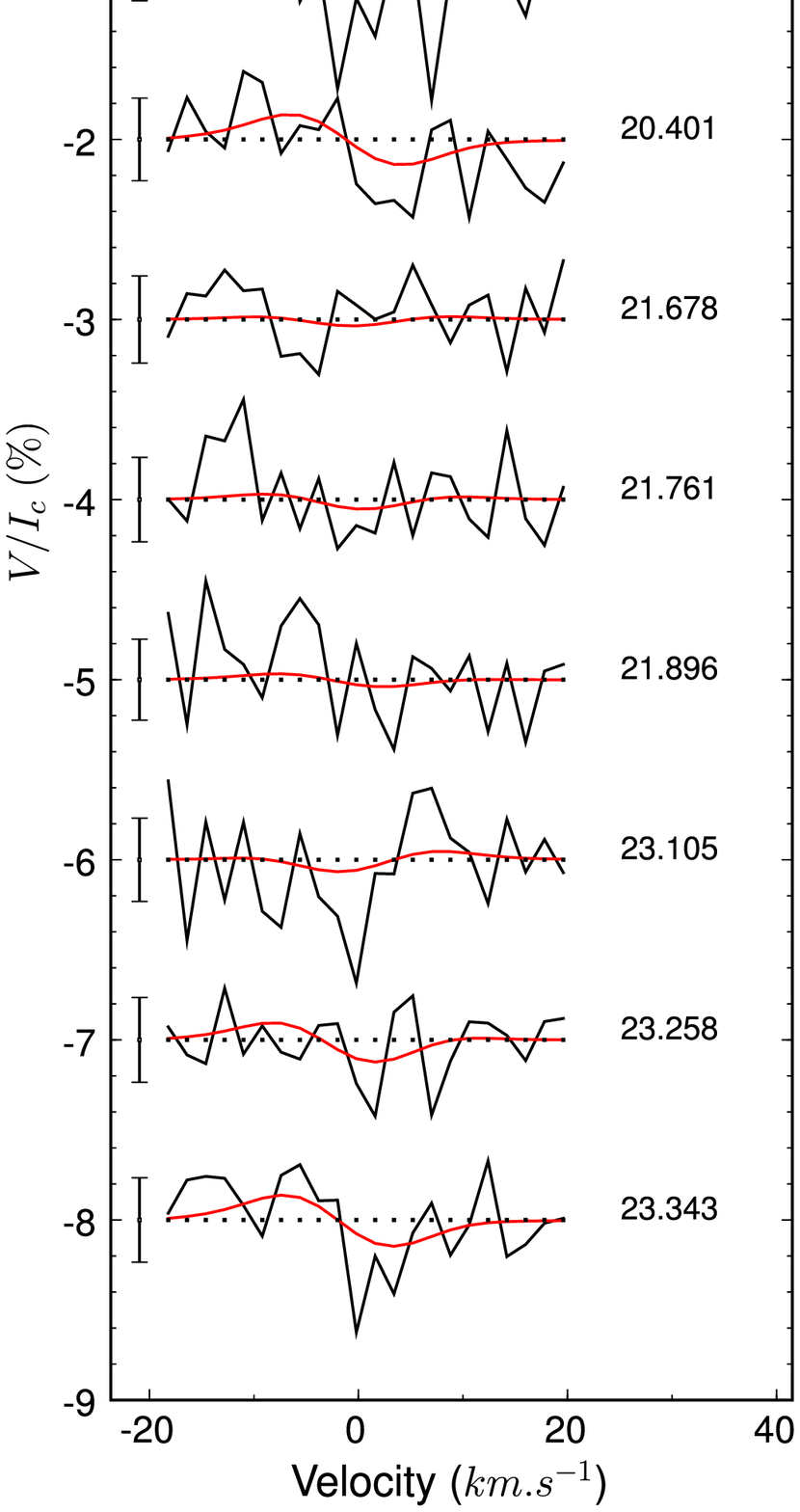}
  \includegraphics[height=0.4\textheight]{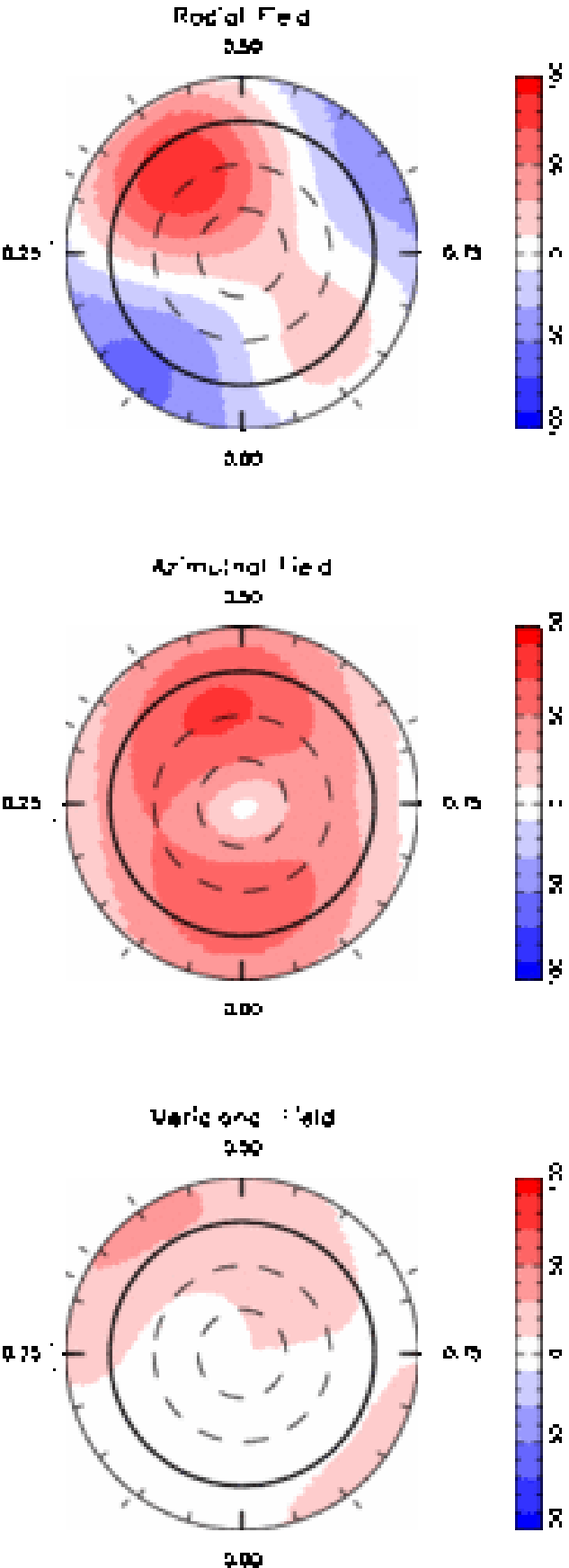}
  \caption[]{2009 spectra and magnetic maps of VB~10. See
  Fig.~\ref{fig:gj51_spec} and \ref{fig:gj51_map} for more details. Data
are
  phased according to the ephemeris ${\rm HJD}=2\,455\,000 + 0.69\,E$.}
  \label{fig:vb10_spec_maps}
\end{figure*}

\section{Discussion and conclusion}
\label{sec:disc}
\begin{table*}
\caption[]{Magnetic quantities derived from our study. For each star,
different observation epochs are presented separately. In columns 2--4 we report
quantities from Table~\ref{tab:sample}, respectively the stellar mass, the
rotation period, and the effective Rossby number. Columns 5, 6 and 7
mention the Stokes $V$ filling factor, the reconstructed magnetic energy and the
average magnetic flux. Columns 8--11 list the percentage of reconstructed
magnetic energy respectively lying in poloidal, dipole (poloidal and $\ell=1$),
quadrupole (poloidal and $\ell=2$), and octupole (poloidal and $\ell=3$) modes.
In column 12, we mention the percentage of magnetic energy reconstructed
in axisymmetric modes (defined as $m < \ell/2$) and the percentage of
poloidal energy in axisymmetric modes.
See section~\ref{sec:techniques-uncert} for a discussion on the robustness of
magnetic map reconstruction and the uncertainties
associated with the derived quantities.
}
 \begin{tabular}{ccccccccccccc}
\hline
Name & Mass & \Prot & $Ro$ & $f_V$ & $\avg{B^2}$ & $\avg{B}$ & pol. &
dipole &
quad. & oct. & axisymm. \\
 & (\msun) & (d) & ($10^{-2}$) & & ($\rm10^5\,G^2$) & (kG) & (\%) &
(\%) & (\%) & (\%) & (\%) \\ 
\hline
GJ~51$^1$ (06) & 0.21 & 1.02 & 1.2 & 0.12 & 38.6 & 1.61 & 99 & 96 &
0 & 2 & 91/91\\
\phantom{GJ~51} (07) &--&--&--& 0.12 & 31.3 & 1.58 & 99 & 92 & 0 & 6 &
77/77\\ 
\phantom{GJ~51} (08) &--&--&--& 0.12 & 32.6 & 1.65 & 97 & 92 & 1 & 3 &
89/89\\ 
GJ~1156$^2$ (07) & 0.14 & 0.49 & 0.5 & 1.0 & 0.06 & 0.05 & 88 & 30 & 26 & 19 &
6/3\\
\phantom{GJ~1156} (08) &--&--&--& 1.0 & 0.19 & 0.11 & 83 & 41 & 28 & 11 &
20/12\\
\phantom{GJ~1156} (09) &--&--&--& 1.0 & 0.13 & 0.10 & 94 & 54 & 24 & 10 &
2/1\\
GJ~1245~B (06) & 0.12 & 0.71 & 0.7 & 0.06 & 0.44 & 0.17 & 80 & 45 & 14 & 13 &
15/9\\
\phantom{GJ~1245~B} (07) &--&--&--& 0.10 & 0.49& 0.18 & 84 & 46 & 27 & 7 &
52/53\\
\phantom{GJ~1245~B} (08) &--&--&--& 0.10 & 0.06 & 0.06 & 85 & 33 & 25 & 19 &
20/18\\
WX~UMa (06) & 0.10 & 0.78 & 0.8 & 0.12 & 16.08 & 0.89 & 97 & 66 & 21 & 6 &
92/92\\
\phantom{WX~UMa} (07) &--&--&--& 0.12 & 24.42 & 0.94 & 97 & 71 & 13 & 3 &
92/94\\
\phantom{WX~UMa} (08) &--&--&--& 0.12 & 23.53 & 1.03 & 97 & 69 & 19 & 6 &
83/85\\
\phantom{WX~UMa} (09) &--&--&--& 0.12 & 37.54 & 1.06 & 96 & 89 & 2 & 2 &
95/96\\
DX~Cnc (07) & 0.10 & 0.46 & 0.5 & 0.20 & 0.17 & 0.11 & 93 & 69 & 11 & 9 &
77/77\\
\phantom{DX~Cnc} (08) &--&--&--& 0.20 & 0.09 & 0.08 & 73 & 31 & 25 & 10 &
49/34\\
\phantom{DX~Cnc} (09) &--&--&--& 0.20 & 0.09 & 0.08 & 62 & 42 & 11 & 4 &
70/61\\
GJ~3622 (08) & 0.09 & 1.5 & 1.5 & 1.0 & 0.04 & 0.05 & 96 & 90 & 7 & -- &
73/72\\ 
\phantom{GJ~3622} (09) &--&--&--& 1.0 & 0.05 & 0.06 & 93 & 84 & 9 & -- & 80/78\\
\hline
 \label{tab:syn}
 \end{tabular}
 \begin{flushleft}
   $^1$ For GJ~51, the imaging process is weakly constrained by our
   data sets due to poor phase coverage (see Sec.~\ref{sec:gj51}).\\
   $^2$ For GJ~1156 the alternative rotation period 0.33~d cannot be
   definitely excluded (see Sec.~\ref{sec:gj1156}). In this case, the
   reconstructed topologies remain   similar for the 3 epochs, and our
   conclusions are not affected. 
 \end{flushleft}
\end{table*}

We present the final part of our exploratory spectropolarimetric
survey of M dwarfs, following D08 and M08b (respectively concentrating on
mid and early M dwarfs) we focus here on the low mass end of our sample.
For 6 stars, it is possible to apply ZDI techniques to the time-series
of circularly polarised spectra and thus to infer the large-scale component of
their magnetic topologies. The properties of the reconstructed topologies of
these stars are presented in Table~\ref{tab:syn}. For the remaining 5 objects,
the data sets do not permit such a study, it is however possible to retrieve
some constraints about their magnetic properties.

Two stars of the subsample (namely GJ~51 and WX~UMa) exhibit large-scale
magnetic fields very similar to those observed by M08b on mid M dwarfs,
i.e. very strong, axisymmetric poloidal and nearly dipolar fields, with
very little temporal variations. For three stars for which we cannot perform a
definitive ZDI reconstruction (GJ~1154~A, GJ~1224 and CN~Leo) our data
strongly suggest similar topologies. From the observations of WX~UMa, we
demonstrate that the timescale of temporal evolution in the magnetic topologies
of these stars can be larger than 3~years, whereas previous observations by M08a
and M08b were based on observations spanning only 1~year.

The other stars for which we can reconstruct the large-scale magnetic
topologies, are clearly different. They are weaker than those of the first
category, and generally feature a significant non-axisymmetric
component, plus a significant toroidal component (although the field is always
predominantly poloidal).
Temporal variability is also noticeable, in particular
for GJ~1245~B our data indicate unambiguously that the magnetic field strength
has dramatically decreased between our 2007 and 2008 observations.
Our conclusions are robust to uncertainties on stellar parameters
(\Prot, $i$ and $\vsini$). When varying these parameters over the width
of their respective error bars (see section~\ref{sec:techniques-uncert}), the
main properties of the reconstructed magnetic topologies remain unchanged.

\begin{figure*}
  \centering
  \includegraphics[height=1.0\textwidth, angle=270]{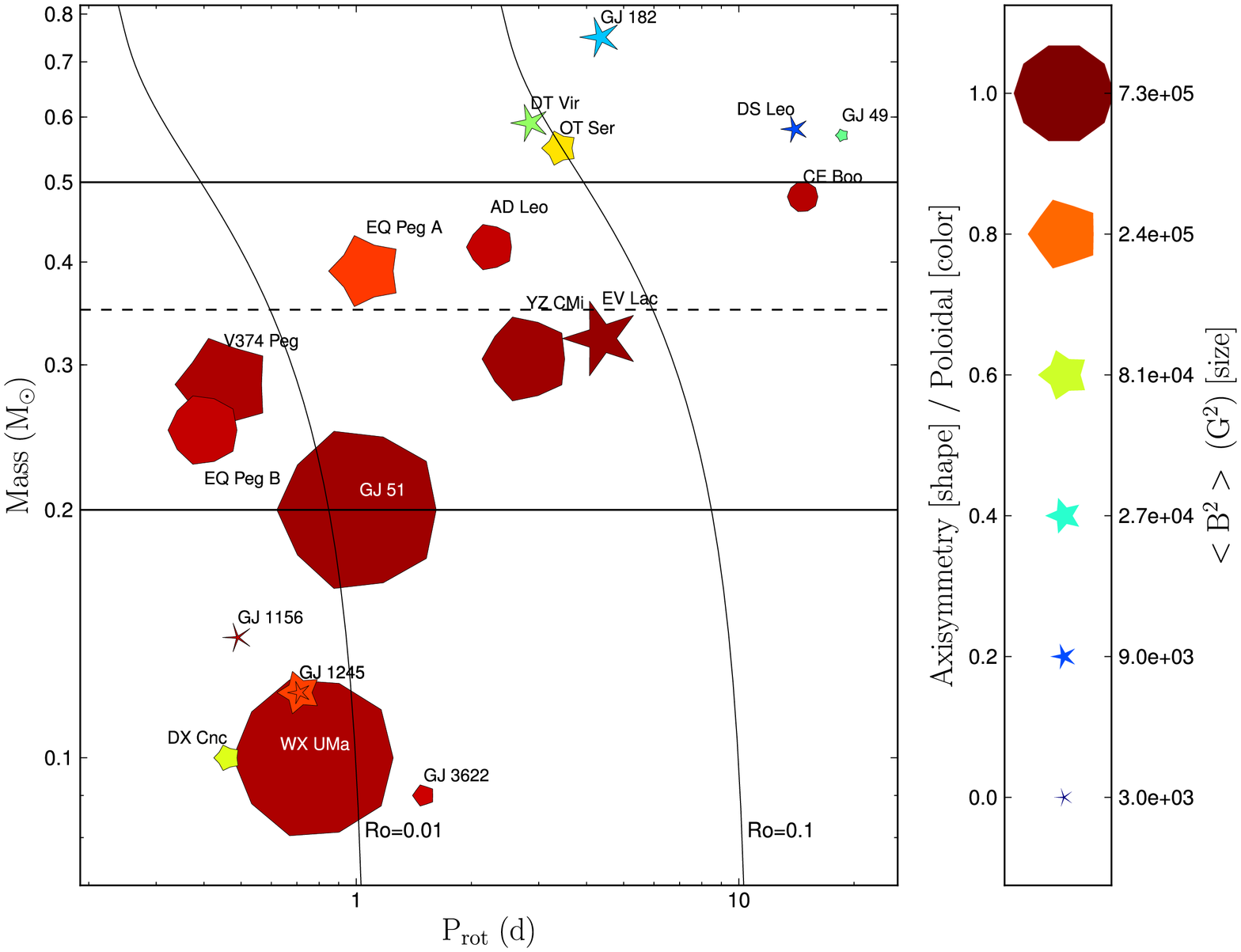}
  \caption{Properties of the magnetic topologies of our sample of M dwarfs as a
  function of rotation period and stellar mass. Larger symbols indicate larger
  magnetic fields while symbol shapes depict the different degrees of
  axisymmetry of the
  reconstructed magnetic field (from decagons for purely axisymmetric fields to
  sharp stars for purely non axisymmetric fields). Colours illustrate the field
  configuration (dark blue for purely toroidal fields, dark red for purely
  poloidal fields and intermediate colours for intermediate configurations).
  Solid lines represent contours of constant Rossby number $Ro=0.1$ and $0.01$
  respectively corresponding approximately to the saturation and
  super-saturation thresholds \citep[e.g.,][]{Pizzolato03}. The theoretical
  full-convection limit ($\mstar \simeq0.35\msun$, \citealt{Chabrier97}) is
  plotted as a horizontal dashed line, and the approximate limits of the three
  stellar groups  discussed in the text are represented as horizontal solid
  lines. Stars with  $\mstar > 0.45~\msun$ are from
  D08, whereas those with $0.25<\mstar<0.45~\msun$ are from M08b. For GJ~1245~B
  symbols corresponding to 2007 and 2008 data sets are superimposed in order to
  emphasize the variability of this object. %
  Uncertainties associated with the plotted magnetic quantities are discussed
  in section~\ref{sec:techniques-uncert}.}
\label{fig:plotMP}
\end{figure*}

These results are presented in a more visual way in Fig.~\ref{fig:plotMP}.
Previous studies by D08 and M08b have revealed strong evidence that a clear
transition occurs at approximately
0.5~\msun\, \ie more or less coincident with the transition to a
fully convective internal structure. The situation here is different, we find
stars with similar stellar parameters that exhibit radically different magnetic
topologies. On Fig.~\ref{fig:plotMP}, WX~UMa is the only star below 0.2~\msun\
to host a mid-M-dwarf-like field. Whereas DX~Cnc and GJ~1245~B are very
close to it in the mass-rotation plane they feature fields with very different
properties. This observation may be explained in several ways. For instance,
another parameter than mass and rotation period, such as stellar age, may play a
role. In our sample we indeed notice that most stars below 0.15~\msun\ that
exhibit a weak complex field belong to a young kinematic population
according to \citealt{Delfosse98} (GJ~1156, DX~Cnc and GJ~3622), whereas those
hosting a strong dipolar field (WX~UMa, GJ~1224 and CN~Leo) belong to older
kinematic populations (old disk and old/young disk). This hypothesis requires
further investigation. One could also imagine, for instance, that the
magnetic fields of very low mass stars may switch between two different states
over time. This hypothesis is supported by the fact that for one of the stars in
the weak and complex field regime (GJ~1245~B) we observe a dramatic variation of
the magnetic flux on a timescale of one year which may indicate that the
magnetic field of these objects may go through chaotic variations and eventually
switch between the two categories of field actually observed. However no such
switch has been observed in our sample. We observe 5 objects in the strong
dipole field category (GJ~51, GJ~1154~A, GJ~1224, CN~Leo and WX~UMa) and 6 in
the weak field category (GJ~1156, GJ~1245~B, DX~Cnc, GJ~3622, VB~8 and VB~10),
indicating that stars would spend as much time in both states. No star is
observed in an intermediate state, suggesting that a putative transition would
be fast. This hypothesis may be investigated through the analysis of long-term
radio monitoring, since radio emission would be presumably strongly impacted by
such a dramatic change of the stellar magnetic field. Recent observations of
ultracool dwarfs reveal a long-term variability of the activity indices
\cite[\eg][]{Antonova07, Berger10} that may support this view.

\begin{figure*}
 \centering
 \includegraphics[angle=270, width=0.50\textwidth]{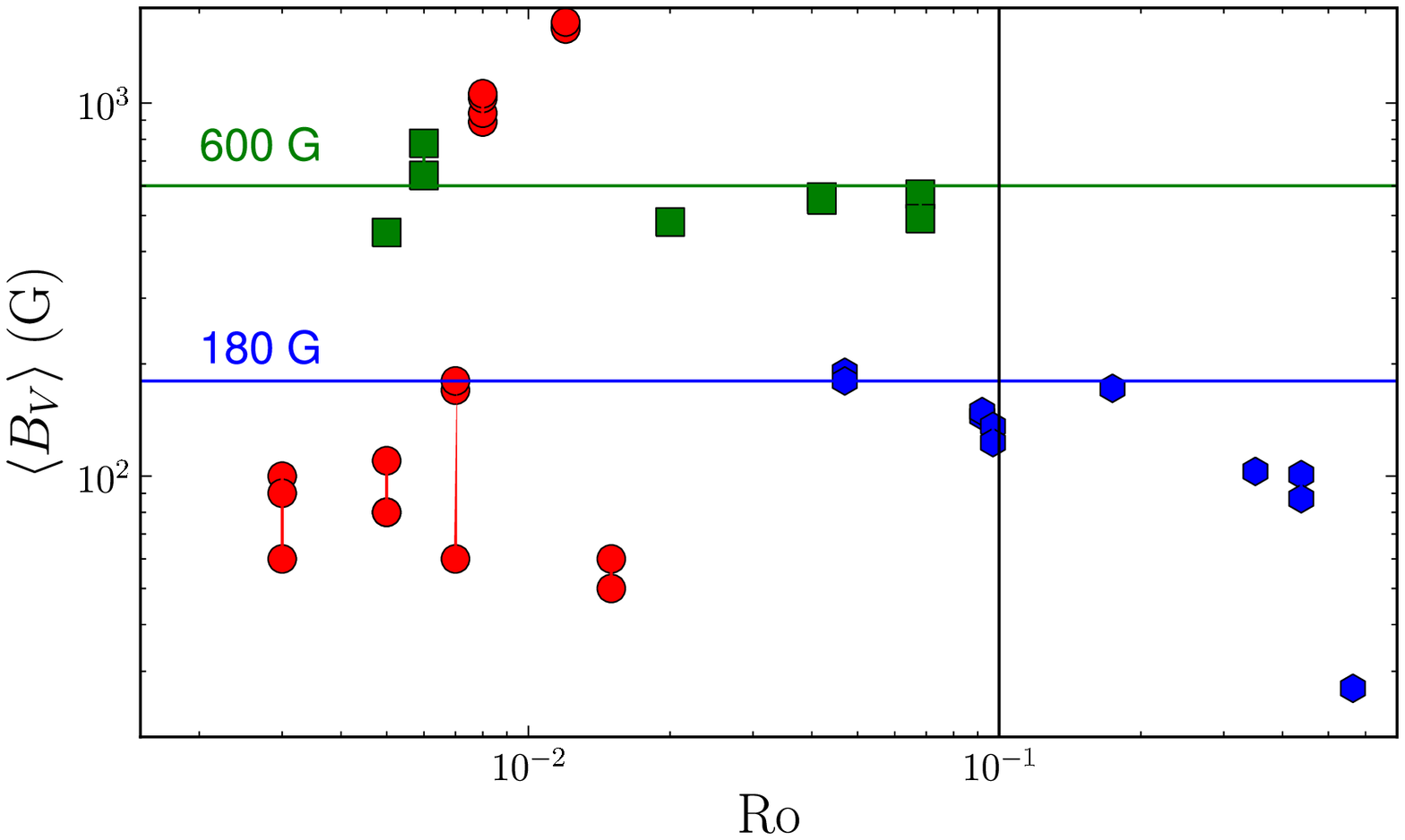}%
 \includegraphics[angle=270, width=0.50\textwidth]{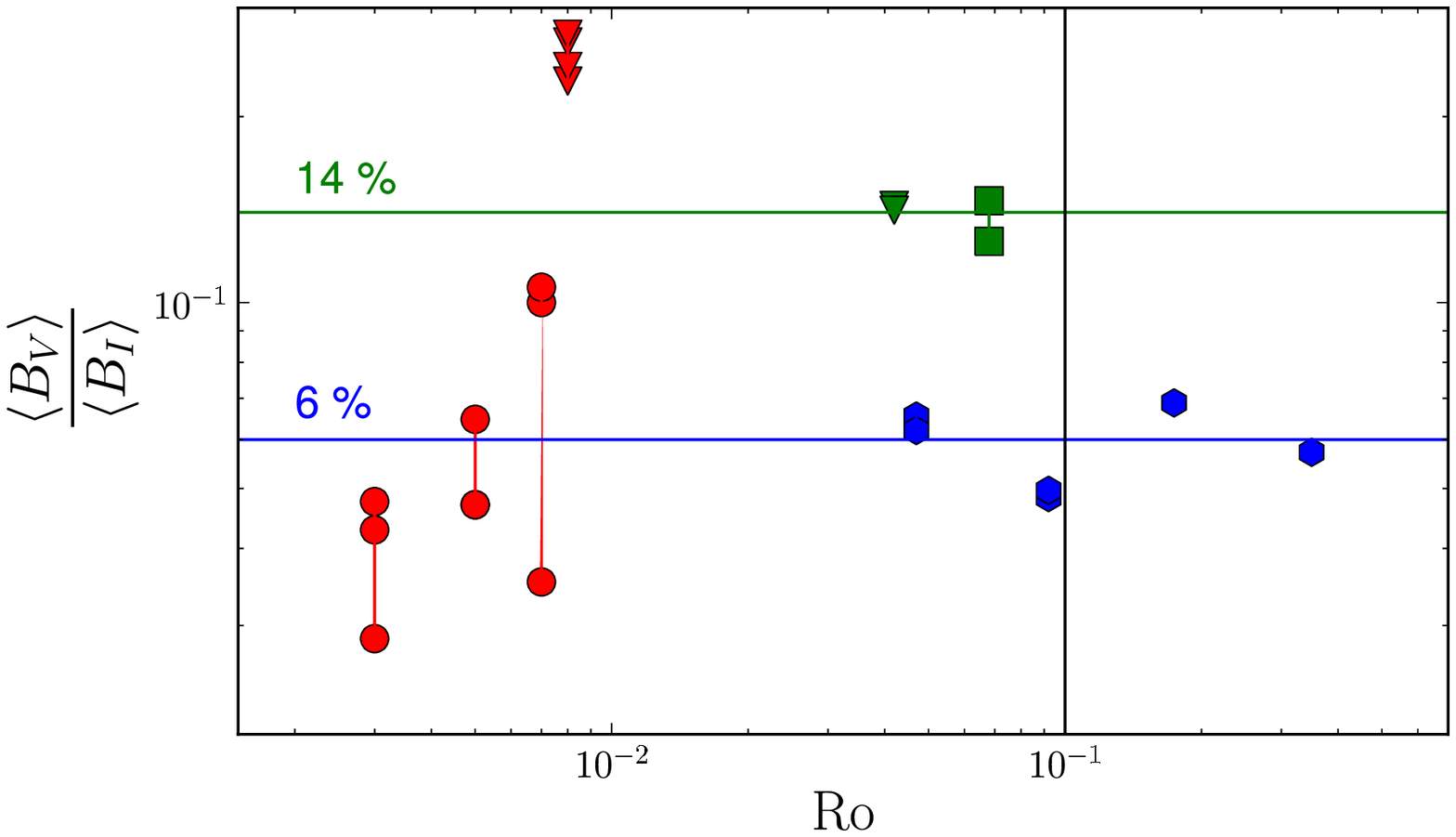}%
 \caption[]{Magnetic flux as a function of Rossby number. On the left
panel magnetic fluxes as measured from Stokes~$V$ spectra and ZDI by M08b
and D08 are mentioned as blue hexagons ($\mstar>0.4~\msun$) and green
squares ($0.2<\mstar<0.4~\msun$). Results from this paper
($\mstar\leq0.2~\msun$) are shown as red circles. On the
right panel the ratio between the magnetic fluxes as recovered from Stokes~$V$
and Stokes~$I$ measurements (whenever available) are shown. Stokes~$I$ magnetic
fluxes are taken from \cite{Johns00}, \cite{Reiners07}, \cite{Reiners09}, and
\cite{Reiners09b} (see Tab.~\ref{tab:sample}). Measurements of different epochs
(whenever available) for Stokes~$V$ are shown connected by a solid line. On
each plot the $Ro=10^{-1}$, corresponding to the saturation level, is depicted
as a vertical solid black line. Triangles denote upper limits. In the left
panel, horizontal lines show the magnetic fluxes corresponding to saturation for
stars with $\mstar>0.4~\msun$ (blue), and $0.2<\mstar<0.4~\msun$ (green). In
the right panel they show the mean fraction of magnetic flux detected in
Stokes~$V$ spectra.}
 \label{fig:RoBsq}
\end{figure*}

On the left panel of Fig.~\ref{fig:RoBsq}, we plot the reconstructed magnetic
flux as a function of the Rossby number. As mentioned in D08, stars with
$\mstar > 0.4~\msun$ follow the expected rotation--magnetic field connection,
with saturation for $Ro \lesssim 0.2$. Whereas for $\mstar > 0.4~\msun$, we only
observe objects in the saturated regime, with a significantly stronger
saturation magnetic flux. All the stars of the late M subsample have Rossby
numbers below $2\times10^{-2}$ and are thus expected to be in the saturated
regime. As in Fig.~\ref{fig:plotMP} the stars studied here can be divided into
two distinct categories. The first one is composed of stars hosting a very
strong magnetic field that lie in the saturated part of the rotation-activity
relation, similar to mid M dwarfs. We notice that significantly higher magnetic
fluxes are reconstructed for WX~UMa and GJ~51 than for mid M dwarfs studied by
M08b, although all these stars seem to lie in the saturated dynamo regime. For
the second category of late M dwarfs (having a weak complex field), we recover
less magnetic flux (even less than for some early M dwarfs studied by D08), in
spite of their very low Rossby number. Super-saturation is unlikely to be the
explanation since stars of both categories have similar Rossby numbers. As
mentioned for Fig.~\ref{fig:plotMP}, we observe no object in an intermediate
state. The aforementioned variability of GJ~1245~B is also prominent in this
plot.

On the right panel of Fig.~\ref{fig:RoBsq}, magnetic flux inferred from
Stokes~$V$ measurements are compared to those derived from Stokes~$I$.
Again, stars with masses below and above 0.4~\msun\ studied by
D08 and M08b clearly form two separate groups. In partly-convective
stars only a few percents of the magnetic flux measured in $I$ is detectable in
$V$ similarly to what is observed for the Sun, while this ratio is
close to 15\% in fully convective ones. This indicates a higher degree of
organization of the field in fully convective mid M dwarfs, with more magnetic
flux in the spherical harmonics of lowest degree. The ratios of about 30\%
plotted for WX~UMa are upper limits (since the flux based on Stokes~$I$ is a
lower limit), it is thus not clear if this star differs from mid M dwarfs in
this respect. This high value however indicates a high degree of organization.
For DX~Cnc, GJ~1245~B and GJ~1156, the ratio of Stokes~$V$ and $I$ fluxes is
closer to the early M dwarfs value. Therefore the magnetic fields of the weak
field category of late M dwarfs share similar properties with that of partly
convective M dwarfs. 

These results on the magnetic topologies of M dwarfs also suggest that
dynamo processes in low-mass main sequence stars and pre-main sequence stars
may be similar. Indeed the first spectropolarimetric results on young stars show
that the fully convective T Tauri star BP~Tau (0.7~\msun) exhibit a strong
large-scale magnetic field \cite[][]{Donati08a}, whereas the more massive partly
convective star V2129~Oph (1.4~\msun) possesses a weaker and more complex
field \cite[][]{Donati07}. This is reminiscent of the transition we observe at
0.5~\msun\ among main sequence M dwarfs. Recent observations, on the fully
convective V2247~Oph (0.35~\msun) reveal a still weaker and more complex
magnetic field \cite[][]{Donati10} which may correspond to the weak field late M
dwarf we present here. 

We do not detect Stokes~$V$ signatures in individual spectra of the two latest
stars of our sample VB~8 and VB~10. However the averaged LSD profile of VB~10
suggests the presence of a toroidal axisymmetric field component on this object.
Further observations may confirm this first spectropolarimetric detection on an
ultra-cool dwarf.

The first spectropolarimetric survey of M dwarfs has already provided dynamo
theorists with strong constraints on the evolution of surface magnetic fields of
M dwarfs across the fully convective divide (D08 and M08b). The results
presented in this paper on the magnetic topologies of late M dwarfs reveal a new
unexpected behaviour below 0.2~\msun. We interpret the fact that objects with
similar stellar parameters host radically different magnetic topologies as a
possible evidence for a switch between two dynamo states (either cyclic or
chaotic). Finally our observations suggest the presence of a large-scale
magnetic field on the M8 dwarf VB~10, featuring a significant toroidal
axisymmetric component, whereas the detection of the magnetic field of VB~8 (M7)
is not possible from our spectropolarimetric data.

\section*{ACKNOWLEDGEMENTS} 
The authors thank the CFHT staff for their valuable help throughout our
observing runs. We are grateful to Jonathan Irwin and Joel Hartman for
providing results prior to publication on the photometric periods of the M
dwarfs studied in this paper. We also thank the referee Gibor Basri for his
fruitful suggestions.


\begin{thebibliography}{}

\bibitem[\protect\citeauthoryear{{Antonova}, {Doyle}, {Hallinan}, {Golden} \&
  {Koen}}{{Antonova} et~al.}{2007}]{Antonova07}
{Antonova} A.,  {Doyle} J.~G.,  {Hallinan} G.,  {Golden} A.,    {Koen} C.,
  2007, \aap, 472, 257

\bibitem[\protect\citeauthoryear{{Babcock}}{{Babcock}}{1961}]{Babcock61}
{Babcock} H.~W.,  1961, \apj, 133, 572

\bibitem[\protect\citeauthoryear{{Bakos}, {Noyes}, {Kov{\'a}cs}, {Stanek},
  {Sasselov} \& {Domsa}}{{Bakos} et~al.}{2004}]{Bakos04}
{Bakos} G.,  {Noyes} R.~W.,  {Kov{\'a}cs} G.,  {Stanek} K.~Z.,  {Sasselov}
  D.~D.,    {Domsa} I.,  2004, \pasp, 116, 266

\bibitem[\protect\citeauthoryear{{Baraffe}, {Chabrier}, {Allard} \&
  {Hauschildt}}{{Baraffe} et~al.}{1998}]{Baraffe98}
{Baraffe} I.,  {Chabrier} G.,  {Allard} F.,    {Hauschildt} P.~H.,  1998, \aap,
  337, 403

\bibitem[\protect\citeauthoryear{{Berger}}{{Berger}}{2006}]{Berger06}
{Berger} E.,  2006, \apj, 648, 629

\bibitem[\protect\citeauthoryear{{Berger}, {Basri}, {Fleming}, {Giampapa},
  {Gizis}, {Liebert}, {Mart{\'{\i}}n}, {Phan-Bao} \& {Rutledge}}{{Berger}
  et~al.}{2010}]{Berger10}
{Berger} E.,  {Basri} G.,  {Fleming} T.~A.,  {Giampapa} M.~S.,  {Gizis} J.~E.,
  {Liebert} J.,  {Mart{\'{\i}}n} E.,  {Phan-Bao} N.,    {Rutledge} R.~E.,
  2010, \apj, 709, 332

\bibitem[\protect\citeauthoryear{{Browning}}{{Browning}}{2008}]{Browning08}
{Browning} M.~K.,  2008, \apj, 676, 1262

\bibitem[\protect\citeauthoryear{{Chabrier} \& {Baraffe}}{{Chabrier} \&
  {Baraffe}}{1997}]{Chabrier97}
{Chabrier} G.,  {Baraffe} I.,  1997, \aap, 327, 1039

\bibitem[\protect\citeauthoryear{{Chabrier} \& {K{\"u}ker}}{{Chabrier} \&
  {K{\"u}ker}}{2006}]{Chabrier06}
{Chabrier} G.,  {K{\"u}ker} M.,  2006, \aap, 446, 1027

\bibitem[\protect\citeauthoryear{{Charbonneau}}{{Charbonneau}}{2005}]{Charbonn%
eau05}
{Charbonneau} P.,  2005, Living Reviews in Solar Physics, 2, 2

\bibitem[\protect\citeauthoryear{{Cowling}}{{Cowling}}{1933}]{Cowling34}
{Cowling} T.~G.,  1933, \mnras, 94, 39

\bibitem[\protect\citeauthoryear{{Cutri}, {Skrutskie}, {van Dyk}, {Beichman},
  {Carpenter}, {Chester}, {Cambresy}, {Evans}, {Fowler}, {Gizis}, {Howard},
  {Huchra}, {Jarrett}, {Kopan}, {Kirkpatrick}, {Light}, {Marsh} \&
  {McCallon}}{{Cutri} et~al.}{2003}]{Cutri03}
{Cutri} R.~M.,  {Skrutskie} M.~F.,  {van Dyk} S.,  {Beichman} C.~A.,
  {Carpenter} J.~M.,  {Chester} T.,  {Cambresy} L.,  {Evans} T.,  {Fowler} J.,
  {Gizis} J.,  {Howard} E.,  {Huchra} J.,  {Jarrett} T.,  {Kopan} E.~L.,
  {Kirkpatrick} J.~D.,  {Light} R.~M.,  {Marsh} K.~A.,    {McCallon} 2003,
  {2MASS All Sky Catalog of point sources.}.
The IRSA 2MASS All-Sky Point Source Catalog, NASA/IPAC Infrared Science
  Archive.~http://irsa.ipac.caltech.edu/applications/Gator/

\bibitem[\protect\citeauthoryear{{Delfosse}, {Forveille}, {Perrier} \&
  {Mayor}}{{Delfosse} et~al.}{1998}]{Delfosse98}
{Delfosse} X.,  {Forveille} T.,  {Perrier} C.,    {Mayor} M.,  1998, \aap, 331,
  581

\bibitem[\protect\citeauthoryear{{Delfosse}, {Forveille}, {S{\'e}gransan},
  {Beuzit}, {Udry}, {Perrier} \& {Mayor}}{{Delfosse} et~al.}{2000}]{Delfosse00}
{Delfosse} X.,  {Forveille} T.,  {S{\'e}gransan} D.,  {Beuzit} J.-L.,  {Udry}
  S.,  {Perrier} C.,    {Mayor} M.,  2000, \aap, 364, 217

\bibitem[\protect\citeauthoryear{{Demory}, {S{\'e}gransan}, {Forveille},
  {Queloz}, {Beuzit}, {Delfosse}, {di Folco}, {Kervella}, {Le Bouquin},
  {Perrier}, {Benisty}, {Duvert}, {Hofmann}, {Lopez} \& {Petrov}}{{Demory}
  et~al.}{2009}]{Demory09}
{Demory} B.,  {S{\'e}gransan} D.,  {Forveille} T.,  {Queloz} D.,  {Beuzit} J.,
  {Delfosse} X.,  {di Folco} E.,  {Kervella} P.,  {Le Bouquin} J.,  {Perrier}
  C.,  {Benisty} M.,  {Duvert} G.,  {Hofmann} K.,  {Lopez} B.,    {Petrov} R.,
  2009, \aap, 505, 205

\bibitem[\protect\citeauthoryear{{Dobler}, {Stix} \& {Brandenburg}}{{Dobler}
  et~al.}{2006}]{Dobler06}
{Dobler} W.,  {Stix} M.,    {Brandenburg} A.,  2006, \apj, 638, 336

\bibitem[\protect\citeauthoryear{{Donati}, {Skelly}, {Bouvier}, {Jardine},
  {Gregory}, {Morin}, {Hussain}, {Dougados}, {M{\'e}nard} \& {Unruh}}{{Donati}
  et~al.}{2010}]{Donati10}
{Donati} J.,  {Skelly} M.~B.,  {Bouvier} J.,  {Jardine} M.~M.,  {Gregory}
  S.~G.,  {Morin} J.,  {Hussain} G.~A.~J.,  {Dougados} C.,  {M{\'e}nard} F.,
  {Unruh} Y.,  2010, \mnras, 402, 1426

\bibitem[\protect\citeauthoryear{{Donati} \& {Brown}}{{Donati} \&
  {Brown}}{1997}]{Donati97b}
{Donati} J.-F.,  {Brown} S.~F.,  1997, \aap, 326, 1135

\bibitem[\protect\citeauthoryear{{Donati}, {Forveille}, {Cameron}, {Barnes},
  {Delfosse}, {Jardine} \& {Valenti}}{{Donati} et~al.}{2006a}]{Donati06}
{Donati} J.-F.,  {Forveille} T.,  {Cameron} A.~C.,  {Barnes} J.~R.,  {Delfosse}
  X.,  {Jardine} M.~M.,    {Valenti} J.~A.,  2006a, Science, 311, 633

\bibitem[\protect\citeauthoryear{{Donati}, {Howarth}, {Jardine}, {Petit},
  {Catala}, {Landstreet}, {Bouret}, {Alecian}, {Barnes}, {Forveille}, {Paletou}
  \& {Manset}}{{Donati} et~al.}{2006b}]{Donati06b}
{Donati} J.-F.,  {Howarth} I.~D.,  {Jardine} M.~M.,  {Petit} P.,  {Catala} C.,
  {Landstreet} J.~D.,  {Bouret} J.-C.,  {Alecian} E.,  {Barnes} J.~R.,
  {Forveille} T.,  {Paletou} F.,    {Manset} N.,  2006b, \mnras, 370, 629

\bibitem[\protect\citeauthoryear{{Donati}, {Jardine}, {Gregory}, {Petit},
  {Bouvier}, {Dougados}, {M{\'e}nard}, {Cameron}, {Harries}, {Jeffers} \&
  {Paletou}}{{Donati} et~al.}{2007}]{Donati07}
{Donati} J.-F.,  {Jardine} M.~M.,  {Gregory} S.~G.,  {Petit} P.,  {Bouvier} J.,
   {Dougados} C.,  {M{\'e}nard} F.,  {Cameron} A.~C.,  {Harries} T.~J.,
  {Jeffers} S.~V.,    {Paletou} F.,  2007, \mnras, 380, 1297

\bibitem[\protect\citeauthoryear{{Donati}, {Jardine}, {Gregory}, {Petit},
  {Paletou}, {Bouvier}, {Dougados}, {M{\'e}nard}, {Cameron}, {Harries},
  {Hussain}, {Unruh}, {Morin}, {Marsden}, {Manset}, {Auri{\`e}re}, {Catala} \&
  {Alecian}}{{Donati} et~al.}{2008a}]{Donati08a}
{Donati} J.-F.,  {Jardine} M.~M.,  {Gregory} S.~G.,  {Petit} P.,  {Paletou} F.,
   {Bouvier} J.,  {Dougados} C.,  {M{\'e}nard} F.,  {Cameron} A.~C.,  {Harries}
  T.~J.,  {Hussain} G.~A.~J.,  {Unruh} Y.,  {Morin} J.,  {Marsden} S.~C.,
  {Manset} N.,  {Auri{\`e}re} M.,  {Catala} C.,    {Alecian} E.,  2008a, \mnras,
  386, 1234

\bibitem[\protect\citeauthoryear{{Donati}, {Morin}, {Petit}, {Delfosse},
  {Forveille}, {Auri{\`e}re}, {Cabanac}, {Dintrans}, {Fares}, {Gastine},
  {Jardine}, {Ligni{\`e}res}, {Paletou}, {Velez} \& {Th{\'e}ado}}{{Donati}
  et~al.}{2008b}]{Donati08b}
{Donati} J.-F.,  {Morin} J.,  {Petit} P.,  {Delfosse} X.,  {Forveille} T.,
  {Auri{\`e}re} M.,  {Cabanac} R.,  {Dintrans} B.,  {Fares} R.,  {Gastine} T.,
  {Jardine} M.~M.,  {Ligni{\`e}res} F.,  {Paletou} F.,  {Velez} J.~C.~R.,
  {Th{\'e}ado} S.,  2008b, \mnras, 390, 545

\bibitem[\protect\citeauthoryear{{Donati}, {Semel}, {Carter}, {Rees} \&
  {Cameron}}{{Donati} et~al.}{1997}]{Donati97}
{Donati} J.-F.,  {Semel} M.,  {Carter} B.~D.,  {Rees} D.~E.,    {Cameron}
  A.~C.,  1997, \mnras, 291, 658

\bibitem[\protect\citeauthoryear{{Durney}, {De Young} \& {Roxburgh}}{{Durney}
  et~al.}{1993}]{Durney93}
{Durney} B.~R.,  {De Young} D.~S.,    {Roxburgh} I.~W.,  1993, Solar Physics,
  145, 207

\bibitem[\protect\citeauthoryear{{ESA}}{{ESA}}{1997}]{ESA97}
{ESA} 1997, VizieR Online Data Catalog, 1239, 0

\bibitem[\protect\citeauthoryear{{Gizis}, {Reid} \& {Hawley}}{{Gizis}
  et~al.}{2002}]{Gizis02}
{Gizis} J.~E.,  {Reid} I.~N.,    {Hawley} S.~L.,  2002, \aj, 123, 3356

\bibitem[\protect\citeauthoryear{{Guedel} \& {Benz}}{{Guedel} \&
  {Benz}}{1993}]{Guedel93}
{Guedel} M.,  {Benz} A.~O.,  1993, \apjl, 405, L63

\bibitem[\protect\citeauthoryear{{Horne}}{{Horne}}{1986}]{Horne86}
{Horne} K.,  1986, \pasp, 98, 609

\bibitem[\protect\citeauthoryear{{Irwin}, {Charbonneau}, {Nutzman} \&
  {Falco}}{{Irwin} et~al.}{2009}]{Irwin09}
{Irwin} J.,  {Charbonneau} D.,  {Nutzman} P.,    {Falco} E.,  2009, in
  {E.~Stempels} ed., American Institute of Physics Conference Series Vol.~1094
  of American Institute of Physics Conference Series, {The MEarth project:
  searching for transiting habitable super-Earth planets around nearby
  M-dwarfs}.
pp 445--448

\bibitem[\protect\citeauthoryear{{Johns-Krull} \& {Valenti}}{{Johns-Krull} \&
  {Valenti}}{1996}]{Johns96}
{Johns-Krull} C.~M.,  {Valenti} J.~A.,  1996, \apjl, 459, L95+

\bibitem[\protect\citeauthoryear{{Johns-Krull} \& {Valenti}}{{Johns-Krull} \&
  {Valenti}}{2000}]{Johns00}
{Johns-Krull} C.~M.,  {Valenti} J.~A.,  2000, in {Pallavicini} R.,  {Micela}
  G.,   {Sciortino} S.,  eds, Stellar Clusters and Associations: Convection,
  Rotation, and Dynamos Vol.~198 of Astronomical Society of the Pacific
  Conference Series, {Measurements of stellar magnetic fields}.
pp 371--+

\bibitem[\protect\citeauthoryear{{Kiraga} \& {Stepien}}{{Kiraga} \&
  {Stepien}}{2007}]{Kiraga07}
{Kiraga} M.,  {Stepien} K.,  2007, Acta Astronomica, 57, 149

\bibitem[\protect\citeauthoryear{{K{\"u}ker} \& {R{\"u}diger}}{{K{\"u}ker} \&
  {R{\"u}diger}}{1999}]{Kuker99}
{K{\"u}ker} M.,  {R{\"u}diger} G.,  1999, \aap, 346, 922

\bibitem[\protect\citeauthoryear{{Kurucz}}{{Kurucz}}{1993}]{Kurucz93}
{Kurucz} R.,  1993, CDROM \#~13 (ATLAS9 atmospheric models) and \#~18 (ATLAS9
  and SYNTHE routines, spectral line database).
Smithsonian Astrophysical Observatory, Washington D.C.

\bibitem[\protect\citeauthoryear{{Larmor}}{{Larmor}}{1919}]{Larmor19}
{Larmor} J.,  1919, Rep. Brit. Assoc. Adv. Sci.

\bibitem[\protect\citeauthoryear{{Leighton}}{{Leighton}}{1969}]{Leighton69}
{Leighton} R.~B.,  1969, \apj, 156, 1

\bibitem[\protect\citeauthoryear{{Marsh}}{{Marsh}}{1989}]{Marsh89}
{Marsh} T.~R.,  1989, \pasp, 101, 1032

\bibitem[\protect\citeauthoryear{{Mohanty} \& {Basri}}{{Mohanty} \&
  {Basri}}{2003}]{Mohanty03}
{Mohanty} S.,  {Basri} G.,  2003, \apj, 583, 451

\bibitem[\protect\citeauthoryear{{Morin}, {Donati}, {Forveille}, {Delfosse},
  {Dobler}, {Petit}, {Jardine}, {Cameron}, {Albert}, {Manset}, {Dintrans},
  {Chabrier} \& {Valenti}}{{Morin} et~al.}{2008a}]{Morin08a}
{Morin} J.,  {Donati} J.-F.,  {Forveille} T.,  {Delfosse} X.,  {Dobler} W.,
  {Petit} P.,  {Jardine} M.~M.,  {Cameron} A.~C.,  {Albert} L.,  {Manset} N.,
  {Dintrans} B.,  {Chabrier} G.,    {Valenti} J.~A.,  2008a, \mnras, 384, 77

\bibitem[\protect\citeauthoryear{{Morin}, {Donati}, {Petit}, {Delfosse},
  {Forveille}, {Albert}, {Auri{\`e}re}, {Cabanac}, {Dintrans}, {Fares},
  {Gastine}, {Jardine}, {Ligni{\`e}res}, {Paletou}, {Ramirez Velez} \&
  {Th{\'e}ado}}{{Morin} et~al.}{2008b}]{Morin08b}
{Morin} J.,  {Donati} J.-F.,  {Petit} P.,  {Delfosse} X.,  {Forveille} T.,
  {Albert} L.,  {Auri{\`e}re} M.,  {Cabanac} R.,  {Dintrans} B.,  {Fares} R.,
  {Gastine} T.,  {Jardine} M.~M.,  {Ligni{\`e}res} F.,  {Paletou} F.,  {Ramirez
  Velez} J.~C.,    {Th{\'e}ado} S.,  2008b, \mnras, 390, 567

\bibitem[\protect\citeauthoryear{{Moutou}, {Donati}, {Savalle}, {Hussain},
  {Alecian}, {Bouchy}, {Catala}, {Collier Cameron}, {Udry} \&
  {Vidal-Madjar}}{{Moutou} et~al.}{2007}]{Moutou07}
{Moutou} C.,  {Donati} J.-F.,  {Savalle} R.,  {Hussain} G.,  {Alecian} E.,
  {Bouchy} F.,  {Catala} C.,  {Collier Cameron} A.,  {Udry} S.,
  {Vidal-Madjar} A.,  2007, \aap, 473, 651

\bibitem[\protect\citeauthoryear{{Nidever}, {Marcy}, {Butler}, {Fischer} \&
  {Vogt}}{{Nidever} et~al.}{2002}]{Nidever02}
{Nidever} D.~L.,  {Marcy} G.~W.,  {Butler} R.~P.,  {Fischer} D.~A.,    {Vogt}
  S.~S.,  2002, \apjs, 141, 503

\bibitem[\protect\citeauthoryear{{Noyes}, {Hartmann}, {Baliunas}, {Duncan} \&
  {Vaughan}}{{Noyes} et~al.}{1984}]{Noyes84}
{Noyes} R.~W.,  {Hartmann} L.~W.,  {Baliunas} S.~L.,  {Duncan} D.~K.,
  {Vaughan} A.~H.,  1984, \apj, 279, 763

\bibitem[\protect\citeauthoryear{{Ossendrijver}}{{Ossendrijver}}{2003}]{Ossend%
rijver03}
{Ossendrijver} M.,  2003, in {Pevtsov} A.~A.,  {Uitenbroek} H.,  eds, Current
  Theoretical Models and Future High Resolution Solar Observations: Preparing
  for ATST Vol.~286 of Astronomical Society of the Pacific Conference Series,
  {The Solar Dynamo: A Challenge for Theory and Observations (Invited review)}.
p.~97

\bibitem[\protect\citeauthoryear{{Parker}}{{Parker}}{1955}]{Parker55}
{Parker} E.~N.,  1955, \apj, 122, 293

\bibitem[\protect\citeauthoryear{{Petit}, {Dintrans}, {Solanki}, {Donati},
  {Auri{\`e}re}, {Ligni{\`e}res}, {Morin}, {Paletou}, {Ramirez Velez}, {Catala}
  \& {Fares}}{{Petit} et~al.}{2008}]{Petit08}
{Petit} P.,  {Dintrans} B.,  {Solanki} S.~K.,  {Donati} J.,  {Auri{\`e}re} M.,
  {Ligni{\`e}res} F.,  {Morin} J.,  {Paletou} F.,  {Ramirez Velez} J.,
  {Catala} C.,    {Fares} R.,  2008, \mnras, 388, 80

\bibitem[\protect\citeauthoryear{{Pizzolato}, {Maggio}, {Micela}, {Sciortino}
  \& {Ventura}}{{Pizzolato} et~al.}{2003}]{Pizzolato03}
{Pizzolato} N.,  {Maggio} A.,  {Micela} G.,  {Sciortino} S.,    {Ventura} P.,
  2003, \aap, 397, 147

\bibitem[\protect\citeauthoryear{{Rees} \& {Semel}}{{Rees} \&
  {Semel}}{1979}]{Rees79}
{Rees} D.~E.,  {Semel} M.~D.,  1979, \aap, 74, 1

\bibitem[\protect\citeauthoryear{{Reid}, {Hawley} \& {Gizis}}{{Reid}
  et~al.}{1995}]{Reid95}
{Reid} I.~N.,  {Hawley} S.~L.,    {Gizis} J.~E.,  1995, \aj, 110, 1838

\bibitem[\protect\citeauthoryear{{Reiners} \& {Basri}}{{Reiners} \&
  {Basri}}{2006}]{Reiners06}
{Reiners} A.,  {Basri} G.,  2006, \apj, 644, 497

\bibitem[\protect\citeauthoryear{{Reiners} \& {Basri}}{{Reiners} \&
  {Basri}}{2007}]{Reiners07}
{Reiners} A.,  {Basri} G.,  2007, \apj, 656, 1121

\bibitem[\protect\citeauthoryear{{Reiners} \& {Basri}}{{Reiners} \&
  {Basri}}{2009}]{Reiners09b}
{Reiners} A.,  {Basri} G.,  2009, \aap, 496, 787

\bibitem[\protect\citeauthoryear{{Reiners}, {Basri} \& {Browning}}{{Reiners}
  et~al.}{2009}]{Reiners09}
{Reiners} A.,  {Basri} G.,    {Browning} M.,  2009, \apj, 692, 538

\bibitem[\protect\citeauthoryear{{Ribas}}{{Ribas}}{2006}]{Ribas06}
{Ribas} I.,  2006, \apss, 304, 89

\bibitem[\protect\citeauthoryear{{Saar} \& {Linsky}}{{Saar} \&
  {Linsky}}{1985}]{Saar85}
{Saar} S.~H.,  {Linsky} J.~L.,  1985, \apj, 299, L47

\bibitem[\protect\citeauthoryear{{Schmitt} \& {Liefke}}{{Schmitt} \&
  {Liefke}}{2004}]{Schmitt04}
{Schmitt} J.~H.~M.~M.,  {Liefke} C.,  2004, \aap, 417, 651

\bibitem[\protect\citeauthoryear{{Semel}, {Donati} \& {Rees}}{{Semel}
  et~al.}{1993}]{Semel93}
{Semel} M.,  {Donati} J.-F.,    {Rees} D.~E.,  1993, \aap, 278, 231

\bibitem[\protect\citeauthoryear{{Spiegel} \& {Zahn}}{{Spiegel} \&
  {Zahn}}{1992}]{Spiegel92}
{Spiegel} E.~A.,  {Zahn} J.-P.,  1992, \aap, 265, 106

\bibitem[\protect\citeauthoryear{{Wade}, {Donati}, {Landstreet} \&
  {Shorlin}}{{Wade} et~al.}{2000}]{Wade00}
{Wade} G.~A.,  {Donati} J.-F.,  {Landstreet} J.~D.,    {Shorlin} S.~L.~S.,
  2000, \mnras, 313, 851

\bibitem[\protect\citeauthoryear{{West}, {Hawley}, {Walkowicz}, {Covey},
  {Silvestri}, {Raymond}, {Harris}, {Munn}, {McGehee}, {Ivezi{\'c}} \&
  {Brinkmann}}{{West} et~al.}{2004}]{West04}
{West} A.~A.,  {Hawley} S.~L.,  {Walkowicz} L.~M.,  {Covey} K.~R.,  {Silvestri}
  N.~M.,  {Raymond} S.~N.,  {Harris} H.~C.,  {Munn} J.~A.,  {McGehee} P.~M.,
  {Ivezi{\'c}} {\v Z}.,    {Brinkmann} J.,  2004, \aj, 128, 426

\end{thebibliography}
\bibliographystyle{mn2e}

\appendix
\section{Journal of observation}
\label{sec:app-journal}

\begin{table*}
\caption[]{Detailed journal of observations for GJ~51. Columns 1--7 list the
UT date, the heliocentric Julian date, the UT time, the exposure time, the
peak signal to noise ratio (per 2.6~\kms\ velocity bin) and the rms noise
level (relative to the unpolarised continuum level and per 1.8~\kms\
velocity bin) in the average circular polarisation profile produced by
Least-Squares Deconvolution (see Sec.~2.2). In column 7 we
indicate the longitudinal field computed from Eq.~1. Column 8 lists
the radial velocities (absolute accuracy $0.10~\kms$, internal accuracy
$0.03~\kms$) associated to each exposure. The rotational cycle $E$ from the
ephemeris ${\rm HJD}=2\,453\,950 + 1.02\,E$ is mentioned in column 9.}
\begin{center}
\begin{tabular}{ccccccccc}
\hline
Date & HJD          & UT & $t_{\rm exp}$ & \sn\ 
& $\sigma_{\rm LSD}$ & $B_{\ell}$ & RV & Cycle \\
           & (2,450,000+) & (h:m:s) &  (s) &          &   (\ptt)
& (G) & (\kms) & \\
\hline
2006 & & & & & & \\
Aug 05 & 3953.13879 & 15:15:58 & 4 $\times$ 450.0 & 121 & 12.6 & -1617 $\pm$
150 & -5.24 & 3.077\\
Aug 07 & 3955.09204 & 14:08:29 & 4 $\times$ 550.0 & 132 & 9.3 & -1126 $\pm$
100 & -5.37 & 4.992\\
Aug 08 & 3956.08931 & 14:04:27 & 4 $\times$ 650.0 & 165 & 7.7 & -1105 $\pm$
91 & -5.43 & 5.970\\
Aug 09 & 3957.08253 & 13:54:37 & 4 $\times$ 500.0 & 139 & 9.0 & -782 $\pm$
94 & -5.56 & 6.944\\
Aug 11 & 3959.09703 & 14:15:18 & 4 $\times$ 500.0 & 149 & 8.2 & -673 $\pm$
92 & -5.80 & 8.919\\
Aug 12 & 3960.10164 & 14:21:51 & 4 $\times$ 500.0 & 128 & 10.1 & -702 $\pm$
103 & -5.74 & 9.904\\
2007 & & & & & & \\
Sep 28 & 4371.97896 & 11:22:59 & 4 $\times$ 840.0 & 171 & 6.7 & -1394 $\pm$
96 & -7.14 & 413.705\\
Sep 30 & 4374.02264 & 12:25:53 & 4 $\times$ 840.0 & 179 & 6.6 & -1389 $\pm$
94 & -7.22 & 415.708\\
Oct 01 & 4374.94925 & 10:40:12 & 4 $\times$ 840.0 & 163 & 7.0 & -2005 $\pm$
111 & -6.12 & 416.617\\
Oct 02 & 4375.82543 & 07:41:54 & 4 $\times$ 840.0 & 159 & 7.0 & -1395 $\pm$
101 & -5.57 & 417.476\\
Oct 02 & 4375.94877 & 10:39:31 & 4 $\times$ 840.0 & 174 & 6.5 & -2011 $\pm$
108 & -5.85 & 417.597\\
Oct 02 & 4376.05346 & 13:10:16 & 4 $\times$ 840.0 & 180 & 6.5 & -1522 $\pm$
97 & -7.37 & 417.699\\
Oct 03 & 4376.82249 & 07:37:40 & 4 $\times$ 840.0 & 163 & 6.8 & -1429 $\pm$
104 & -5.52 & 418.453\\
Oct 03 & 4376.94931 & 10:40:18 & 4 $\times$ 900.0 & 195 & 5.7 & -2076 $\pm$
108 & -5.55 & 418.578\\
Oct 03 & 4377.05178 & 13:07:51 & 4 $\times$ 900.0 & 198 & 5.7 & -1693 $\pm$
95 & -6.87 & 418.678\\
2008 & & & & & & \\
Oct 14 & 4753.82296 & 07:38:32 & 4 $\times$ 815.0 & 170 & 7.5 & -1372 $\pm$
90 & -7.16 & 788.062\\
Oct 14 & 4753.86728 & 08:42:22 & 4 $\times$ 815.0 & 183 & 6.9 & -1049 $\pm$
79 & -7.29 & 788.105\\
Oct 14 & 4753.96099 & 10:57:19 & 4 $\times$ 815.0 & 157 & 8.0 & -619 $\pm$
82 & -6.63 & 788.197\\
Oct 14 & 4754.05403 & 13:11:17 & 4 $\times$ 815.0 & 159 & 7.6 & -516 $\pm$
81 & -6.24 & 788.288\\
Oct 17 & 4756.85209 & 08:20:35 & 4 $\times$ 815.0 & 119 & 10.4 & -1525 $\pm$
105 & -6.13 & 791.031\\
Oct 20 & 4759.77630 & 06:31:33 & 4 $\times$ 815.0 & 176 & 7.0 & -1600 $\pm$
84 & -5.63 & 793.898\\
Oct 20 & 4759.85136 & 08:19:38 & 4 $\times$ 815.0 & 159 & 8.0 & -1699 $\pm$
95 & -6.13 & 793.972\\
Oct 20 & 4759.94401 & 10:33:04 & 4 $\times$ 815.0 & 152 & 8.3 & -1521 $\pm$
93 & -6.84 & 794.063\\
Oct 20 & 4760.03093 & 12:38:13 & 4 $\times$ 815.0 & 167 & 7.7 & -1072 $\pm$
85 & -7.34 & 794.148\\
\hline

\label{tab:app-gj51}
\end{tabular}
\end{center}
\end{table*}

\begin{table*}
\caption[]{Detailed journal of observations for GJ~1156.
See Table~\ref{tab:app-gj51} for more details. Data are phased according to the
ephemeris ${\rm HJD}=2\,453\,950 + 0.491\,E$.}
\begin{center}
\begin{tabular}{ccccccccc}
\hline
Date & HJD          & UT & $t_{\rm exp}$ & \sn\ 
& $\sigma_{\rm LSD}$ & $B_{\ell}$ & RV & Cycle \\
           & (2,450,000+) & (h:m:s) &  (s)  &         &   (\ptt)
& (G) & (\kms) & \\
\hline
2007 &&&&&&&&\\
Mar 02 & 4162.05198 & 13:08:00 & 4 $\times$ 600.0 & 157 & 8.5 & 84 $\pm$ 67
& 5.86 & 431.878\\
Mar 03 & 4163.05912 & 13:18:14 & 4 $\times$ 420.0 & 120 & 11.0 & 177 $\pm$
81 & 5.86 & 433.929\\
Mar 04 & 4164.04908 & 13:03:43 & 4 $\times$ 700.0 & 181 & 6.8 & 157 $\pm$ 59
& 5.83 & 435.945\\
Mar 05 & 4165.08932 & 14:01:37 & 4 $\times$ 600.0 & 165 & 7.7 & -15 $\pm$ 65
& 5.80 & 438.064\\
Mar 06 & 4166.03032 & 12:36:37 & 4 $\times$ 500.0 & 143 & 9.3 & 91 $\pm$ 73
& 5.89 & 439.980\\
Mar 07 & 4167.00828 & 12:04:50 & 4 $\times$ 600.0 & 163 & 7.9 & -2 $\pm$ 74
& 6.50 & 441.972\\
2008 &&&&&&&&\\
Jan 20 & 4485.98764 & 11:39:04 & 4 $\times$ 540.0 & 158 & 8.5 & -264 $\pm$
63 & 5.72 & 1091.625\\
Jan 20 & 4486.14332 & 15:23:14 & 4 $\times$ 540.0 & 156 & 8.4 & 148 $\pm$ 61
& 5.95 & 1091.942\\
Jan 21 & 4486.98190 & 11:30:41 & 4 $\times$ 540.0 & 127 & 10.4 & -224 $\pm$
74 & 5.61 & 1093.649\\
Jan 21 & 4487.14392 & 15:23:59 & 4 $\times$ 540.0 & 151 & 8.8 & 26 $\pm$ 65
& 5.95 & 1093.979\\
Jan 22 & 4488.09159 & 14:08:30 & 4 $\times$ 610.0 & 150 & 8.8 & 80 $\pm$ 65
& 5.83 & 1095.910\\
2009 &&&&&&&&\\
Jan 08 & 4840.01516 & 12:20:05 & 4 $\times$ 815.0 & 195 & 6.5 & -187 $\pm$
48 & 5.78 & 1812.658\\
Jan 08 & 4840.10476 & 14:29:06 & 4 $\times$ 815.0 & 187 & 6.5 & 55 $\pm$ 48
& 5.73 & 1812.841\\
Jan 08 & 4840.17094 & 16:04:23 & 4 $\times$ 815.0 & 166 & 8.8 & 168 $\pm$ 67
& 5.83 & 1812.975\\
Jan 09 & 4841.03123 & 12:43:05 & 4 $\times$ 815.0 & 167 & 7.7 & -66 $\pm$ 55
& 5.10 & 1814.728\\
Jan 09 & 4841.09873 & 14:20:17 & 4 $\times$ 815.0 & 188 & 6.8 & 44 $\pm$ 50
& 5.77 & 1814.865\\
Jan 09 & 4841.16680 & 15:58:18 & 4 $\times$ 815.0 & 184 & 6.7 & 88 $\pm$ 49
& 5.97 & 1815.004\\
Jan 10 & 4842.02100 & 12:28:15 & 4 $\times$ 815.0 & 184 & 6.8 & -105 $\pm$
49 & 5.76 & 1816.743\\
Jan 10 & 4842.09687 & 14:17:29 & 4 $\times$ 815.0 & 183 & 6.9 & 81 $\pm$ 51
& 5.73 & 1816.898\\
Jan 10 & 4842.16403 & 15:54:11 & 4 $\times$ 815.0 & 187 & 6.7 & 135 $\pm$ 50
& 5.94 & 1817.035\\
\hline
\label{tab:app-gj1156}
\end{tabular}
\end{center}
\end{table*}

\begin{table*}
\caption[]{Detailed journal of observations for GJ~1245~B.
See Table~\ref{tab:app-gj51} for more details. Data are phased according to the
ephemeris ${\rm HJD}=2\,453\,950 + 0.71\,E$.}
\begin{center}
\begin{tabular}{ccccccccc}
\hline
Date & HJD          & UT & $t_{\rm exp}$ & \sn\ 
& $\sigma_{\rm LSD}$ & $B_{\ell}$ & RV & Cycle \\
           & (2,450,000+) & (h:m:s) &  (s)  &         &   (\ptt)
& (G) & (\kms) & \\
\hline
2006 & & & & & & & & \\
Aug 05 & 3952.96705 & 11:05:26 & 4 $\times$ 630.0 & 166 & 8.6 & -251 $\pm$
33 & 5.29 & 4.179\\ 
Aug 08 & 3955.95886 & 10:53:38 & 4 $\times$ 700.0 & 191 & 7.1 & -79 $\pm$ 26
& 5.40 & 8.393\\ 
Aug 09 & 3956.96600 & 11:03:55 & 4 $\times$ 600.0 & 175 & 7.7 & 186 $\pm$ 28
& 5.56 & 9.811\\ 
Aug 10 & 3957.95942 & 10:54:27 & 4 $\times$ 700.0 & 185 & 7.5 & -179 $\pm$
27 & 5.28 & 11.210\\ 
Aug 11 & 3958.96340 & 11:00:11 & 4 $\times$ 700.0 & 185 & 7.5 & 145 $\pm$ 27
& 5.52 & 12.625\\ 
Aug 12 & 3959.96518 & 11:02:45 & 4 $\times$ 700.0 & 158 & 8.8 & -133 $\pm$
29 & 5.48 & 14.035\\ 
2007 & & & & & & & & \\
Sep 30 & 4373.89061 & 09:17:56 & 4 $\times$ 840.0 & 197 & 6.8 & -107 $\pm$
27 & 5.30 & 597.029\\ 
Oct 01 & 4374.86550 & 08:41:51 & 4 $\times$ 840.0 & 199 & 6.6 & 148 $\pm$ 27
& 5.47 & 598.402\\ 
Oct 02 & 4375.72600 & 05:21:04 & 4 $\times$ 840.0 & 214 & 6.4 & 63 $\pm$ 26
& 5.51 & 599.614\\ 
Oct 02 & 4375.86847 & 08:46:14 & 4 $\times$ 840.0 & 182 & 7.4 & -136 $\pm$
29 & 5.26 & 599.815\\ 
Oct 03 & 4376.72595 & 05:21:05 & 4 $\times$ 900.0 & 226 & 5.7 & -179 $\pm$
25 & 5.27 & 601.022\\ 
Oct 03 & 4376.86630 & 08:43:13 & 4 $\times$ 900.0 & 218 & 6.0 & 108 $\pm$ 26
& 5.46 & 601.220\\ 
2008 & & & & & & & & \\
Aug 20 & 4698.75329 & 05:57:45 & 4 $\times$ 715.0 & 187 & 7.3 & -44 $\pm$ 25
& 5.38 & 1054.582\\ 
Aug 20 & 4698.84202 & 08:05:31 & 4 $\times$ 715.0 & 191 & 7.3 & -64 $\pm$ 26
& 5.34 & 1054.707\\ 
Aug 20 & 4698.99653 & 11:48:01 & 4 $\times$ 715.0 & 171 & 8.2 & 45 $\pm$ 28
& 5.52 & 1054.925\\ 
Aug 21 & 4699.75240 & 05:56:29 & 4 $\times$ 715.0 & 194 & 7.2 & 56 $\pm$ 25
& 5.52 & 1055.989\\ 
Aug 21 & 4699.87910 & 08:58:56 & 4 $\times$ 715.0 & 186 & 7.2 & 26 $\pm$ 26
& 5.53 & 1056.168\\ 
Aug 21 & 4699.99952 & 11:52:21 & 4 $\times$ 715.0 & 176 & 8.0 & 70 $\pm$ 27
& 5.60 & 1056.337\\ 
Aug 22 & 4700.74540 & 05:46:26 & 4 $\times$ 715.0 & 150 & 9.3 & 74 $\pm$ 32
& 5.44 & 1057.388\\ 
Aug 22 & 4700.87247 & 08:49:25 & 4 $\times$ 715.0 & 153 & 9.3 & -138 $\pm$
31 & 5.29 & 1057.567\\ 
Aug 22 & 4700.99137 & 11:40:38 & 4 $\times$ 715.0 & 151 & 9.6 & -47 $\pm$ 32
& 5.45 & 1057.734\\ 
Aug 22 & 4701.02897 & 12:34:46 & 4 $\times$ 715.0 & 138 & 10.1 & -41 $\pm$
33 & 5.48 & 1057.787\\ 
\hline
\label{tab:app-gj1245b}
\end{tabular}
\end{center}
\end{table*}

\begin{table*}
\caption[]{Detailed journal of observations for WX~UMa.
See Table~\ref{tab:app-gj51} for more details. Data are phased according to the
ephemeris ${\rm HJD}=2\,453\,850 + 0.78\,E$.}
\begin{center}
\begin{tabular}{ccccccccc}
\hline
Date & HJD          & UT & $t_{\rm exp}$ & \sn\ 
& $\sigma_{\rm LSD}$ & $B_{\ell}$ & RV & Cycle \\
           & (2,450,000+) & (h:m:s) &  (s) &          &   (\ptt)
& (G) & (\kms) & \\
\hline
2006 & & & & & & \\ 
Jun 09 & 3895.77128 & 06:31:12 & 4 $\times$ 600.0 & 86 & 16.1 & -1128 $\pm$
213 & 70.39 & 0.476\\
Jun 09 & 3895.83862 & 08:08:10 & 4 $\times$ 600.0 & 67 & 19.8 & -914 $\pm$
207 & 70.60 & 0.562\\
Jun 10 & 3896.77760 & 06:40:22 & 4 $\times$ 600.0 & 137 & 11.9 & -1816 $\pm$
162 & 70.59 & 1.766\\
Jun 10 & 3896.84666 & 08:19:49 & 4 $\times$ 600.0 & 120 & 11.0 & -2150 $\pm$
198 & 70.78 & 1.855\\
Jun 11 & 3897.76780 & 06:26:19 & 4 $\times$ 600.0 & 142 & 9.6 & -2053 $\pm$
174 & 68.97 & 3.036\\
Jun 11 & 3897.83460 & 08:02:31 & 4 $\times$ 600.0 & 130 & 10.4 & -1731 $\pm$
119 & 70.21 & 3.121\\
Jun 12 & 3898.74596 & 05:54:56 & 4 $\times$ 600.0 & 122 & 11.2 & -1147 $\pm$
101 & 70.03 & 4.290\\
Jun 12 & 3898.81286 & 07:31:16 & 4 $\times$ 600.0 & 120 & 11.2 & -1110 $\pm$
103 & 70.43 & 4.375\\
2007 & & & & & & \\ 
Mar 02 & 4161.97902 & 11:26:01 & 4 $\times$ 700.0 & 143 & 8.8 & -1825 $\pm$
116 & 69.91 & 341.768\\
Mar 03 & 4162.99705 & 11:51:58 & 4 $\times$ 600.0 & 115 & 11.8 & -1340 $\pm$
124 & 70.04 & 343.073\\
Mar 04 & 4163.97675 & 11:22:44 & 4 $\times$ 700.0 & 154 & 8.4 & -1581 $\pm$
105 & 69.93 & 344.329\\
Mar 05 & 4165.01491 & 12:17:42 & 4 $\times$ 700.0 & 152 & 8.4 & -2357 $\pm$
124 & 69.90 & 345.660\\
Mar 07 & 4166.94370 & 10:35:10 & 4 $\times$ 560.0 & 137 & 10.0 & -1265 $\pm$
107 & 70.03 & 348.133\\
Mar 08 & 4168.03986 & 12:53:39 & 4 $\times$ 600.0 & 139 & 9.3 & -2175 $\pm$
125 & 69.92 & 349.538\\
2008 & & & & & & \\ 
Jan 19 & 4484.92606 & 10:10:43 & 4 $\times$ 560.0 & 63 & 21.4 & -1416 $\pm$
206 & 70.56 & 755.803\\
Jan 20 & 4486.11357 & 14:40:40 & 4 $\times$ 500.0 & 116 & 12.0 & -1725 $\pm$
115 & 70.01 & 757.325\\
Jan 21 & 4486.92292 & 10:06:06 & 4 $\times$ 540.0 & 123 & 11.2 & -2135 $\pm$
123 & 69.97 & 758.363\\
Jan 21 & 4487.11353 & 14:40:34 & 4 $\times$ 520.0 & 129 & 10.4 & -1968 $\pm$
114 & 70.05 & 758.607\\
2009 & & & & & & \\ 
Jan 11 & 4843.01564 & 12:20:02 & 4 $\times$ 715.0 & 116 & 11.9 & -1147 $\pm$
108 & 70.24 & 1214.892\\
Jan 11 & 4843.11846 & 14:48:06 & 4 $\times$ 715.0 & 113 & 12.5 & -1151 $\pm$
114 & 70.02 & 1215.024\\
Jan 12 & 4843.94502 & 10:38:18 & 4 $\times$ 715.0 & 132 & 10.0 & -1670 $\pm$
100 & 69.86 & 1216.083\\
Jan 12 & 4844.04478 & 13:01:57 & 4 $\times$ 715.0 & 113 & 12.2 & -1707 $\pm$
110 & 70.04 & 1216.211\\
Jan 12 & 4844.14627 & 15:28:05 & 4 $\times$ 715.0 & 133 & 10.0 & -1772 $\pm$
98 & 69.91 & 1216.341\\
Jan 13 & 4844.94269 & 10:34:54 & 4 $\times$ 715.0 & 133 & 10.0 & -1638 $\pm$
101 & 69.87 & 1217.362\\
Jan 13 & 4845.09115 & 14:08:40 & 4 $\times$ 715.0 & 146 & 9.0 & -1864 $\pm$
101 & 69.76 & 1217.553\\
Jan 13 & 4845.16697 & 15:57:51 & 4 $\times$ 715.0 & 153 & 8.4 & -1735 $\pm$
101 & 69.62 & 1217.650\\
Jan 14 & 4845.98095 & 11:29:57 & 4 $\times$ 715.0 & 145 & 9.2 & -1274 $\pm$
90 & 69.50 & 1218.694\\
Jan 14 & 4846.08129 & 13:54:25 & 4 $\times$ 715.0 & 163 & 8.0 & -1366 $\pm$
86 & 69.57 & 1218.822\\
Jan 14 & 4846.17078 & 16:03:17 & 4 $\times$ 715.0 & 158 & 8.4 & -1130 $\pm$
89 & 69.75 & 1218.937\\
\hline
\label{tab:app-gj412b}
\end{tabular}
\end{center}
\end{table*}

\begin{table*}
\caption[]{Detailed journal of observations for DX~Cnc.
See Table~\ref{tab:app-gj51} for more details. Data are phased according to the
ephemeris ${\rm HJD}=2\,453\,950 + 0.46\,E$.}
\begin{center}
\begin{tabular}{ccccccccc}
\hline
Date & HJD          & UT & $t_{\rm exp}$ & \sn\ 
& $\sigma_{\rm LSD}$ & $B_{\ell}$ & RV & Cycle \\
           & (2,450,000+) & (h:m:s) &  (s) &          &   (\ptt)
& (G) & (\kms) & \\
\hline
2007 & & & & & & \\ 
Mar 02 & 4161.94084 & 10:30:16 & 4 $\times$ 700.0 & 165 & 8.8 & 145 $\pm$ 49
& 10.60 & 460.741\\
Mar 04 & 4163.93885 & 10:27:33 & 4 $\times$ 700.0 & 170 & 8.6 & 158 $\pm$ 47
& 10.52 & 465.084\\
Mar 05 & 4164.94533 & 10:36:57 & 4 $\times$ 700.0 & 179 & 8.3 & -13 $\pm$ 46
& 10.65 & 467.272\\
Mar 06 & 4165.93747 & 10:25:43 & 4 $\times$ 560.0 & 119 & 12.3 & 209 $\pm$
65 & 10.54 & 469.429\\
Mar 07 & 4166.91257 & 09:49:56 & 4 $\times$ 560.0 & 139 & 11.1 & 161 $\pm$
55 & 10.46 & 471.549\\
2008 & & & & & & \\ 
Jan 18 & 4483.86547 & 08:40:11 & 4 $\times$ 520.0 & 116 & 12.7 & 81 $\pm$ 61
& 10.85 & 1160.577\\
Jan 19 & 4484.86323 & 08:36:58 & 4 $\times$ 560.0 & 141 & 10.6 & 34 $\pm$ 52
& 10.88 & 1162.746\\
Jan 19 & 4485.04885 & 13:04:15 & 4 $\times$ 570.0 & 90 & 16.7 & 86 $\pm$ 79
& 11.00 & 1163.150\\
Jan 20 & 4486.05051 & 13:06:39 & 4 $\times$ 600.0 & 161 & 9.4 & 11 $\pm$ 46
& 10.70 & 1165.327\\
Jan 21 & 4486.86360 & 08:37:30 & 4 $\times$ 560.0 & 138 & 10.9 & 135 $\pm$
54 & 10.43 & 1167.095\\
Jan 21 & 4487.04831 & 13:03:28 & 4 $\times$ 700.0 & 160 & 9.8 & 131 $\pm$ 58
& 9.94 & 1167.496\\
Jan 22 & 4487.98223 & 11:28:19 & 4 $\times$ 610.0 & 152 & 9.5 & 166 $\pm$ 54
& 9.29 & 1169.527\\
2009 & & & & & & \\ 
Feb 13 & 4875.88471 & 09:08:28 & 4 $\times$ 800.0 & 122 & 12.3 & 86 $\pm$ 56
& 10.76 & 2012.793\\
Feb 14 & 4876.87737 & 08:57:56 & 4 $\times$ 800.0 & 137 & 11.6 & 149 $\pm$
52 & 10.80 & 2014.951\\
Feb 14 & 4877.00055 & 11:55:19 & 4 $\times$ 800.0 & 106 & 14.8 & -4 $\pm$ 65
& 10.68 & 2015.219\\
Feb 15 & 4877.75183 & 05:57:12 & 4 $\times$ 800.0 & 164 & 9.2 & 48 $\pm$ 43
& 10.66 & 2016.852\\
Feb 15 & 4877.87789 & 08:58:44 & 4 $\times$ 800.0 & 180 & 8.5 & 13 $\pm$ 40
& 10.49 & 2017.126\\
Feb 15 & 4878.01424 & 12:15:05 & 4 $\times$ 800.0 & 157 & 9.4 & 110 $\pm$ 44
& 10.58 & 2017.422\\
Feb 16 & 4878.76253 & 06:12:40 & 4 $\times$ 800.0 & 187 & 7.8 & 54 $\pm$ 37
& 10.61 & 2019.049\\
Feb 16 & 4878.88556 & 09:09:50 & 4 $\times$ 800.0 & 185 & 7.9 & 78 $\pm$ 36
& 10.78 & 2019.316\\
Feb 16 & 4879.00624 & 12:03:37 & 4 $\times$ 800.0 & 177 & 8.4 & 71 $\pm$ 38
& 10.71 & 2019.579\\
\hline
\label{tab:app-gj1111}
\end{tabular}
\end{center}
\end{table*}

\begin{table*}
\caption[]{Detailed journal of observations for GJ~3622.
See Table~\ref{tab:app-gj51} for more details. Data are phased according to the
ephemeris ${\rm HJD}=2\,453\,950 + 1.5\,E$.}
\begin{center}
\begin{tabular}{ccccccccc}
\hline
Date & HJD          & UT & $t_{\rm exp}$ & \sn\ 
& $\sigma_{\rm LSD}$ & $B_{\ell}$ & RV & Cycle \\
           & (2,450,000+) & (h:m:s) &  (s)  &         &   (\ptt)
& (G) & (\kms) & \\
\hline
2008 &&&&&&&&\\
Mar 20 & 4545.85137 & 08:17:36 & 4 $\times$ 990.0 & 159 & 9.3 & -50
$\pm$ 17 & 2.20 & 397.234\\
Mar 21 & 4546.94418 & 10:31:17 & 4 $\times$ 990.0 & 162 & 9.1 & -2 $\pm$
18 & 1.77 & 397.963\\
Mar 22 & 4548.00749 & 12:02:30 & 4 $\times$ 990.0 & 154 & 9.5 & -20
$\pm$ 18 & 1.88 & 398.672\\
Mar 23 & 4548.92606 & 10:05:17 & 4 $\times$ 990.0 & 156 & 9.4 & -72
$\pm$ 17 & 2.33 & 399.284\\
Mar 25 & 4550.82345 & 07:37:37 & 4 $\times$ 990.0 & 152 & 9.8 & -60
$\pm$ 18 & 2.57 & 400.549\\
Mar 26 & 4551.82107 & 07:34:14 & 4 $\times$ 990.0 & 167 & 8.9 & -56
$\pm$ 16 & 2.53 & 401.214\\
Mar 27 & 4552.91578 & 09:50:41 & 4 $\times$ 990.0 & 150 & 10.1 & 7 $\pm$
18 & 2.46 & 401.944\\
Mar 28 & 4553.86120 & 08:32:08 & 4 $\times$ 990.0 & 128 & 11.1 & -5
$\pm$ 20 & 2.41 & 402.574\\
2009 &&&&&&&&\\
Feb 13 & 4875.93524 & 10:18:39 & 4 $\times$ 900.0 & 101 & 15.0 & -22
$\pm$ 26 & 2.35 & 617.290\\
Feb 15 & 4877.93049 & 10:11:42 & 4 $\times$ 900.0 & 134 & 11.0 & -73
$\pm$ 20 & 2.44 & 618.620\\
Feb 16 & 4878.93305 & 10:15:21 & 4 $\times$ 900.0 & 151 & 10.0 & -42
$\pm$ 18 & 2.32 & 619.289\\
Feb 16 & 4879.06535 & 13:25:51 & 4 $\times$ 900.0 & 135 & 11.3 & -15
$\pm$ 19 & 2.37 & 619.377\\
Feb 17 & 4879.91581 & 09:50:29 & 4 $\times$ 900.0 & 162 & 9.3 & 13 $\pm$
17 & 2.32 & 619.944\\
Feb 17 & 4880.05123 & 13:05:29 & 4 $\times$ 900.0 & 151 & 10.0 & -18
$\pm$ 18 & 2.42 & 620.034\\
\hline
\label{tab:app-gj3622}
\end{tabular}
\end{center}
\end{table*}

\begin{table*}
\caption[]{Detailed journal of observations for GJ~1154~A.
See Table~\ref{tab:app-gj51} for more details. The rotational cycle is not
indicated since no rotation period could be derived for this star.}
\begin{center}
\begin{tabular}{cccccccc}
\hline
Date & HJD          & UT & $t_{\rm exp}$ & \sn\ 
& $\sigma_{\rm LSD}$ & $B_{\ell}$ & RV \\
           & (2,450,000+) & (h:m:s) &  (s) &          &   (\ptt)
& (G) & (\kms) \\
\hline
2007 &&&&&&&\\
Mar 02 & 4162.01888 & 12:19:34 & 4 $\times$ 600.0 & 150 & 7.9 & -825 $\pm$
63 & -12.76 \\
Mar 03 & 4163.03213 & 12:38:35 & 4 $\times$ 500.0 & 118 & 10.5 & -672 $\pm$
69 & -12.79 \\
Mar 04 & 4164.01521 & 12:14:11 & 4 $\times$ 600.0 & 167 & 7.0 & -718 $\pm$
55 & -12.77 \\
Mar 05 & 4165.05486 & 13:11:23 & 4 $\times$ 600.0 & 147 & 7.2 & -612 $\pm$
55 & -13.07 \\
Mar 06 & 4166.00262 & 11:55:57 & 4 $\times$ 500.0 & 152 & 8.1 & -660 $\pm$
59 & -12.82 \\
Mar 07 & 4166.97810 & 11:20:36 & 4 $\times$ 500.0 & 140 & 8.6 & -800 $\pm$
62 & -12.78 \\
2008 &&&&&&&\\
Jan 19 & 4485.01555 & 12:18:53 & 4 $\times$ 540.0 & 86 & 15.9 & -576 $\pm$
85 & -12.81 \\
Jan 20 & 4486.01710 & 12:20:59 & 4 $\times$ 500.0 & 147 & 8.2 & -776 $\pm$
57 & -12.89 \\
Jan 21 & 4487.01179 & 12:13:12 & 4 $\times$ 540.0 & 130 & 9.3 & -721 $\pm$
61 & -12.76 \\
Jan 22 & 4488.01494 & 12:17:37 & 4 $\times$ 610.0 & 154 & 8.3 & -727 $\pm$
61 & -13.22 \\
\hline
\label{tab:app-gj1154a}
\end{tabular}
\end{center}
\end{table*}

\begin{table*}
\caption[]{Detailed journal of observations for GJ~1224.
See Table~\ref{tab:app-gj51} for more details. The rotational cycle is not
indicated since no rotation period could be derived for this star.}
\begin{center}
\begin{tabular}{cccccccc}
\hline
Date & HJD          & UT & $t_{\rm exp}$ & \sn\ 
& $\sigma_{\rm LSD}$ & $B_{\ell}$ & RV \\
           & (2,450,000+) & (h:m:s) &  (s) &          &   (\ptt)
& (G) & (\kms) \\
\hline
2008 &&&&&&&\\
Jun 25 & 4643.02829 & 12:33:21 & 4 $\times$ 800.0 & 176 & 6.2 & -535 $\pm$
36 & -32.75 \\
Jun 28 & 4645.88511 & 09:07:15 & 4 $\times$ 800.0 & 51 & 19.7 & -546 $\pm$
95 & -32.69 \\
Jun 28 & 4645.92645 & 10:06:46 & 4 $\times$ 800.0 & 86 & 14.1 & -631 $\pm$
67 & -32.62 \\
Jun 29 & 4646.94450 & 10:32:47 & 4 $\times$ 800.0 & 147 & 7.9 & -550 $\pm$
42 & -32.67 \\
Jul 02 & 4649.94447 & 10:32:50 & 4 $\times$ 800.0 & 155 & 7.3 & -632 $\pm$
41 & -32.72 \\
Jul 03 & 4650.91640 & 09:52:27 & 4 $\times$ 800.0 & 171 & 6.6 & -550 $\pm$
37 & -32.69 \\
Jul 04 & 4651.85383 & 08:22:23 & 4 $\times$ 800.0 & 171 & 6.3 & -556 $\pm$
37 & -32.64 \\
Jul 05 & 4652.86038 & 08:31:51 & 4 $\times$ 800.0 & 176 & 6.1 & -617 $\pm$
37 & -32.69 \\
Jul 06 & 4653.83070 & 07:49:09 & 4 $\times$ 800.0 & 185 & 5.8 & -545 $\pm$
34 & -32.64 \\
Jul 07 & 4654.89004 & 09:14:39 & 4 $\times$ 800.0 & 166 & 6.6 & -573 $\pm$
38 & -32.65 \\
Jul 08 & 4655.84785 & 08:13:57 & 4 $\times$ 800.0 & 172 & 6.4 & -523 $\pm$
37 & -32.63 \\
Jul 09 & 4656.96336 & 11:00:20 & 4 $\times$ 800.0 & 172 & 6.4 & -500 $\pm$
40 & -32.72 \\
\hline
\label{tab:app-gj1224}
\end{tabular}
\end{center}
\end{table*}

\begin{table*}
\caption[]{Detailed journal of observations for CN~Leo.
See Table~\ref{tab:app-gj51} for more details. The rotational cycle is not
indicated since no rotation period could be derived for this star.}
\begin{center}
\begin{tabular}{cccccccc}
\hline
Date & HJD          & UT & $t_{\rm exp}$ & \sn\ 
& $\sigma_{\rm LSD}$ & $B_{\ell}$ & RV \\
           & (2,450,000+) & (h:m:s) &  (s) &          &   (\ptt)
& (G) & (\kms) \\
\hline
2008 &&&&&&&\\
Mar 21 & 4546.99345 & 11:43:06 & 4 $\times$ 800.0 & 179 & 5.6 & -651 $\pm$
45 & 19.66 \\
Mar 22 & 4547.91616 & 09:51:50 & 4 $\times$ 800.0 & 172 & 6.7 & -635 $\pm$
48 & 19.68 \\
Mar 22 & 4547.95840 & 10:52:40 & 4 $\times$ 800.0 & 186 & 6.1 & -708 $\pm$
46 & 19.63 \\
Mar 23 & 4548.87485 & 08:52:23 & 4 $\times$ 800.0 & 245 & 4.5 & -772 $\pm$
45 & 19.55 \\
\hline
\label{tab:app-cnleo}
\end{tabular}
\end{center}
\end{table*}

\begin{table*}
\caption[]{Detailed journal of observations for VB~8.
See Table~\ref{tab:app-gj51} for more details. The rotational cycle is not
indicated since no rotation period could be derived for this star.}
\begin{center}
\begin{tabular}{cccccccc}
\hline
Date & HJD          & UT & $t_{\rm exp}$ & \sn\ 
& $\sigma_{\rm LSD}$ & $B_{\ell}$ & RV \\
           & (2,450,000+) & (h:m:s) &  (s) &          &   (\ptt)
& (G) & (\kms) \\
\hline
2009 &&&&&&&\\
May 03 & 4954.89250 & 09:18:10 & 4 $\times$ 1154.0 & 104 & 16.5 & 24
$\pm$ 29 & 15.41\\
May 03 & 4954.99885 & 11:51:19 & 4 $\times$ 1154.0 & 107 & 15.8 & -15
$\pm$ 28 & 15.24\\
May 03 & 4955.10357 & 14:22:06 & 4 $\times$ 1154.0 & 107 & 15.7 & -52
$\pm$ 28 & 15.39\\
May 04 & 4955.90408 & 09:34:47 & 4 $\times$ 1154.0 & 83 & 20.0 & 113
$\pm$ 36 & 15.63\\
May 04 & 4956.00991 & 12:07:11 & 4 $\times$ 1154.0 & 94 & 17.9 & 49
$\pm$ 32 & 15.41\\
May 04 & 4956.11013 & 14:31:29 & 4 $\times$ 1154.0 & 98 & 17.0 & -25
$\pm$ 30 & 15.31\\
May 05 & 4956.90261 & 09:32:37 & 4 $\times$ 1154.0 & 99 & 16.9 & 77
$\pm$ 30 & 15.31\\
May 05 & 4957.00469 & 11:59:36 & 4 $\times$ 1154.0 & 102 & 16.2 & 87
$\pm$ 29 & 15.32\\
May 05 & 4957.11128 & 14:33:05 & 4 $\times$ 1154.0 & 103 & 16.3 & 5
$\pm$ 29 & 15.51\\
\hline
\label{tab:app-vb8}
\end{tabular}
\end{center}
\end{table*}

\begin{table*}
\caption[]{Detailed journal of observations for VB~10.
See Table~\ref{tab:app-gj51} for more details. The rotational cycle is
given for the most probable period according to the ephemeris
${\rm HJD}=2\,455\,000 + 0.69\,E$.}
\begin{center}
\begin{tabular}{ccccccccc}
\hline
Date & HJD          & UT & $t_{\rm exp}$ & \sn\ 
& $\sigma_{\rm LSD}$ & $B_{\ell}$ & RV & Cycle\\
           & (2,450,000+) & (h:m:s) &  (s) &          &   (\ptt)
& (G) & (\kms) &\\
\hline
2009 &&&&&&&\\
Jul 01 & 5013.96290 & 10:58:52 & 4 $\times$ 1154.0 & 68 & 25.7 & 94 $\pm$ 65
& 36.16 & 20.236\\
Jul 01 & 5014.01972 & 12:20:41 & 4 $\times$ 1154.0 & 77 & 23.0 & 92 $\pm$ 57
& 36.25 & 20.318\\
Jul 01 & 5014.07652 & 13:42:29 & 4 $\times$ 1154.0 & 76 & 23.0 & 145 $\pm$
57 & 36.03 & 20.401\\
Jul 02 & 5014.95765 & 10:51:17 & 4 $\times$ 1154.0 & 71 & 24.2 & 16 $\pm$ 59
& 36.25 & 21.678\\
Jul 02 & 5015.01489 & 12:13:42 & 4 $\times$ 1154.0 & 74 & 23.4 & 68 $\pm$ 57
& 36.37 & 21.761\\
Jul 02 & 5015.10842 & 14:28:23 & 4 $\times$ 1154.0 & 80 & 22.5 & 93 $\pm$ 58
& 36.13 & 21.896\\
Jul 03 & 5015.94279 & 10:29:52 & 4 $\times$ 1154.0 & 74 & 23.1 & -69 $\pm$
59 & 36.27 & 23.105\\
Jul 03 & 5016.04819 & 13:01:38 & 4 $\times$ 1154.0 & 75 & 23.6 & -4 $\pm$ 56
& 36.51 & 23.258\\
Jul 03 & 5016.10641 & 14:25:28 & 4 $\times$ 1154.0 & 74 & 23.5 & 92 $\pm$ 61
& 36.13 & 23.343\\
\hline
\label{tab:app-vb10}
\end{tabular}
\end{center}
\end{table*}

\end{document}